\documentclass[a4paper,11pt,twoside]{article}
\usepackage{rotating, amsmath, amsfonts, amssymb, mathrsfs, theorem, calc, graphicx, color, enumerate, tabularx, bbm, mathtools, dashrule, multirow, makecell, geometry,fancyhdr}
\usepackage[UKenglish]{babel}
\usepackage[T1]{fontenc}
\usepackage[round,longnamesfirst]{natbib}
\definecolor{dblue}{rgb}{0,0,0.3}
\usepackage{hyperref} 
\hypersetup{
  colorlinks=true,
  linkcolor=dblue,
  anchorcolor=dblue,
  citecolor=dblue,
  filecolor=dblue,
  menucolor=dblue,
  urlcolor=dblue,
  pdftitle={NMGKDFEv3},
  pdfauthor={Vitaliy Oryshchenko},
  pdfcreator={\LaTeX\ with package \flqq hyperref\frqq}, 
  final=true,
  breaklinks=true
  }
\newcommand{\abs}[1]{\left\lvert#1\right\rvert}
\newcommand{\I}{\mathbbm{1}}           
\newcommand{\Kummer}{\prescript{}{1}{F}^{}_{1}}
\DeclareMathOperator{\E}{E}
\DeclareMathOperator{\Var}{Var}
\DeclareMathOperator{\MISE}{MISE}
\DeclareMathOperator{\ISB}{ISB}
\DeclareMathOperator{\IVar}{IV}
\DeclareMathOperator{\ISE}{ISE}
\DeclareMathOperator{\IQR}{IQR}
\DeclareMathOperator{\OF}{OF}
\DeclareMathOperator{\sinc}{sinc}
\DeclareMathOperator{\Si}{Si}
\theoremstyle{plain} 
\newtheorem{theorem}{Theorem} 
\numberwithin{equation}{section}
\newcounter{ContdTable}
\begin{document}

\title{Exact mean integrated squared error and bandwidth selection for kernel distribution function estimators}
\author{Vitaliy Oryshchenko\thanks{Address for correspondence: 2.068 Arthur Lewis Building, Department of Economics, School of Social Sciences, University of Manchester, Oxford Road, Manchester M13 9PL, United Kingdom. E-mail: \href{mailto:v.oryshchenko@cantab.net}{v.oryshchenko@cantab.net}.}\\ University of Manchester }
\date{\today}

\maketitle

\begin{abstract} 
An exact, closed form, and easy to compute expression for the mean integrated squared error (MISE) of a kernel estimator of a normal mixture cumulative distribution function is derived for the class of arbitrary order Gaussian-based kernels. 
Comparisons are made with MISE of the empirical distribution function, the infeasible minimum MISE of kernel estimators, and the asymptotically optimal second order uniform kernel. The results afford straightforward extensions to other classes of kernel functions and distributions. The analysis also offers a guide on when to use higher order kernels in distribution function estimation.

A simple plug-in method of simultaneously selecting the optimal bandwidth and kernel order is proposed based on a non-asymptotic approximation of the unknown distribution by a normal mixture. A simulation study shows that the method works well in finite samples, thus providing a viable alternative to existing bandwidth selection procedures. 

\vspace*{0.3\baselineskip}
\noindent{\bf Key Words and Phrases:} finite sample, Gaussian-based kernel, uniform kernel, normal mixture, plug-in rule, smoothing.

\vspace*{0.3\baselineskip}
\noindent{\bf MSC 2010 subject classifications}: 62G05
\end{abstract}

\section{Introduction}
Let $X_{1},\ldots,X_{n}$ be a sample of independent, identically distributed random variables with an absolutely continuous distribution function (cdf) $F$ and density $f$. The kernel estimator of $F$ (KDFE) at a point $x$ is 
\begin{equation}\label{Eq.KDFE.def}
\widehat{F}(x;h)=\frac{1}{n}\sum_{i=1}^{n}K\left((x-X_{i})/h\right), 
\end{equation}
where $K(z)=\int_{-\infty}^{z}k(x)\mathrm{d}x$ is the kernel, with $k$ being symmetric about the origin and integrating to unity, and $h=h_{n}\geq0$ is the bandwidth sequence which determines the degree of smoothing \citep{nadaraya1964b,watson1964}.  
A kernel $k$ satisfying $\mu_{0}(k)=1$, $\mu_{j}(k)=0$ for $j=1,\ldots,2r-1$, and $\mu_{2r}(k)<\infty$, where $\mu_{j}(k)=\int_{-\infty}^{\infty}x^{j}k(x)\mathrm{d}x$, is called a $(2r)^{th}$ order kernel. The empirical distribution function (EDF), $F_{n}(x)=n^{-1}\sum_{i=1}^{n}\I\{X_{i}\leq x\}$, coincides with $\widehat{F}(x;0)$ except at $x=X_{i}$, $i=1,\ldots,n$. 

It is well known that under very mild conditions $\widehat{F}$ is a uniformly strongly consistent and asymptotically normal estimator of $F$ \citep{nadaraya1964b,watson1964,yamato1973}. Relative to EDF, KDFE is an asymptotically more efficient estimator in the mean integrated squared error and Hodges-Lehmann sense  \citep{reiss1981,falk1983,swanepoel1988}. Smoothness of the kernel estimates and the reduction in MISE are the two main reasons to prefer KDFE. It is also reasonable to expect that replacing the EDF with the KDFE will improve performance of the resultant estimators and test statistics. For example, this is the case for quantile estimation \citep{azzalini1981,falk1984}. 

The mean integrated squared error, $\MISE[\widehat{F}(\cdot;h)]=\E[\int_{-\infty}^{\infty}[\widehat{F}(x;h)-F(x)]^{2}\mathrm{d}x]$, is a commonly used global measure of performance of KDFE. The optimal bandwidth, $h^{\ast}$, is then defined as the value minimising $\MISE[\widehat{F}(\cdot;h)]$ over $h\geq0$. Since $F(x)$ and hence $\MISE[\widehat{F}(\cdot;h)]$ are of course unknown, existing feasible bandwidth selection methods rely either on cross-validation \citep{bowman1998} or an asymptotic approximation to MISE. The latter class includes simple rule of thumb as well as single- or multistage plug-in estimators; see, e.g., \citet{altman1995}, \citet{polansky2000}, and \citet{tenreiro2006}. However, these may underperform in situations where the asymptotic approximation to MISE is poor in finite samples. 

Exact MISE expressions derived in this paper can be used to define an alternative `plug-in' method of bandwidth selection which has the advantage that the optimal kernel order can be selected simultaneously (Section \ref{Sec:bandwidth.selection}). The proposed method is based on the non-asymptotic approximation of the unknown distribution by a finite normal mixture, but does not involve asymptotic approximations to MISE. It may thus outcompete the methods based on asymptotic approximations in small samples provided the normal mixture approximation to the underlying cdf is good. 

To obtain exact finite sample MISE expressions of practical interest, it is necessary to restrict attention to specific classes of distributions and kernel functions. For the case of density estimation, exact MISE has been derived in \citet{fryer1976} for the normal distribution and Gaussian kernel, and later extended to the class of finite normal mixture distributions and Gaussian-based kernels of \citet{wand1990} in \citet{marron1992}, hereinafter MW, and to the class of polynomial kernels in \citet{hansen2005}. 
This paper extends the results of MW to kernel estimation of distribution functions, restricting attention to finite normal mixture (NM) distributions and arbitrary order Gaussian-based kernels. The latter are convenient as the convolutions with normal density have explicit closed form expressions. Gaussian-based kernels also result in KDFE with MISE very close to the infeasible minimum MISE for distributions close to normal and generally perform remarkably well in very small and large samples. 

The Gaussian-based kernels of \citet{wand1990} considered in this paper arise as asymptotically optimal smooth, i.e., infinitely continuously differentiable, kernels for the problem of minimising the integrated squared $\nu^{th}$ derivative of $k$ subject to the side conditions given in \citet[Section 4]{granovsky1991}, in the limit as $\nu\to\infty$. However, it is not known whether they possess any optimality properties for the problem of estimating a distribution function, wherein the quantity $\psi(k)=2\int xK(x)k(x)\mathrm{d}x$ can be regarded as a measure of asymptotic performance of different kernel functions; see \citet{falk1983}, \citet{jones1990}, and Sections \ref{Sec:asymptotic.MISE} and \ref{sec:uniform.kernel}. An asymptotically optimal kernel maximises $\psi(k)$ subject to certain conditions, and kernels for which $\psi(k)<0$ are not worth considering as in this case EDF is asymptotically more efficient. \citet{jones1990} shows that the uniform $k$ is optimal among the class of second order kernels. Maximisation of $\psi(k)$ over a larger class of kernels is generally impossible without further restrictions. \citet{falk1983} and \citet{mammitzsch1984} show that there is no kernel which maximises $\psi(k)$ over the class of square integrable arbitrary order kernels with support $[-1,1]$, but that certain polynomial kernels are nearly optimal. 

Note that there is no reason for an asymptotically optimal kernel to retain its optimality in finite samples. Comparisons of the uniform kernel with the asymptotically suboptimal second order Gaussian kernel performed in Section \ref{sec:uniform.kernel} confirm the resultant asymptotic loss of efficiency, but also indicate that such a loss is negligible for most practical purposes. The same has been pointed out in \citet{jones1990} and the literature on kernel density estimation. Furthermore, the second order Gaussian kernel can outperform the uniform kernel in sample sizes of practical interest.
The general consensus is, thus, that the choice of kernel may be based on other considerations, such as smoothness.
One of the benefits of using smooth kernels is that the resultant estimators inherit these smoothness properties. This may be a desirable feature in applications. Another argument in favour of smooth kernels is the apparent mismatch between the assumptions required to derive the asymptotic optimality results and the properties of the resultant estimators. For example, to prove the optimality of the uniform kernel, $F$ is assumed to possess two square integrable derivatives, whereas the resultant estimator $\widehat{F}$ is not twice differentiable. 

The outline of the paper is as follows. Expressions for the exact MISE components, integrated squared bias (ISB) and integrated variance (IV), are given in Section \ref{Sec:main.results}. The proofs are given in Appendix \ref{App:Proofs}, and alternative, computationally convenient expressions are given in Appendix \ref{App:computational.aspects}. For comparison, expressions for the asymptotic MISE and MISE with the infinite order kernel are also given for the special case of NM distributions. 
A brief analysis of MISE is provided in Section \ref{sec:analysis.of.mise}, where the comparisons are made with the empirical distribution function and the infeasible minimum MISE of kernel estimators \citep{abdous1993}. 
Section \ref{sec:uniform.kernel} gives exact MISE expressions for the uniform kernel and NM distributions. 
Section \ref{Sec:bandwidth.selection} discusses the proposed plug-in bandwidth selection method and its performance in small samples, evaluated using a simulation study. 
Performance of the simple normal reference and Silverman's rule of thumb bandwidths is discussed in Appendix \ref{Sec:NRR.bandwidth}. Section \ref{sec:conclusions} concludes.

\section{Main results}\label{Sec:main.results}
In what follows, $\phi(x)$ and $\Phi(x)$ denote the standard normal density and distribution functions, respectively; $\phi(x;\mu,\sigma^{2})=\phi((x-\mu)/\sigma)/\sigma$. The derivatives of $g(x)$ with respect to $x$ are denoted by $g^{(r)}(x)=\mathrm{d}^{r}g(x)/\mathrm{d}x^{r}$, $r=1,2,\ldots$; $g^{(0)}(x)=g(x)$, and for $r=-1,-2,\ldots$, $g^{(r)}(x)$ denotes the antiderivatives. The first four antiderivaties of $\phi(x)$ that appear in the results below are 
$\phi^{(-1)}(x)=\Phi(x)$, $\phi^{(-2)}(x)=\phi(x)+x\Phi(x)$, 
$\phi^{(-3)}(x)=x\phi(x)/2 + (x^{2}+1)\Phi(x)/2$, and 
$\phi^{(-4)}(x)=(x^{2}+2)\phi(x)/6+ (x^{3}+3x)\Phi(x)/6$. 

\subsection{Normal mixture distributions}
The class of finite $m$-component normal mixture distributions considered in this paper is defined by the density function
\begin{equation}\label{Eq:Def:NM.dens}
f(x) = \sum_{j=1}^{m}w_{j}\phi(x;\mu_{j},\sigma_{j}^{2}),
\end{equation}
where $-\infty<\mu_{j}<\infty$, $\sigma_{j}>0$, and $w_{j}>0$ for all $j=1,\ldots,m$, and $\sum_{j=1}^{m}w_{j}=1$. The  corresponding distribution function is $F(x)=\int_{-\infty}^{x}f(z)\mathrm{d}z$. 

The NM class \eqref{Eq:Def:NM.dens} is sufficiently large to be of practical interest as the examples in Figure \ref{Fig:MWNMdens1-15} demonstrate; see also examples in, e.g., \citet{mclachlan2000}. The results derived in this paper can be used to study the finite sample performance of kernel estimators of a broad variety of distribution functions well approximated by normal mixtures. One important exception which will require separate treatment is the class of distributions on the bounded support with non-zero densities at the boundaries. 

Of course, if the true density is of the form \eqref{Eq:Def:NM.dens}, better estimators of $F$ than \eqref{Eq.KDFE.def} may exist. However, the purpose of this paper is not to analyse such estimators, but rather to provide a tool which can usefully complement asymptotic analysis and simulation studies. Nevertheless, as shown in Section \ref{Sec:bandwidth.selection}, even when the true distribution is known to be the normal mixture \eqref{Eq:Def:NM.dens}, and even when the true number of mixture components is known, smoothed estimators can have a significantly smaller MISE than parametric normal mixture cdf estimators. 

\subsection{Gaussian-based kernels}
For $r=1,2,3,\ldots$, the $(2r)^{th}$ order Gaussian-based kernels for density estimation are
\begin{equation}
\label{Eq:Def:Gaussian.Kernels.PDF}
g_{2r}(x) = \frac{(-1)^{r}\phi^{(2r-1)}(x)}{2^{r-1}(r-1)!x} = \sum_{s=0}^{r-1}\frac{(-1)^{s}}{2^{s}s!}\phi^{(2s)}(x);
\end{equation}
see \citet[Section 2]{wand1990}. 
The corresponding Gaussian-based kernels of order $2r$ for cdf estimation are obtained by integrating $g_{2r}$, \textit{viz.} 
\begin{equation}
\label{Eq:Def:Gaussian.Kernels.CDF}
G_{2r}(x) =\int_{-\infty}^{x}g_{2r}(z)\mathrm{d}z = \sum_{s=0}^{r-1}\frac{(-1)^{s}}{2^{s}s!}\phi^{(2s-1)}(x). 
\end{equation}
Kernels $G_{2r}(x)$ are of the form $G_{2r}(x)=\Phi(x)+P_{r}(x)\phi(x)$, where $P_{r}(x)$ are polynomials in $x$; 
for example, $P_{1}(x)=0$, $P_{2}(x) = x/2$, $P_{3}(x) = (-x^{3}+7x)/8$, and $P_{4}(x) = (x^{5}-16x^{3}+57x)/48$.
When $r$ is large, expression \eqref{Eq:Def:Gaussian.Kernels.CDF.c} in Appendix \ref{App:computational.aspects} can be used to compute the terms $P_{r}(x)\phi(x)$ recursively. 

To obtain the limiting kernel as $r\to\infty$, let $g_{2}^{\ast}=g_{2}$ and $G_{2}^{\ast}=G_{2}$, and for $r>1$, define the rescaled kernels as
\begin{equation}
\label{Eq:Def:Rescaled.Gaussian.Kernels}
g_{2r}^{\ast}(x) =  \frac{1}{\sqrt{2r-2}}g_{2r}\left(\frac{x}{\sqrt{2r-2}}\right) \qquad \text{and}\qquad
G_{2r}^{\ast}(x) = G_{2r}\left(\frac{x}{\sqrt{2r-2}}\right).
\end{equation}
Then the corresponding infinite order kernels are $g_{\infty}^{\ast}(x)=\lim_{r\to\infty}g_{2r}^{\ast}(x)=\sinc(x)/\pi$, where $\sinc(x)=\sin(x)/x$ for $x\neq0$ and $\sinc(0)=1$ is the cardinal sine function; see, e.g., \citet[Theorem 3]{hansen2005}. Thus we can define $G_{\infty}^{\ast}(x)=\Si(x)/\pi+1/2$, where $\Si(x)=\int_{0}^{x}\sinc(z)\mathrm{d}z$ is the sine integral \citep[Section 2.2]{chacon2014}. The rescaling \eqref{Eq:Def:Rescaled.Gaussian.Kernels} is only necessary to obtain the limiting kernel; it has no effect on MISE computations for a finite $r$ as the results for $G_{2r}^{\ast}$ can be obtained from those for $G_{2r}$ by rescaling the bandwidth $h$. 

Since kernels $G_{2r}$ of order greater than two are not monotone (kernels $g_{2r}$ take negative values), the resultant estimates may not themselves be distribution functions. However, if necessary, the estimates can be corrected by rearrangement \citep{chernozhukov2009} or the methods described in \citet{glad2003}. 
The rearrangement (which is effectively sorting) is particularly simple to use,  and  the MISE of the rearranged estimator can be at most equal to, and is often strictly smaller than the MISE of the original estimator. 

\subsection{Exact MISE}
Theorem \ref{Thm:KCDFE.Exact.MISE.0} gives general expressions for the integrated squared bias, $\ISB[\widehat{F}(\cdot;h)]=\int_{-\infty}^{\infty}\{\E[\widehat{F}(x;h)]-F(x)\}^{2}\mathrm{d}x$, and the integrated variance, $\IVar[\widehat{F}(\cdot;h)] = \int_{-\infty}^{\infty}\Var[\widehat{F}(x;h)]\mathrm{d}x$, of the kernel cdf estimator \eqref{Eq.KDFE.def}; $\MISE[\widehat{F}(\cdot;h)]=\ISB[\widehat{F}(\cdot;h)]+\IVar[\widehat{F}(\cdot;h)]$. These results can then be specialised to different classes of kernels and distributions and will be useful in their own right. 

Let $(f\ast g)(x)=\int_{-\infty}^{\infty}f(t)g(x-t)\mathrm{d}t$ and 
 $(f\star g)(x)=\int_{-\infty}^{\infty}f(t)g(t+x)\mathrm{d}t$ denote the convolution and the cross-correlation of functions $f$ and $g$ respectively.  Also let $k_{h}(x)=k(x/h)/h$. 

\begin{theorem}\label{Thm:KCDFE.Exact.MISE.0}
Let $X_{1},\ldots,X_{n}$ be a random sample from a distribution with a square integrable density $f$, and $k$ be a symmetric, square integrable kernel, such that $\int_{-\infty}^{\infty}k(t)\mathrm{d}t=1$ and $\lim_{t\to\infty}t^{2}k(t)=0$. Then for $h>0$, the integrated squared bias and integrated variance of KDFE \eqref{Eq.KDFE.def} are 
\begin{align}
\label{Thm:KCDFE.Exact.MISE.0.ISB}
\ISB[\widehat{F}(\cdot;h)]  & = -\xi^{(-2)}(0), \qquad \text{where} \qquad   \xi(u)=(\eta\star\eta)(u)-2(\eta\star f)(u)+(f\star f)(u), \\
\intertext{$\eta(u)=(k_{h}\ast f)(u)$, $\xi(u)$ is symmetric about the origin and integrates to zero over the real line, and}
\label{Thm:KCDFE.Exact.MISE.0.IVar}
\IVar[\widehat{F}(\cdot;h)] & = -\frac{h}{n}\psi(k)+\frac{1}{n}(\eta\star\eta)^{(-2)}(0).
\end{align}
\end{theorem}
The proof given in Appendix \ref{App:Proofs} is based on interchanging the order of integration by Fubini theorem after a judiciously chosen change of coordinates.

Theorem \ref{Thm:Exact.MISE} specialises the above results to the arbitrary (finite) order Gaussian-based kernels and normal mixture distributions. Let $\widehat{F}_{2r}$ denote the kernel estimator \eqref{Eq.KDFE.def} of the distribution function $F$ using the $(2r)^{th}$ order kernel \eqref{Eq:Def:Gaussian.Kernels.CDF}. Also let $\OF(2n)$ denote the odd factorial, i.e., for $n\geq1$, $\OF(2n)=\prod_{i=1}^{n}(2i-1)$, $\OF(-2n)=(-1)^{n}/\OF(2n)$, $\OF(0)=1$, and for $n$ odd, $\OF(n)=0$. 

\begin{theorem}\label{Thm:Exact.MISE} 
Let $X_{1},\ldots,X_{n}$ be a random sample from an $m$-component normal mixture distribution \eqref{Eq:Def:NM.dens}, and $K=G_{2r}$ be the $(2r)^{th}$-order Gaussian-based kernel \eqref{Eq:Def:Gaussian.Kernels.CDF}. Then for $h>0$, $r=1,2,3,\ldots$, 
\begin{equation}\label{Thm1.ISB}
\ISB[\widehat{F}_{2r}(\cdot;h)]   =  -\sum_{s=0}^{r-1}\sum_{t=0}^{r-1}\frac{(-1)^{s+t}}{2^{s+t}s!t!}V(h;s+t,2) + 2\sum_{s=0}^{r-1}\frac{(-1)^{s}}{2^{s}s!}V(h;s,1) - V(h;0,0), 
\end{equation}
\begin{equation}\label{Thm1.IVar}
\IVar[\widehat{F}_{2r}(\cdot;h)]  = -\frac{h}{n}\psi(g_{2r}) + \frac{1}{n}\sum_{s=0}^{r-1}\sum_{t=0}^{r-1}\frac{(-1)^{s+t}}{2^{s+t}s!t!}V(h;s+t,2),
\end{equation}
where
\begin{equation}\label{Thm1.V.def}
V(h;p,q) = h^{2p}\sum_{i=1}^{m}\sum_{j=1}^{m}w_{i}w_{j}\sigma_{ij,q}^{1-2p}\phi^{(2p-2)}\left(\frac{\mu_{j}-\mu_{i}}{\sigma_{ij,q}}\right),
\end{equation}
$\sigma_{ij,q}=\sqrt{\sigma_{i}^{2}+\sigma_{j}^{2}+qh^{2}}$, and 
\begin{equation}\label{Thm1.C1.def}
\psi(g_{2r}) = -\frac{1}{\sqrt{\pi}}\sum_{s=0}^{r-1}\sum_{t=0}^{r-1}\frac{\OF(2s+2t-2)}{2^{2s+2t}s!t!}
 =\frac{\Gamma(2r-3/2)}{\pi\Gamma(2r-1)}
+\sum_{s=0}^{r-2}\frac{\Gamma(r+s-1/2)}{\pi\Gamma(r+s+1)}I_{1/2}(r,s+1),
\end{equation}
where  $I_{z}(\alpha,\beta)$ denotes the regularized incomplete beta function and it is understood that the sum over $s$ in the second expression  is zero when $r=1$. As $r\to\infty$, 
\begin{equation}\label{Thm1.C1.large.r.approx}
\psi(g_{2r})  = \frac{1}{\pi\sqrt{2r-2}} +\frac{\sqrt{2}}{\pi^{3/2}(4r-3)}+O(r^{-3/2}).
\end{equation}
\end{theorem}
The proof given in Appendix \ref{App:Proofs} is based on Theorem \ref{Thm:KCDFE.Exact.MISE.0} and the convolution formulae in \citet{aldershof1995}. Alternative, computationally convenient expressions for  $\MISE[\widehat{F}_{2r}(\cdot;h)]$ are given in Appendix \ref{App:computational.aspects}. All the quantities can be computed recursively, which is particularly useful when $r$ is large. The only special function that needs to be evaluated is the standard normal cdf. The minimiser of MISE, $h^{\ast}_{e}$, can be obtained by standard numerical optimisation techniques with the caveat that there may be multiple local minima. Existence of the global minimiser $h^{\ast}_{e}=\mathop{\mathrm{argmin}}_{h>0}\MISE[\widehat{F}_{2r}(\cdot;h)]$ follows from \citet[Theorem 1]{tenreiro2006}. 

It is evident from \eqref{Thm1.C1.def} that for the Gaussian-based kernels $\psi(g_{2r})>0$ for all $r=1,2,3,\ldots$,
and $\sqrt{2r-2}\psi(g_{2r})\to1/\pi$ as $r\to\infty$; cf. second term in \eqref{Inf.Kern.MISE.2}.
This property of Gaussian-based kernels implies that asymptotically the KDFE $\widehat{F}_{2r}$ provides a second order improvement in MISE relative to EDF; cf. the second term in \eqref{Eq:Asy.MISE}.

A special case of Theorem \ref{Thm:Exact.MISE} worth stating separately is the second order Gaussian kernel  which is commonly used in practice. With $r=1$ the expressions simplify to 
$\ISB[\widehat{F}_{2}(\cdot;h)]  =  -U(h;2) + 2U(h;1) - U(h;0)$ and 
$\IVar[\widehat{F}_{2}(\cdot;h)] =  -h/(n\sqrt{\pi}) + U(h;2)/n$,
where
\begin{equation}\label{Thm1.Cor.2ndOrderKenr.U.def}
U(h;q) = \sum_{i=1}^{m}\sum_{j=1}^{m}w_{i}w_{j}\left[\sigma_{ij,q}\phi\left(\frac{\mu_{i}-\mu_{j}}{\sigma_{ij,q}}\right) + (\mu_{i}-\mu_{j})\Phi\left(\frac{\mu_{i}-\mu_{j}}{\sigma_{ij,q}}\right)\right].
\end{equation}

Expressions in Theorem \ref{Thm:Exact.MISE} are also valid with $h=0$. This recovers the well-known result that the EDF is an unbiased estimator, $\ISB[\widehat{F}_{2r}(\cdot;0)]  = 0$, with integrated variance $\IVar[\widehat{F}_{2r}(\cdot;0)] = V_{F}/n$, where 
\begin{equation}\label{Thm1.Cor2.V0}
V_{F} = \sum_{i=1}^{m}\sum_{j=1}^{m}w_{i}w_{j}\left[\sigma_{ij,0}\phi\left(\frac{\mu_{i}-\mu_{j}}{\sigma_{ij,0}}\right) + (\mu_{i}-\mu_{j})\Phi\left(\frac{\mu_{i}-\mu_{j}}{\sigma_{ij,0}}\right)\right] 
 = \int_{-\infty}^{\infty}F(x)\left[1-F(x)\right]\mathrm{d}x.
\end{equation}

\subsection{Asymptotic MISE}\label{Sec:asymptotic.MISE}
Let $k$ be a general symmetric $(2r)^{th}$ order kernel. Then, under the standard smoothness and integrability conditions on $F$ which are satisfied by NM distributions \eqref{Eq:Def:NM.dens}, as $h\to0$,
\begin{equation}\label{Eq:Asy.MISE}
\MISE[\widehat{F}(\cdot;h)] = \frac{1}{n}V_{F}-\frac{h}{n}\psi(k)+\frac{\mu_{2r}(k)^{2}}{(2r)!^{2}}R(F^{(2r)})h^{4r}+o(hn^{-1}+h^{4r}),
\end{equation} 
where $R(g)=\int_{-\infty}^{\infty}g(x)^{2}\mathrm{d}x$ for any square integrable function $g$. This result follows by straightforward Taylor series manipulations as in, e.g., \citet{azzalini1981} or \citet[Sections 2.1, 9.2]{rao1983}. The idea of using higher order kernels as a bias reduction technique originates at least as far back as \citet{bartlett1963}.

Provided $\psi(k)>0$, the asymptotically optimal bandwidth minimising the leading terms in \eqref{Eq:Asy.MISE} is $h_{a}^{\ast}=\varsigma n^{-1/(4r-1)}$, where $\varsigma=\{(2r)!^{2}\psi(k)/[4r\mu_{2r}(k)^{2}R(F^{(2r)})]\}^{1/(4r-1)}$, and asymptotically, the MISE at $h_{a}^{\ast}$ is 
\begin{equation}\label{Eq:Asy.MISE.opt.bw}
\MISE[\widehat{F}(\cdot;h_{a}^{\ast})] = n^{-1}V_{F}
-\varsigma\psi(k)[1-(4r)^{-1}]n^{-4r/(4r-1)} +o(n^{-4r/(4r-1)}).
\end{equation} 

For the Gaussian-based kernels $G_{2r}$, 
$\varsigma = [\psi(g_{2r})2^{2r-2}r!(r-1)!/R(F^{(2r)})]^{1/(4r-1)}$, and for the NM distribution \eqref{Eq:Def:NM.dens},
$R(F^{(2r)}) = -\sum_{i=1}^{m}\sum_{j=1}^{m}w_{i}w_{j}\sigma_{ij,0}^{1-4r}\phi^{(4r-2)}\left((\mu_{i}-\mu_{j})/\sigma_{ij,0}\right)$.

\subsection{Infinite order kernel}
Exact MISE of a KDFE with the sinc kernel has been derived in \citet{abdous1993} and \citet{chacon2014}. For the NM distribution the absolute square of the characteristic function is $\abs{\varphi_{f}(t)}^{2}= \sum_{i=1}^{m}\sum_{j=1}^{m}w_{i}w_{j}\cos\left[(\mu_{i}-\mu_{j})t\right]\exp[-(\sigma_{i}^{2}+\sigma_{j}^{2})t^{2}/2]$. Thus, the MISE is 
\begin{equation}\label{Inf.Kern.MISE.2}
\MISE[\widehat{F}_{\infty}(\cdot;h)]
 = \frac{1}{n}V_{F} -\frac{h}{n\pi}
+\frac{1}{\pi}\left(1+\frac{1}{n}\right)
\sum_{i=1}^{m}\sum_{j=1}^{m}w_{i}w_{j}I(h;\mu_{i}-\mu_{j},\bar{\sigma}_{ij}),
\end{equation}
where $V_{F}$ is defined in \eqref{Thm1.Cor2.V0},  $I(h;\mu,\sigma) = \sigma\int_{\sigma/h}^{\infty} \cos(\mu t/\sigma)t^{-2}\exp(-t^{2})\mathrm{d}t$, and $\bar{\sigma}_{ij}=(\sigma_{i}^{2}+\sigma_{j}^{2})^{1/2}/\sqrt{2}$. 
Note that $I(h;0,\sigma) =h\exp(-\sigma^{2}/h^{2}) -2\sigma\sqrt{\pi}[1-\Phi(\sqrt{2}\sigma/h)]$. If $\mu\neq0$, numerical integration techniques such as the Gauss-Kronrod quadrature  can be used to evaluate $I(h;\mu,\sigma)$. 

The optimal bandwidth solves $\abs{\varphi_{f}(1/h^{\ast})}^{2}=1/(n+1)$. For the normal distribution the solution is $h^{\ast}=\sigma/\sqrt{\ln(n+1)}$. In general, however, there does not appear to be a way of obtaining a closed form solution for $h^{\ast}$, and it has to be found using numerical techniques with the caveat that the solution may not be unique. Existence of the global minimiser of \eqref{Inf.Kern.MISE.2} has been established in \citet[Theorem 3]{chacon2014}; see also related discussion in \citet{glad2007}. 

\subsection{Analysis of MISE}\label{sec:analysis.of.mise}
\begin{figure}[!htbp]\centering
\begin{tabular}{ccc}
\#1: Gaussian & \#2: Skewed unimodal & \#3: Strongly skewed  \\
  \includegraphics[width=0.3\linewidth]{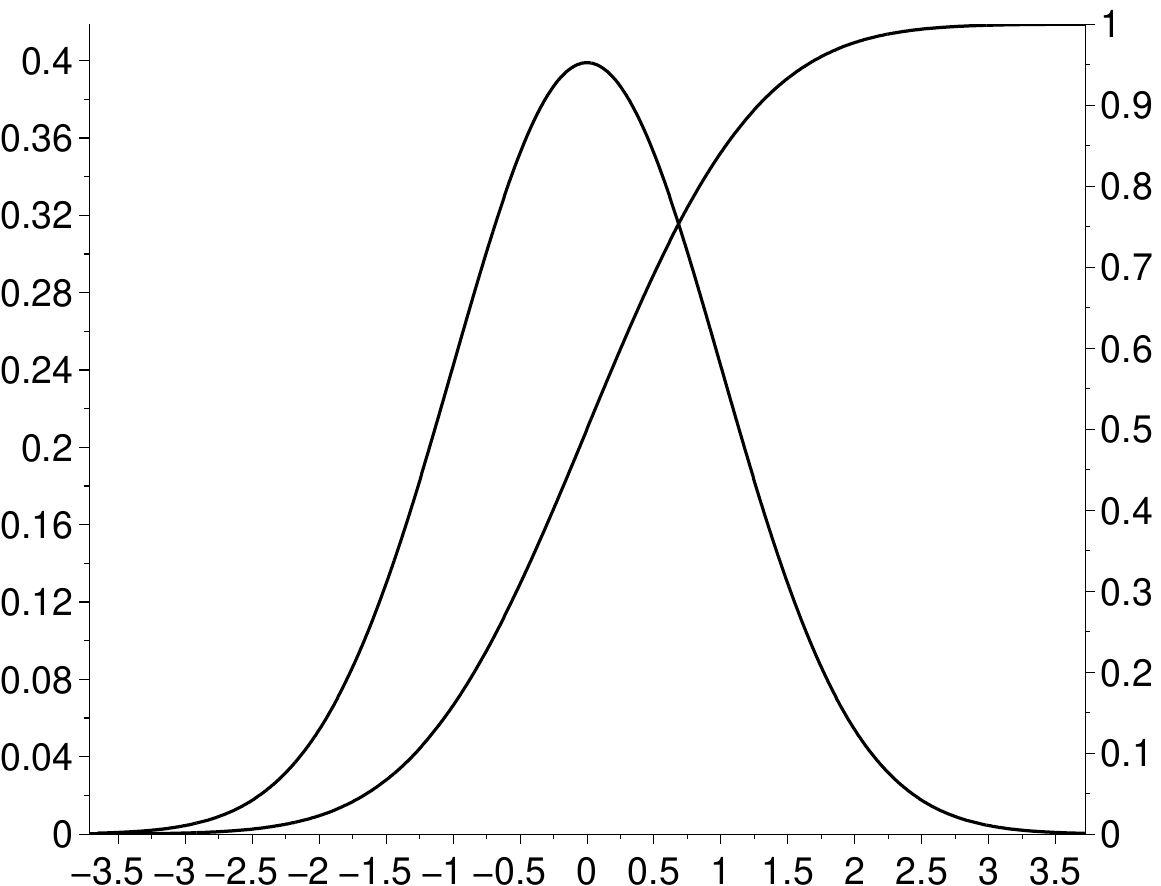} & 
  \includegraphics[width=0.3\linewidth]{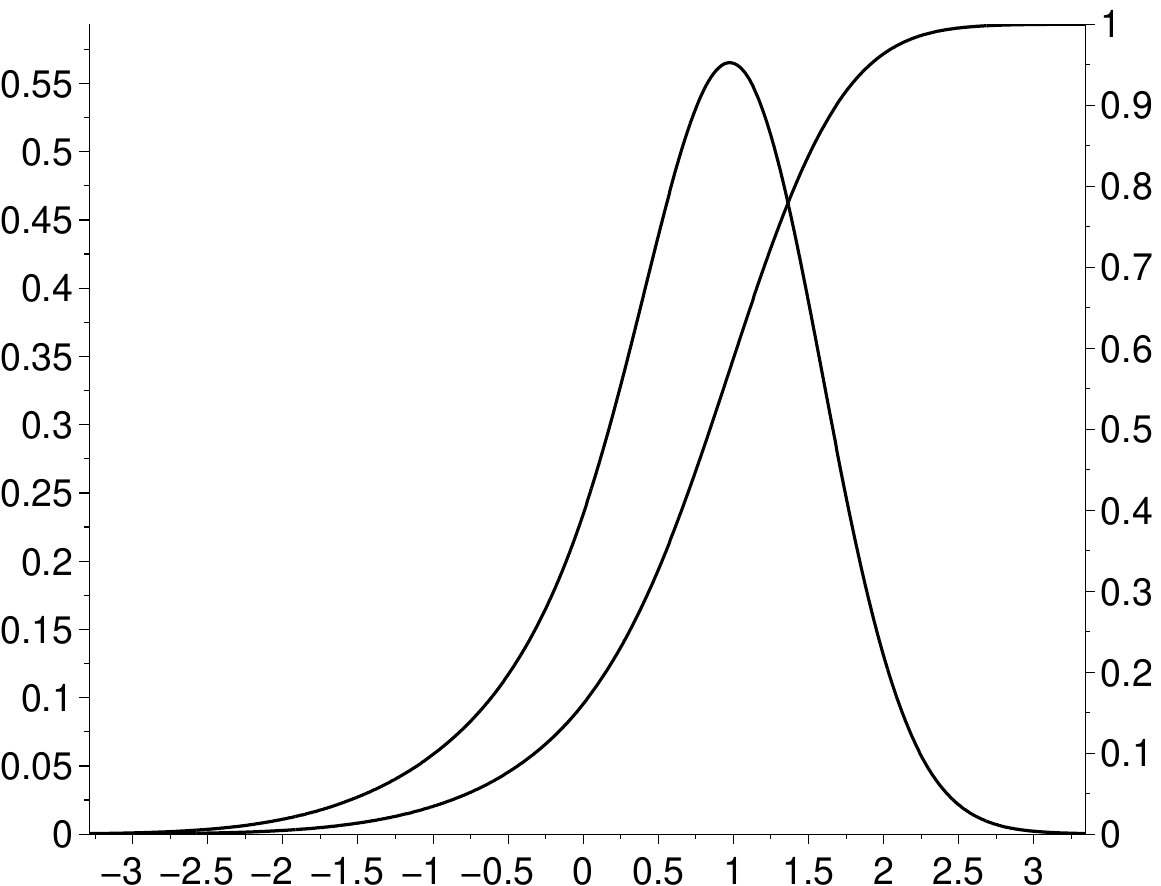} & 
  \includegraphics[width=0.3\linewidth]{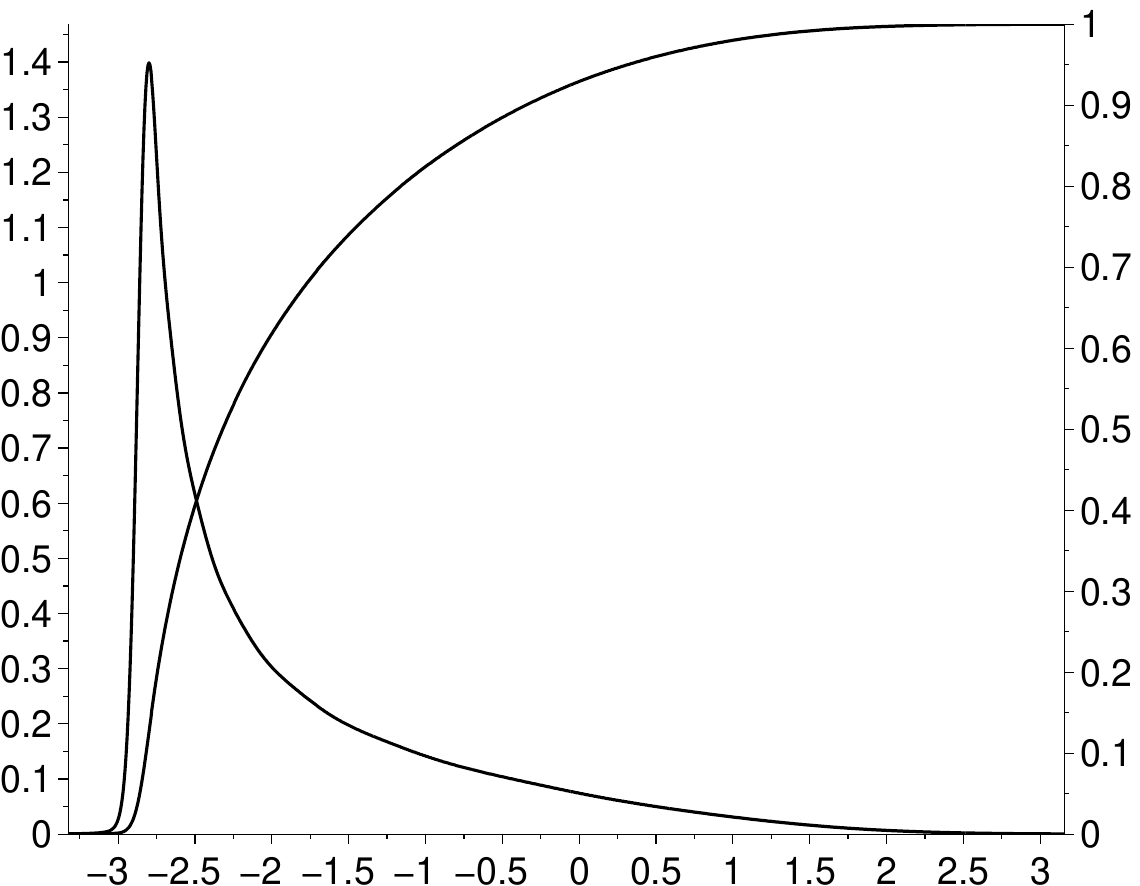} \\ 
\#4: Kurtotic unimodal & \#5: Outlier & \#6: Bimodal  \\
  \includegraphics[width=0.3\linewidth]{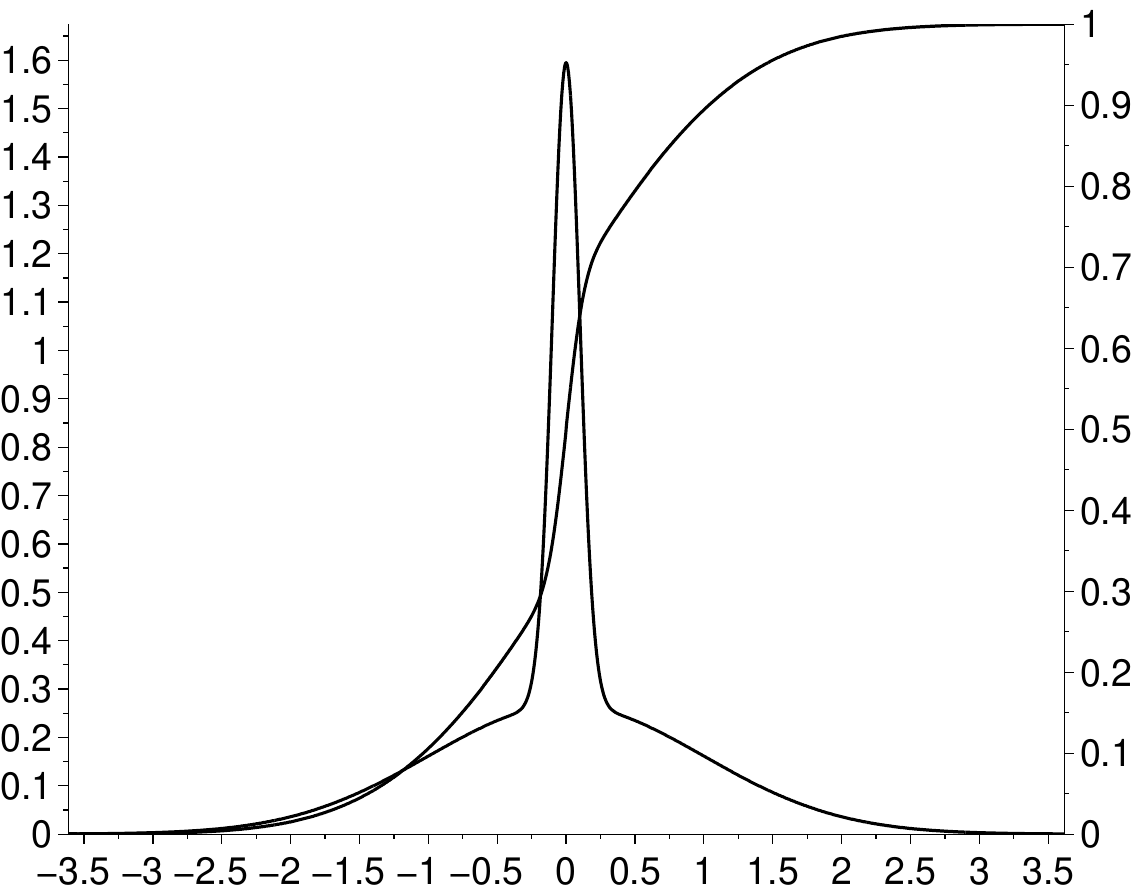} & 
  \includegraphics[width=0.3\linewidth]{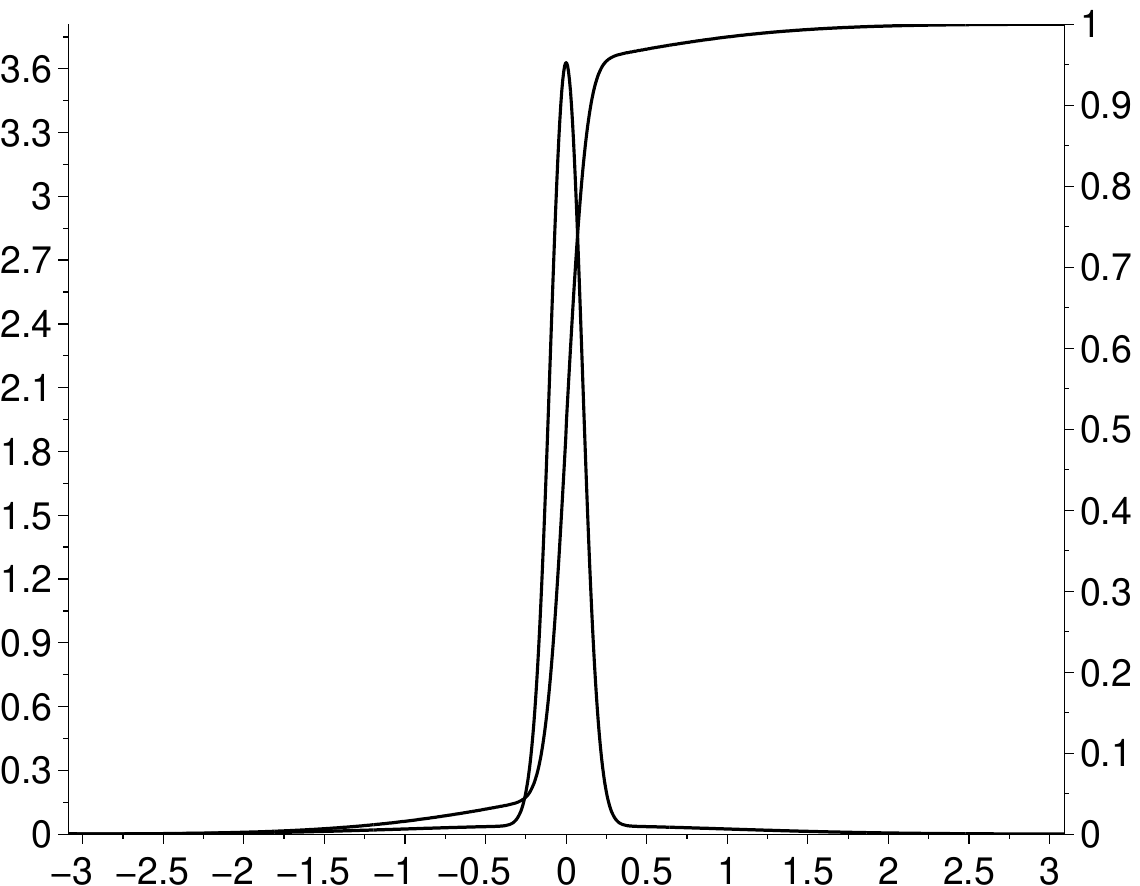} & 
  \includegraphics[width=0.3\linewidth]{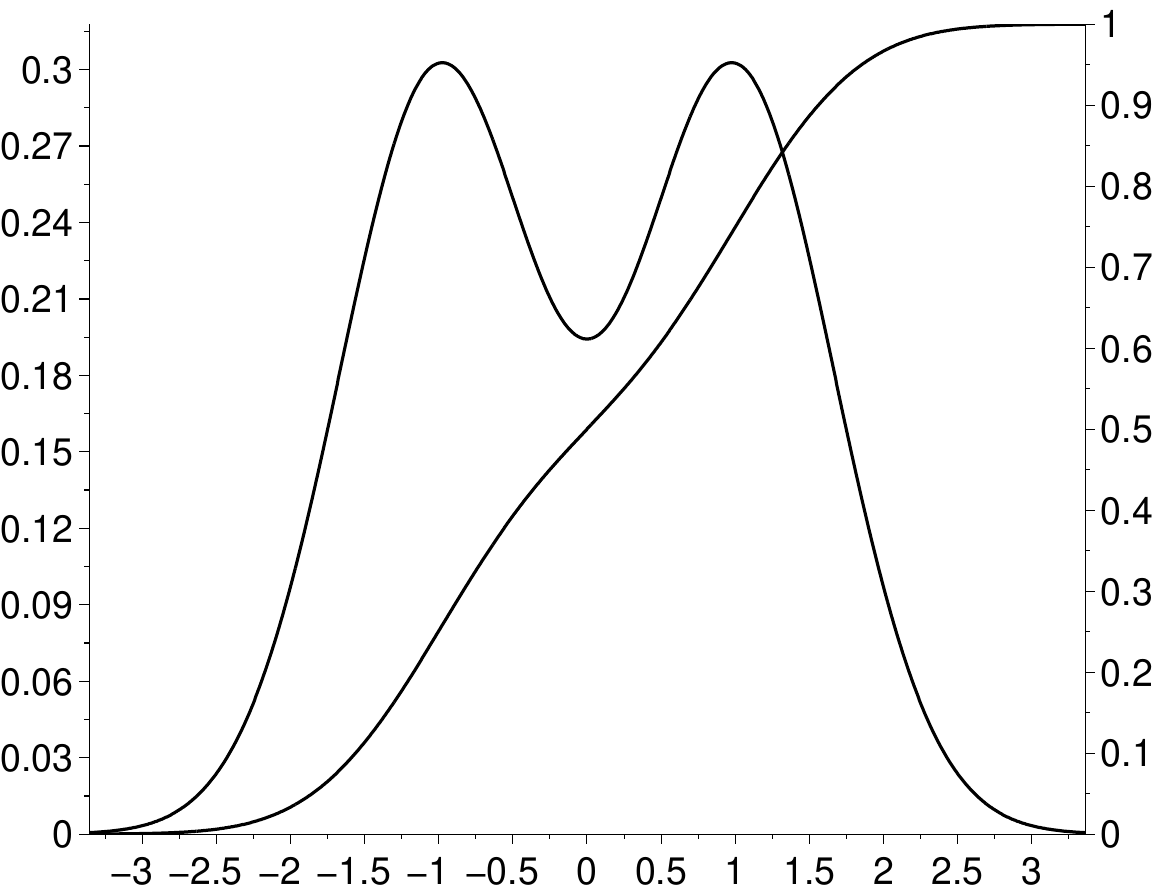} \\ 
\#7: Separated bimodal & \#8: Skewed bimodal & \#9: Trimodal  \\
  \includegraphics[width=0.3\linewidth]{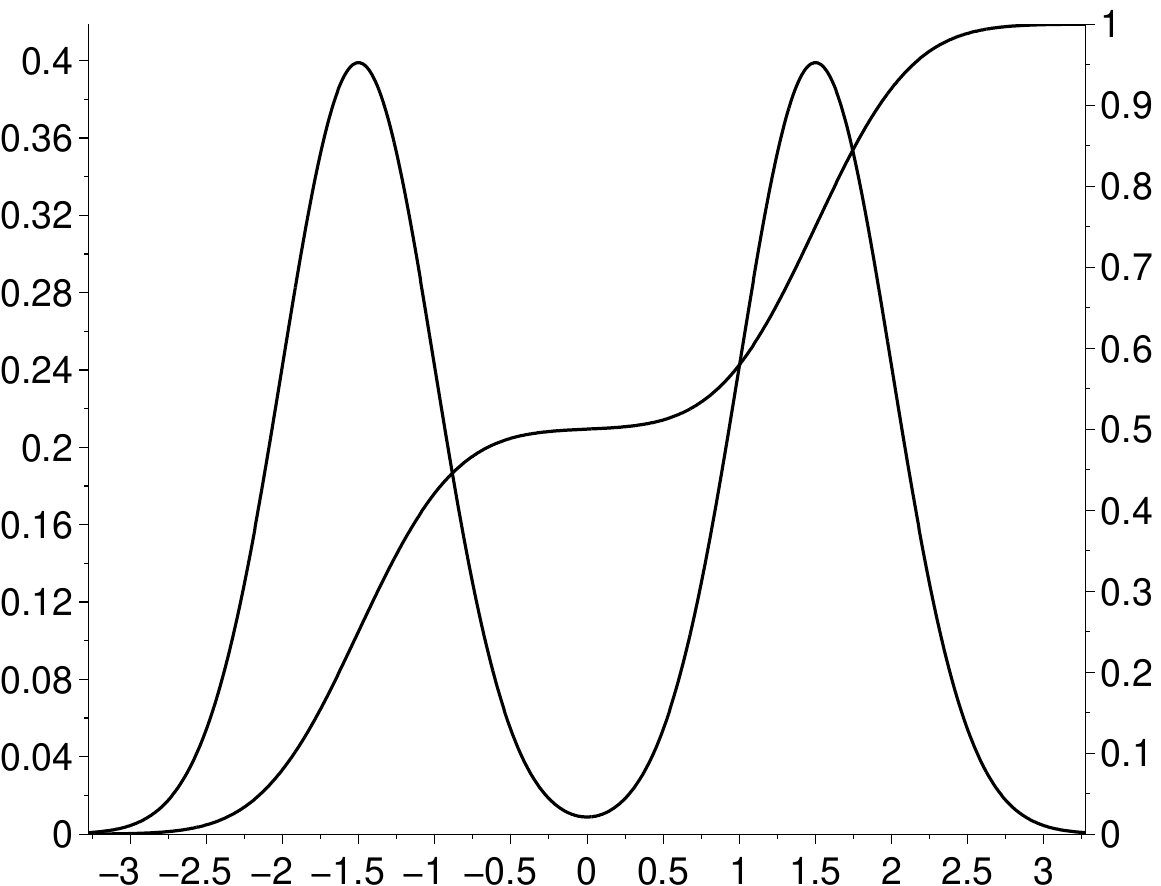} & 
  \includegraphics[width=0.3\linewidth]{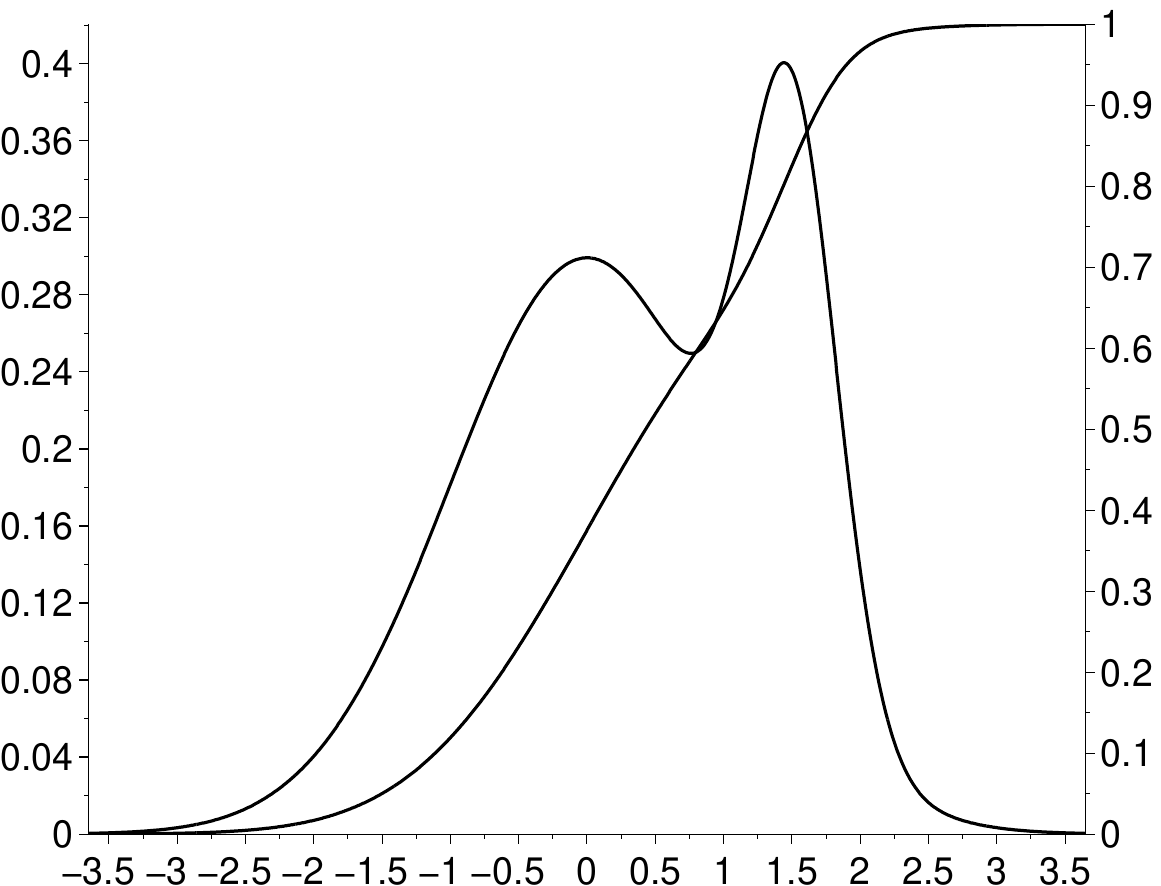} & 
  \includegraphics[width=0.3\linewidth]{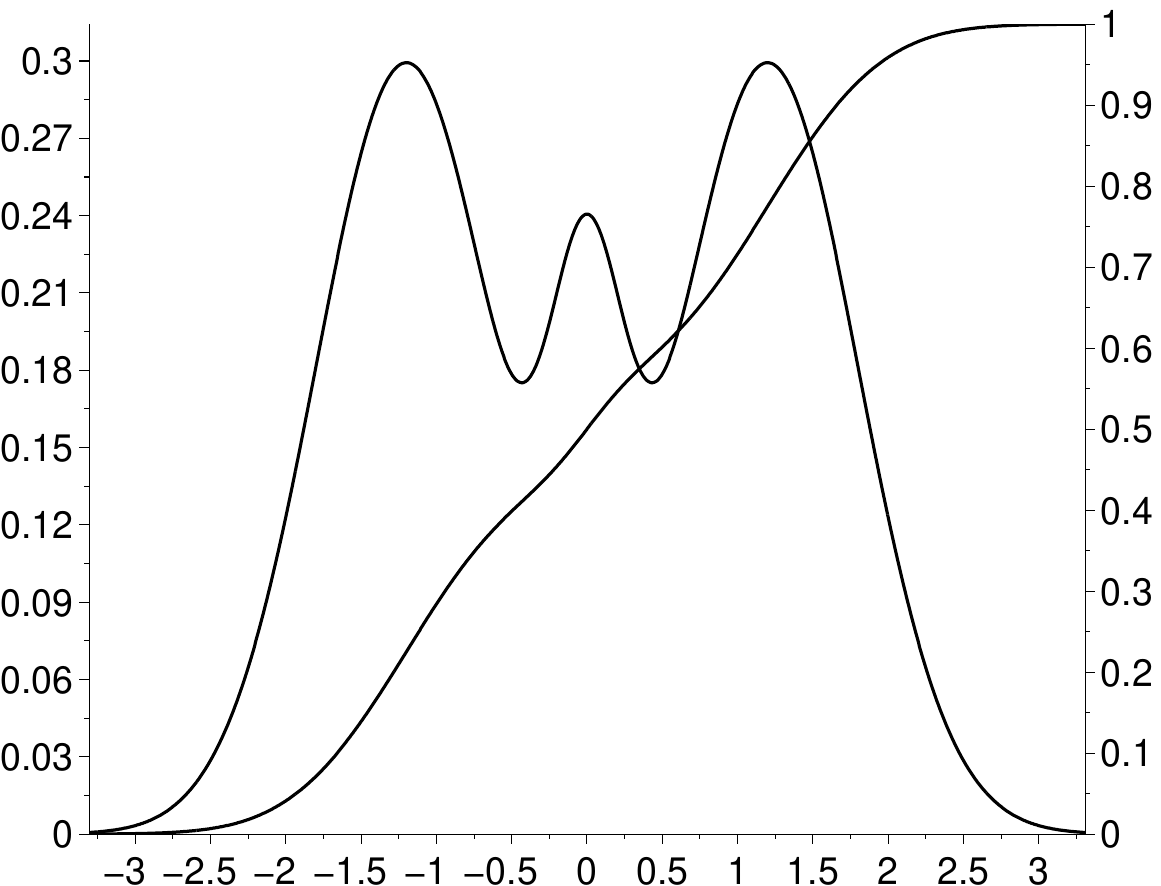} \\ 
\#10: Claw & \#11: Double claw & \#12: Asymmetric claw  \\
  \includegraphics[width=0.3\linewidth]{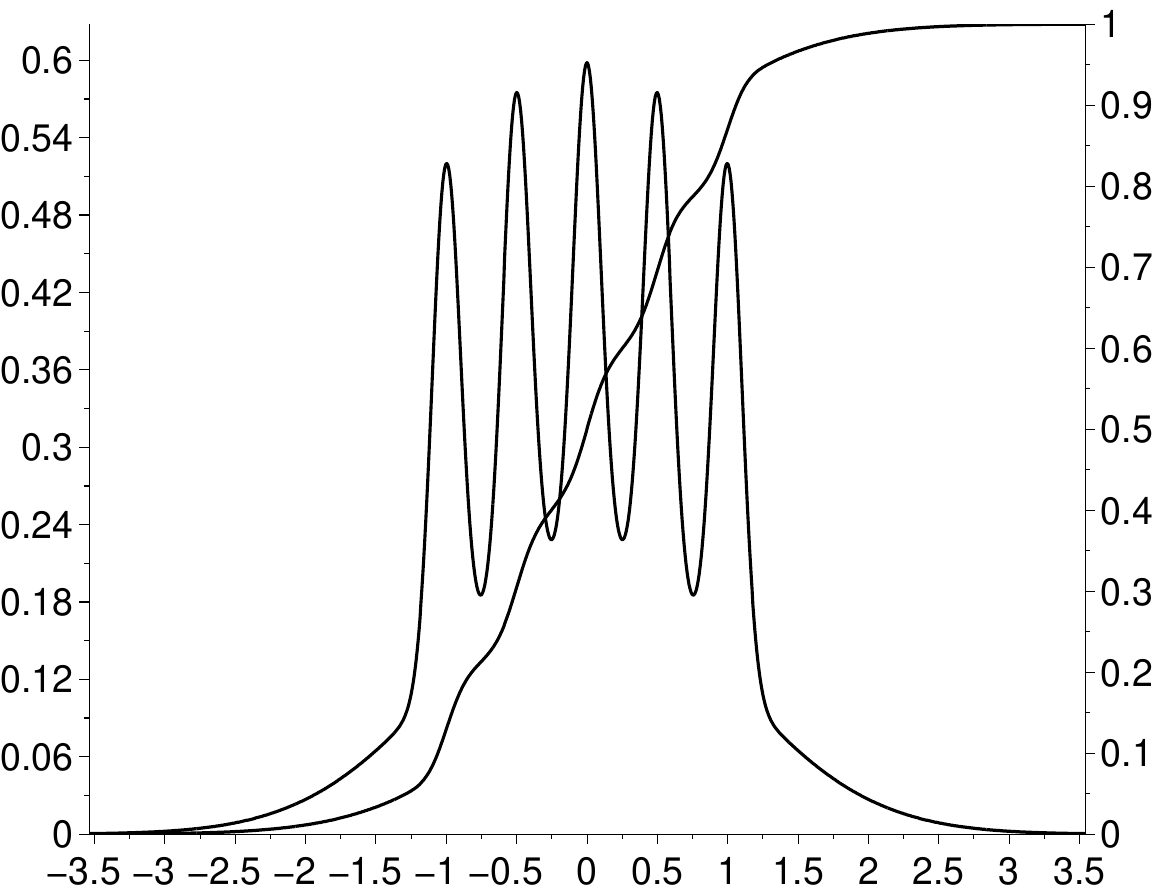} & 
  \includegraphics[width=0.3\linewidth]{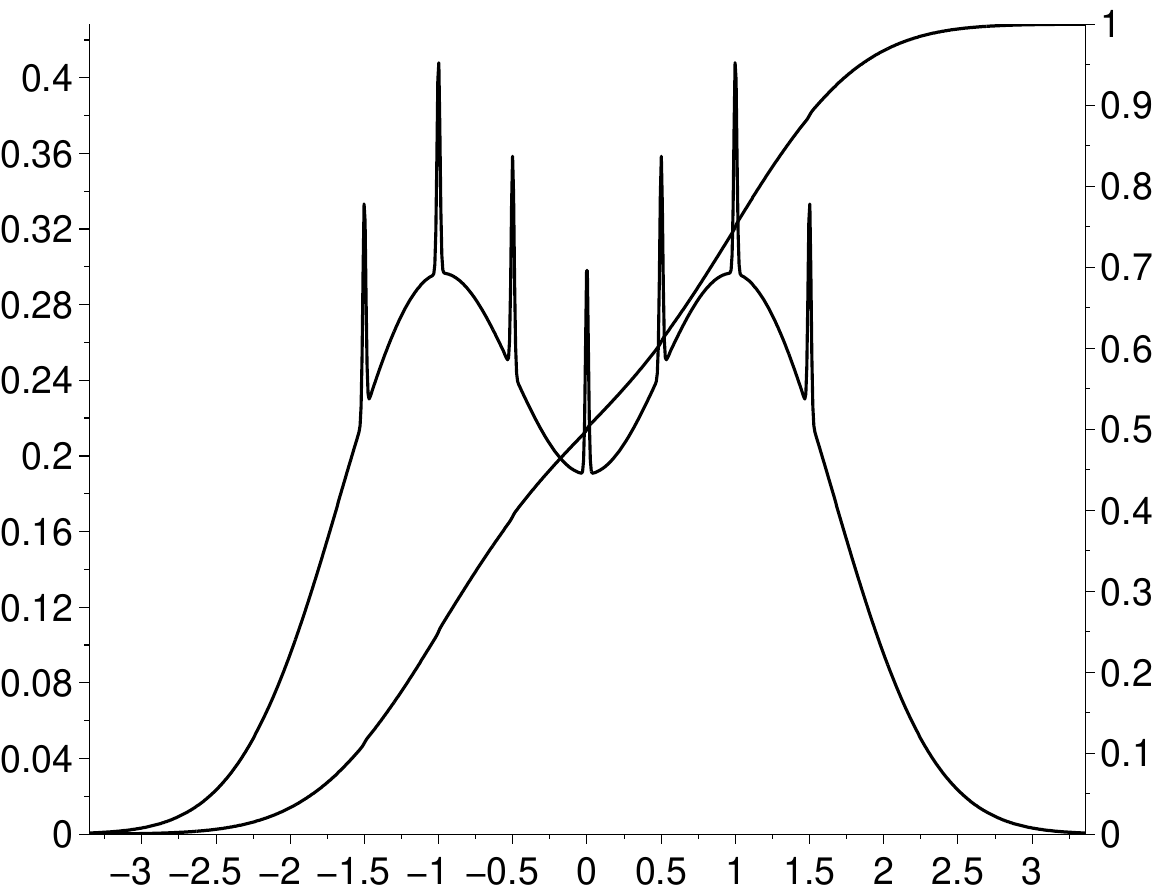} & 
  \includegraphics[width=0.3\linewidth]{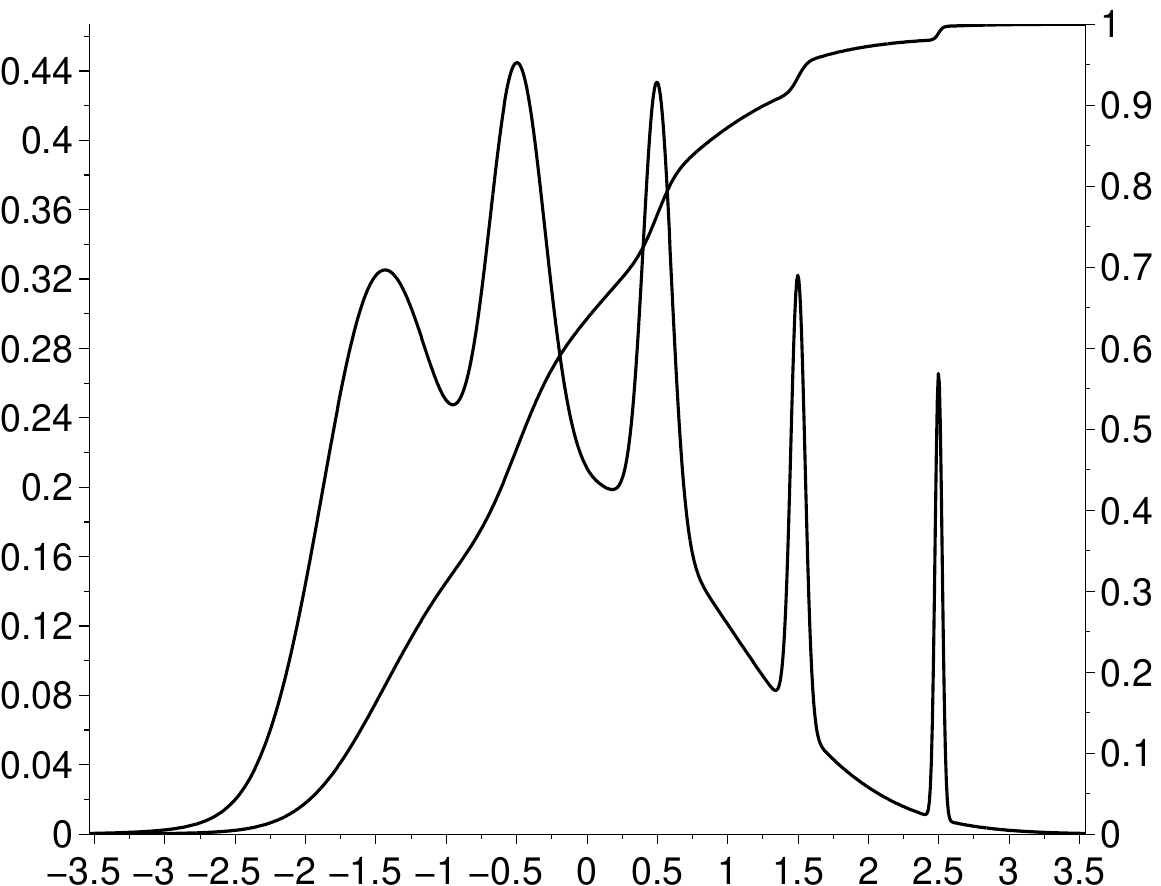} \\ 
\#13: Asymmetric double claw & \#14: Smooth comb & \#15: Discrete comb  \\
  \includegraphics[width=0.3\linewidth]{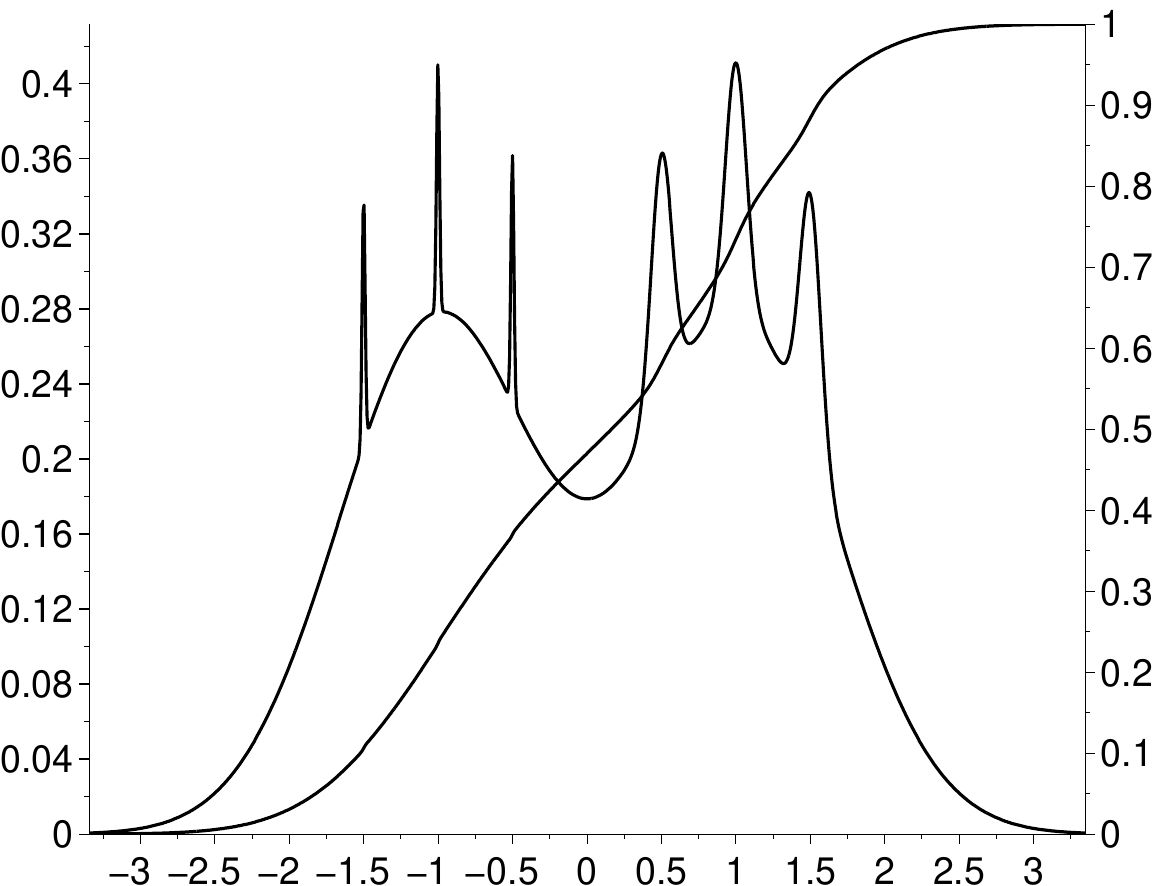} & 
  \includegraphics[width=0.3\linewidth]{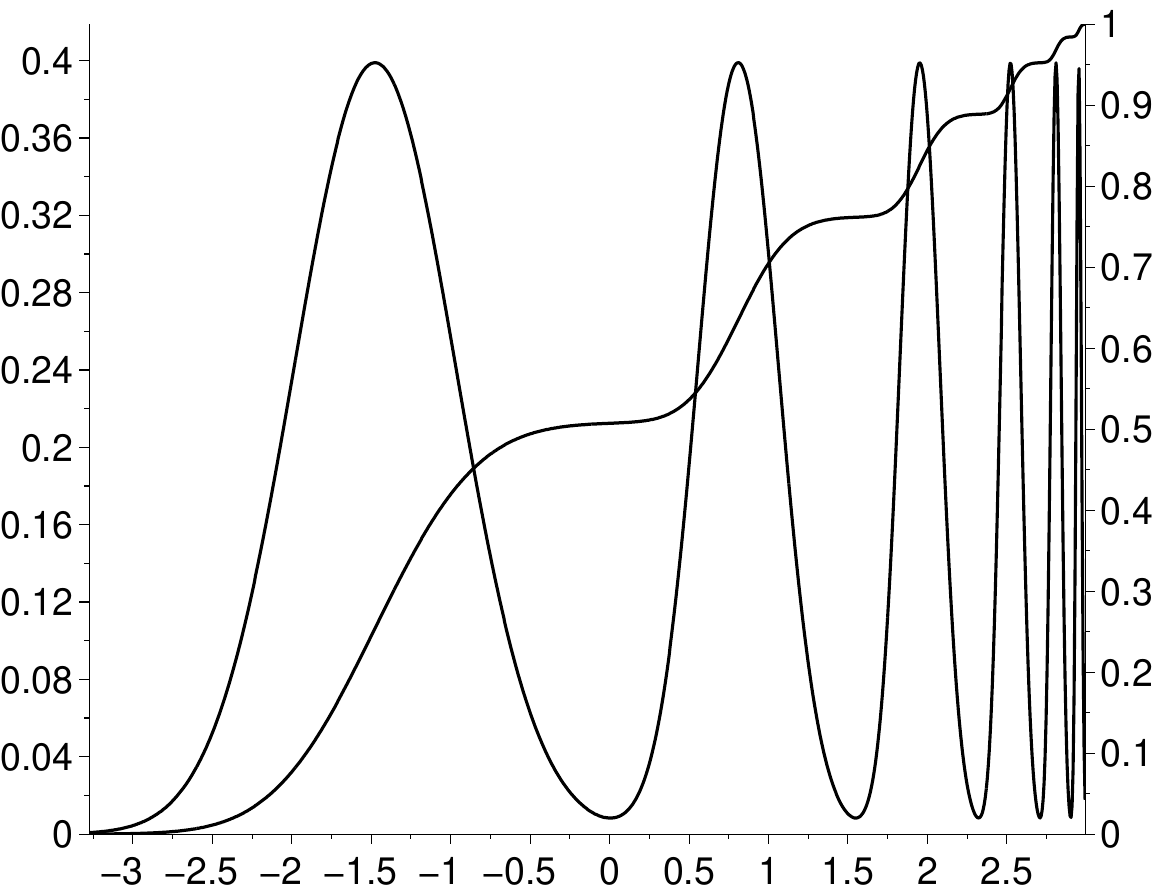} & 
  \includegraphics[width=0.3\linewidth]{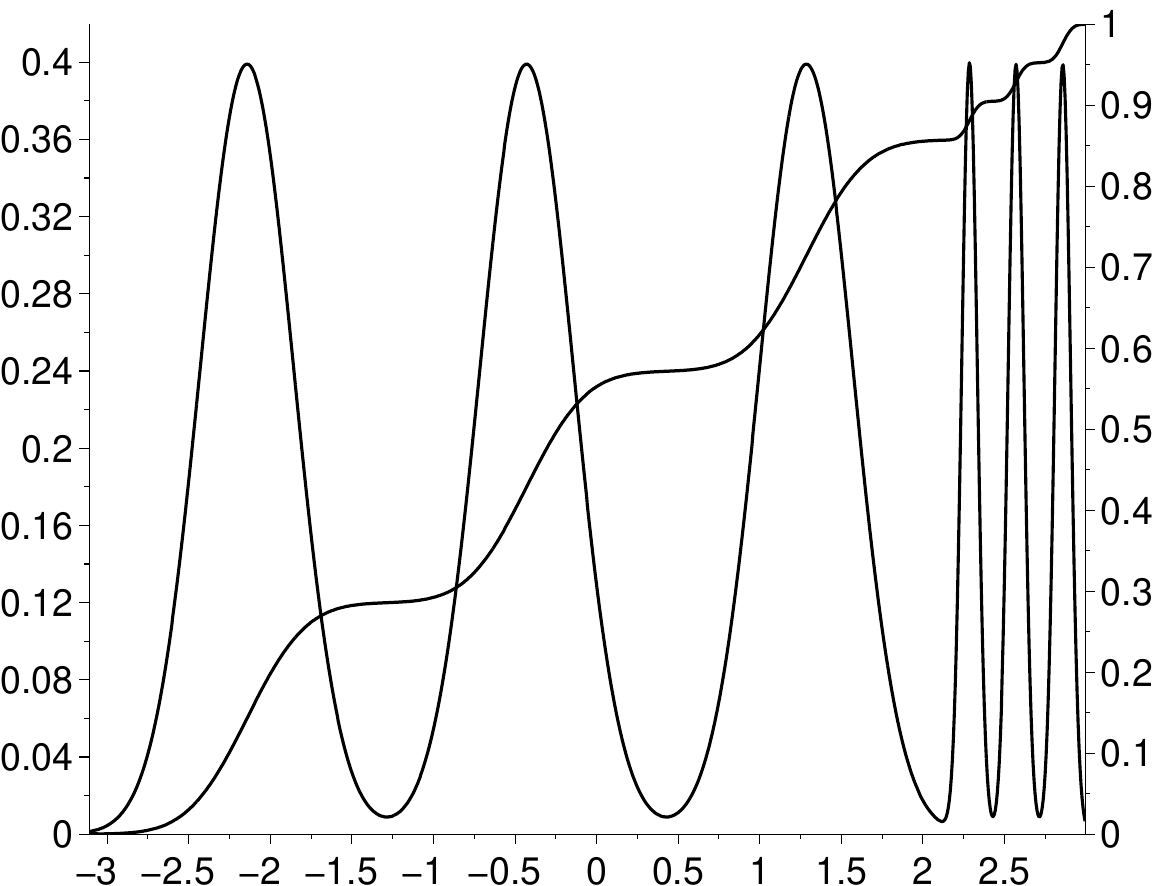} \\[1ex]
\multicolumn{3}{l}{Vertical axes: left--density, right--cdf.}
\end{tabular}
\caption{Selected normal mixture distributions}
\label{Fig:MWNMdens1-15}
\end{figure}

This section provides a brief analysis of MISE using the fifteen NM distributions shown in Figure \ref{Fig:MWNMdens1-15} as examples; see \citet[Table 1]{marron1992} the definitions of these mixtures. 

Since MISE itself is not a unitless quantity, it is natural to perform comparisons relative to the MISE of the EDF; hereinafter the relative MISE, in percentages. Indeed, if MISE of a kernel estimator is larger than that of the EDF, KDFE loses much of its appeal, even though a case can still be made for the benefits offered by smoothness alone. 
On the other hand, if an estimator achieves (or is reasonably close to) the infeasible minimum MISE,
\begin{equation*}
\MISE^{\ast} = \frac{1}{2\pi}\int_{-\infty}^{\infty}t^{-2}\abs{\varphi_{f}(t)}^{2}(1-\abs{\varphi_{f}(t)}^{2})\left[1+(n-1)\abs{\varphi_{f}(t)}^{2}\right]^{-1}\mathrm{d}t
\end{equation*}
\citep[Proposition 2]{abdous1993}, one can be satisfied that no further improvements are possible (or are of practical interest).
Relative $\MISE^{\ast}$ is shown as dashed lines in Figure \ref{Fig:RelOptMISE.1} (left vertical axes). One immediate observation to be made is that for some distributions the best achievable reduction in MISE is quite small; e.g., for distributions \#3\&4 and sample sizes more than about one thousand, no more than 2-3\% reduction is possible. Nonetheless, for the small sample sizes the available improvement in MISE is substantial. Of course, any such improvement comes from a decrease in variance at the cost of introducing a non-zero bias. 

\newcommand{\FigRelOptMISEcaption}{Optimal kernel order and MISE}
\begin{figure}[!ht]\centering
\begin{tabular}{@{}>{\small}c@{\hspace{1mm}}@{\hspace{1mm}}>{\small}c@{\hspace{1mm}}@{\hspace{1mm}}>{\small}c@{}}
\#1: Gaussian & \#2: Skewed unimodal         & \#3: Strongly skewed         \\
\includegraphics[width=0.32\linewidth]{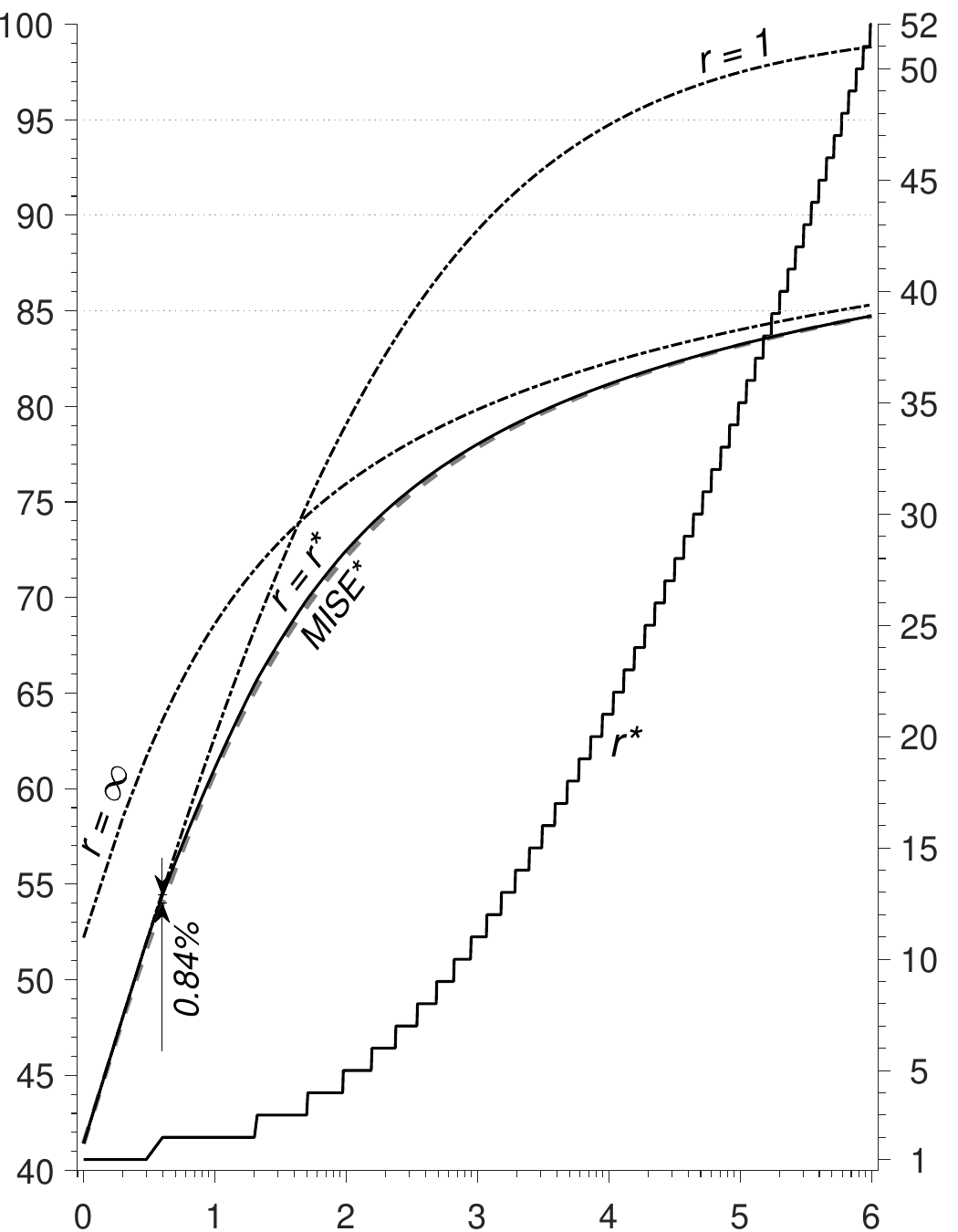} & 
\includegraphics[width=0.32\linewidth]{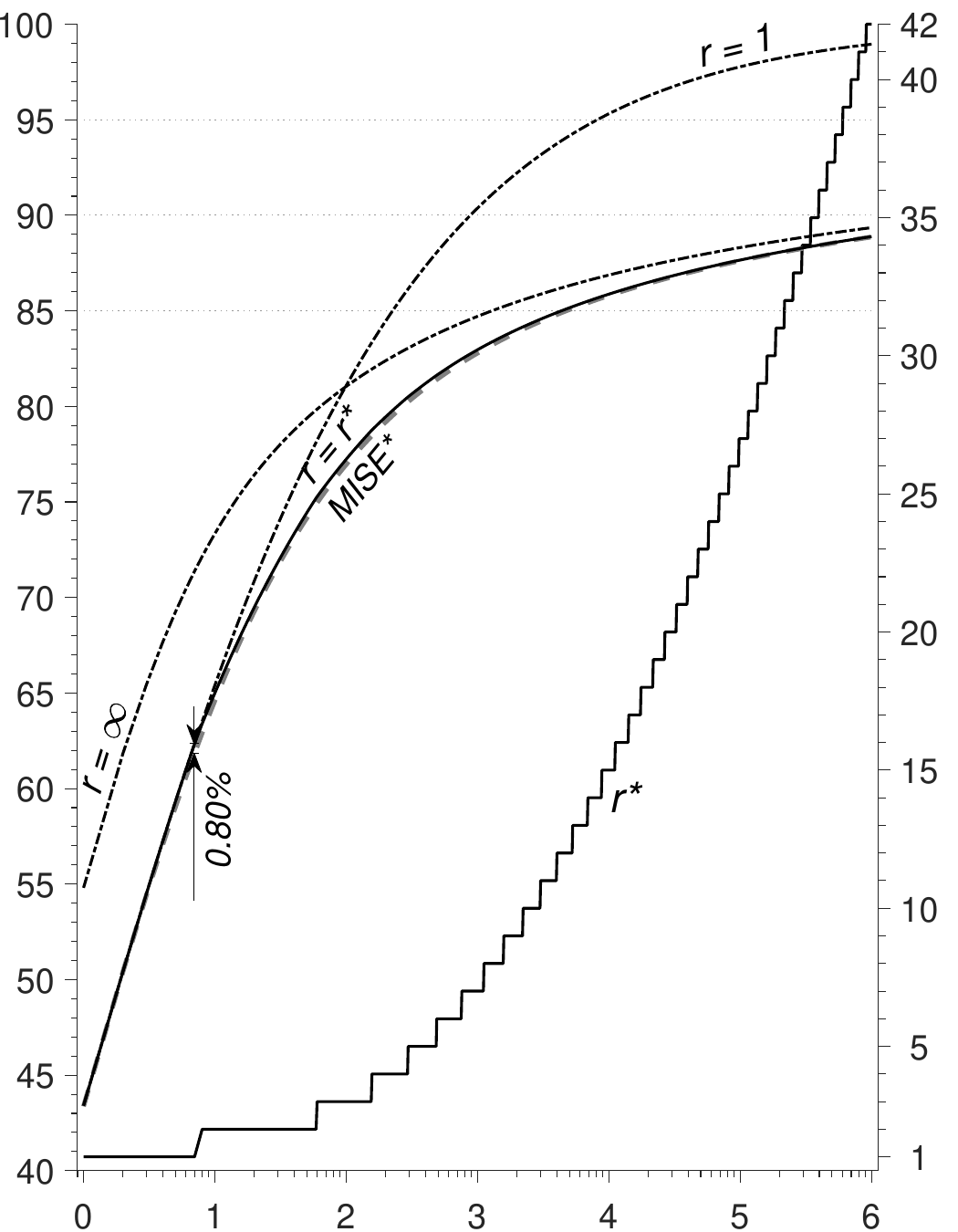} & 
\includegraphics[width=0.32\linewidth]{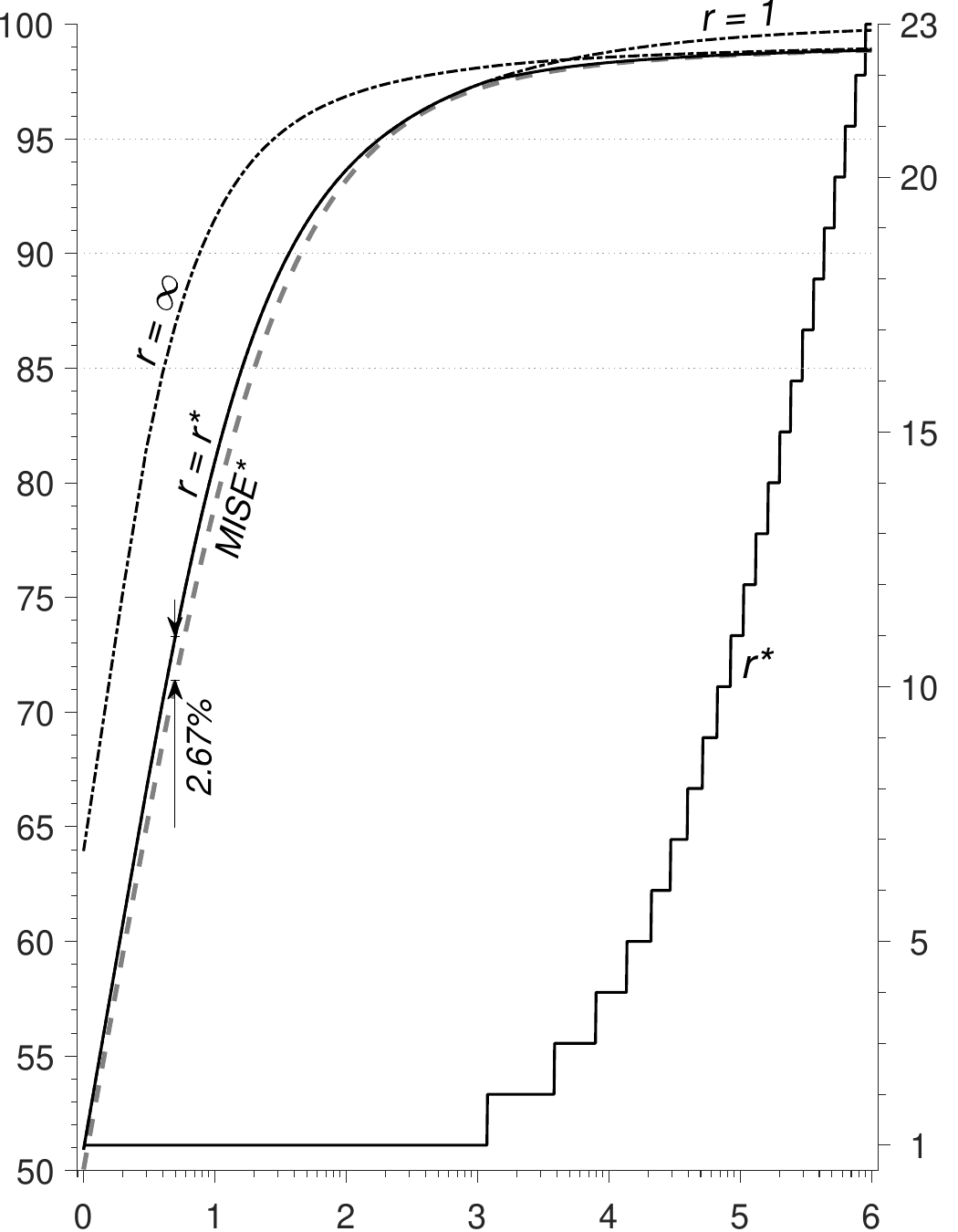} \\[3mm]
\#4: Kurtotic unimodal       & \#5: Outlier  & \#6: Bimodal                  \\
\includegraphics[width=0.32\linewidth]{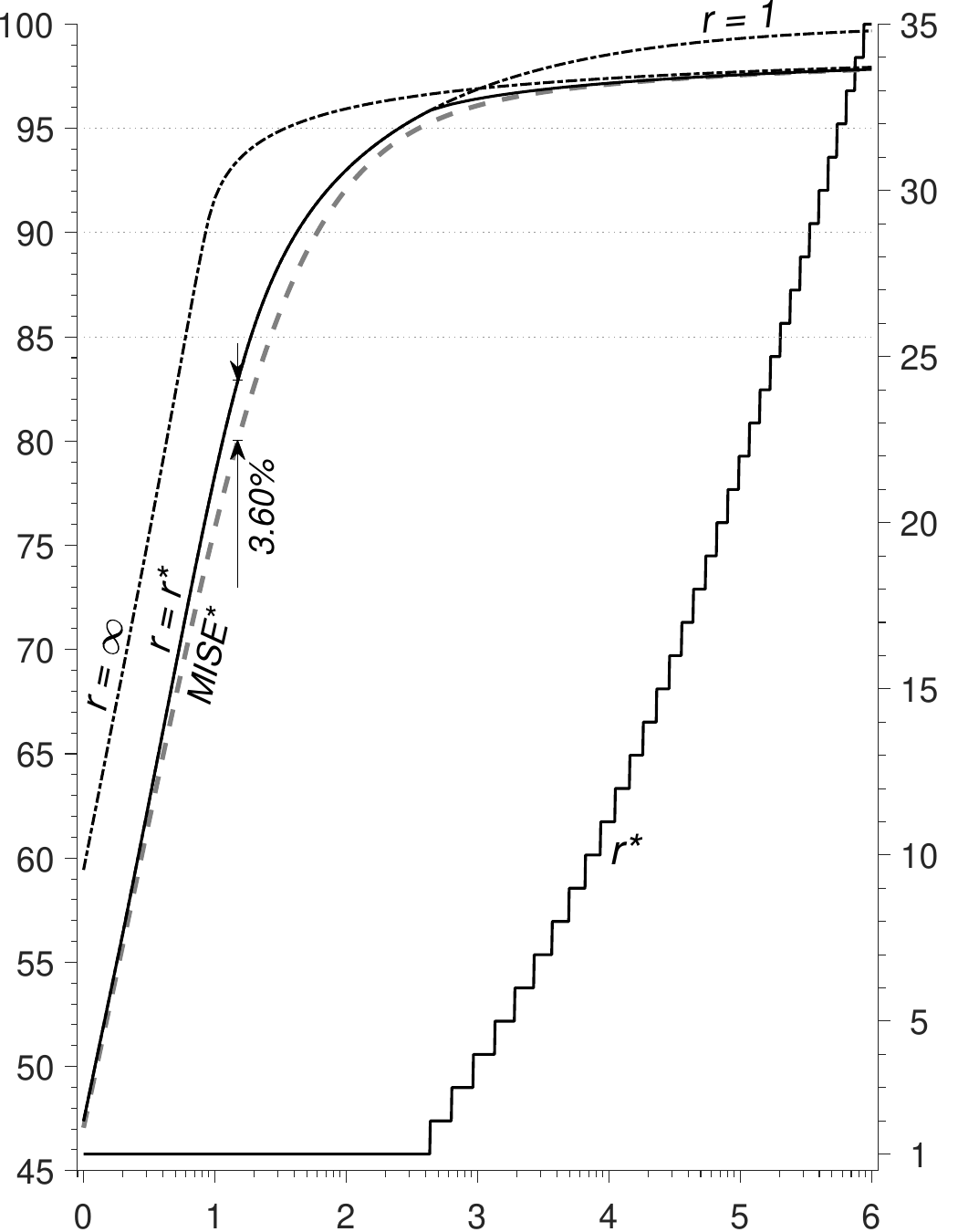} &
\includegraphics[width=0.32\linewidth]{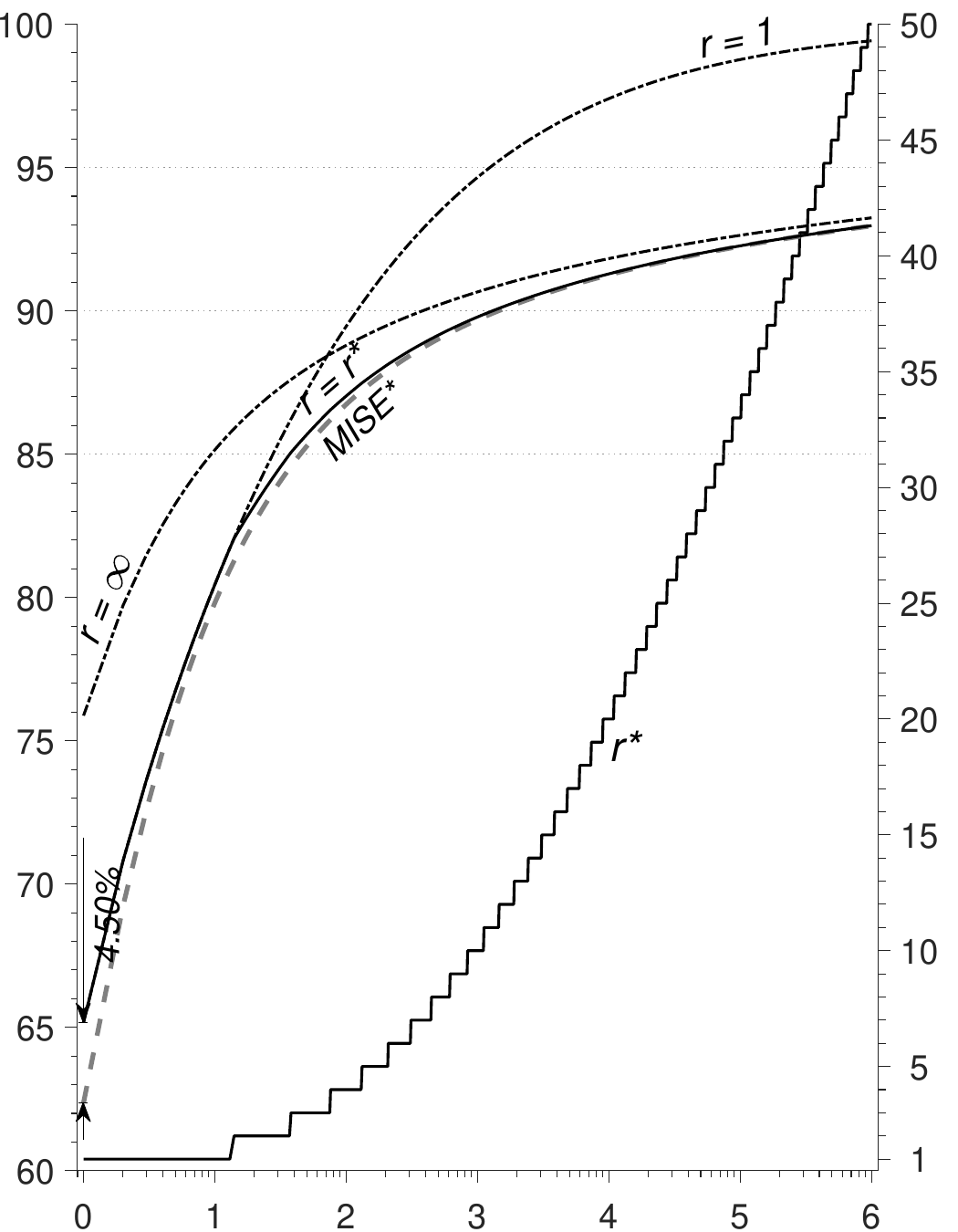} &
\includegraphics[width=0.32\linewidth]{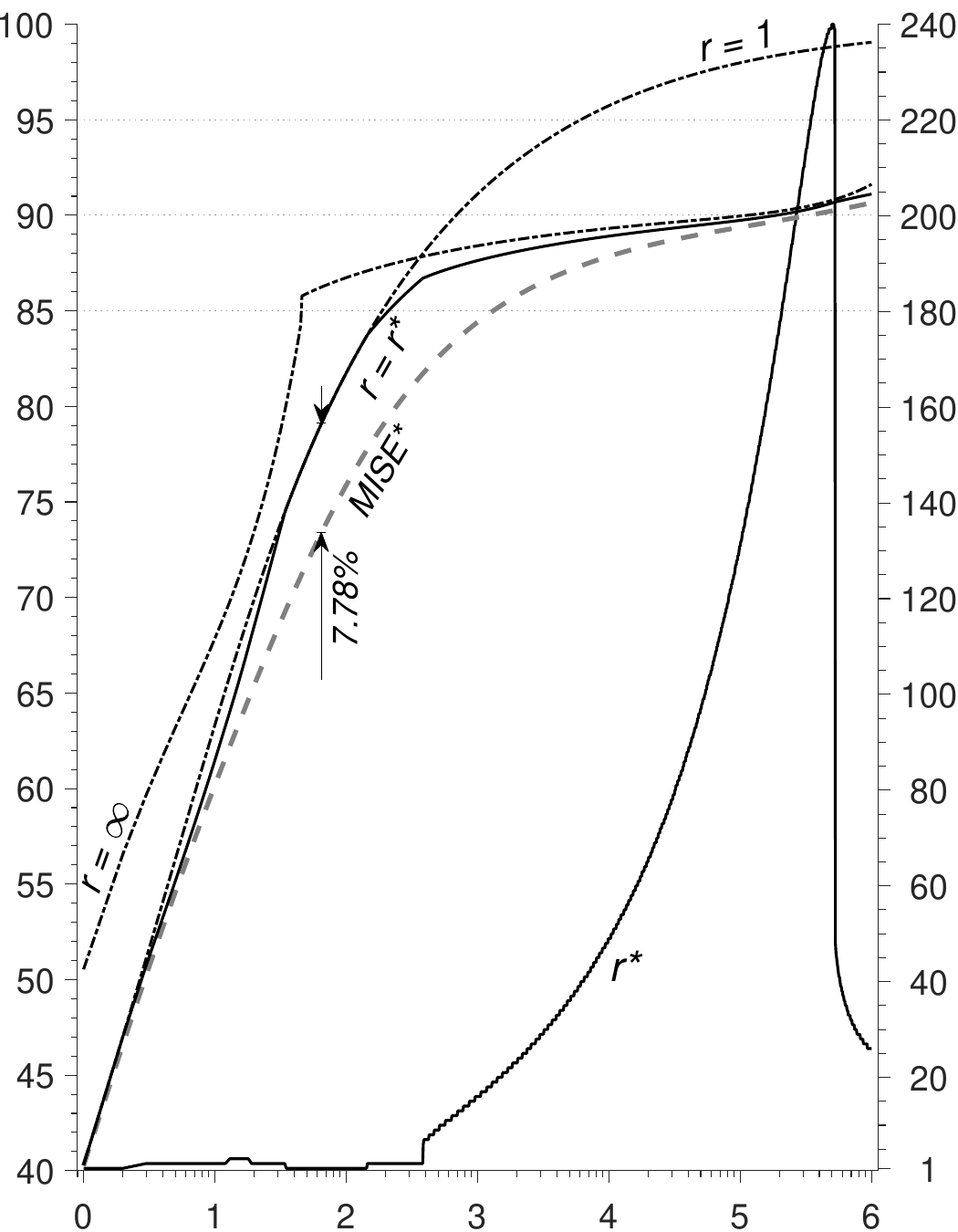} \\[3mm]
\end{tabular}
\begin{flushleft}
\footnotesize     
\vspace*{-0.5\baselineskip}
\noindent\textbf{Legend.} Horizontal axes: common logarithm of the sample size, $\log_{10}(n)$. 
Right vertical axes: \hdashrule[0.5ex][c]{20pt}{0.5pt}{}$\; r^{\ast}$, optimal $r$, integer $\geq1$. 
Left vertical axes, \%: 
\hdashrule[0.5ex][c]{25pt}{0.5pt}{5pt 2pt 1pt 2pt}\;$r=1$, 
\hdashrule[0.5ex][c]{25pt}{0.5pt}{5pt 2pt 1pt 2pt}\;$r=\infty$, and 
\hdashrule[0.5ex][c]{20pt}{0.5pt}{}\;$r=r^{\ast}$---minimum relative $\MISE$ with 2\textsuperscript{nd} ($r=1$), infinite, and optimal ($r=r^{\ast}$) order kernels, respectively; 
\hdashrule[0.5ex][c]{21pt}{1pt}{5pt 2pt}\;$\MISE^{\ast}$---minimum achievable relative MISE (infeasible).
\end{flushleft}
\vspace*{-1.2\baselineskip}
\caption{\FigRelOptMISEcaption}
\label{Fig:RelOptMISE.1}
\end{figure}

\renewcommand{\theHfigure}{cont.b.\arabic{figure}}
\renewcommand{\thefigure}{\arabic{figure} (Continued)}
\addtocounter{figure}{-1}
\begin{figure}[!htbp]\centering
\begin{tabular}{@{}>{\small}c@{\hspace{1mm}}@{\hspace{1mm}}>{\small}c@{\hspace{1mm}}@{\hspace{1mm}}>{\small}c@{}}
 \#7: Separated bimodal     & \#8: Skewed bimodal          & \#9: Trimodal      \\
\includegraphics[width=0.32\linewidth]{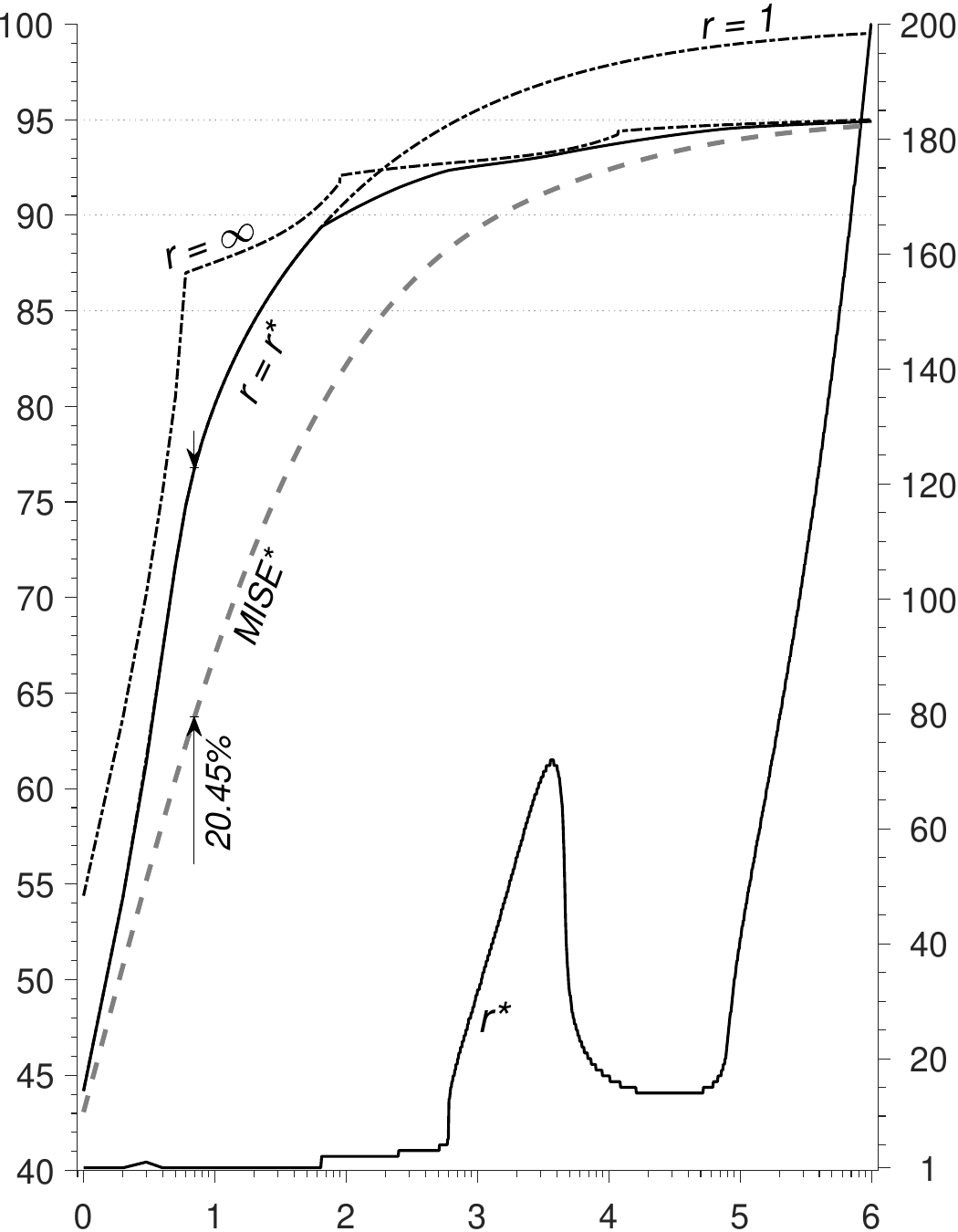} &
\includegraphics[width=0.32\linewidth]{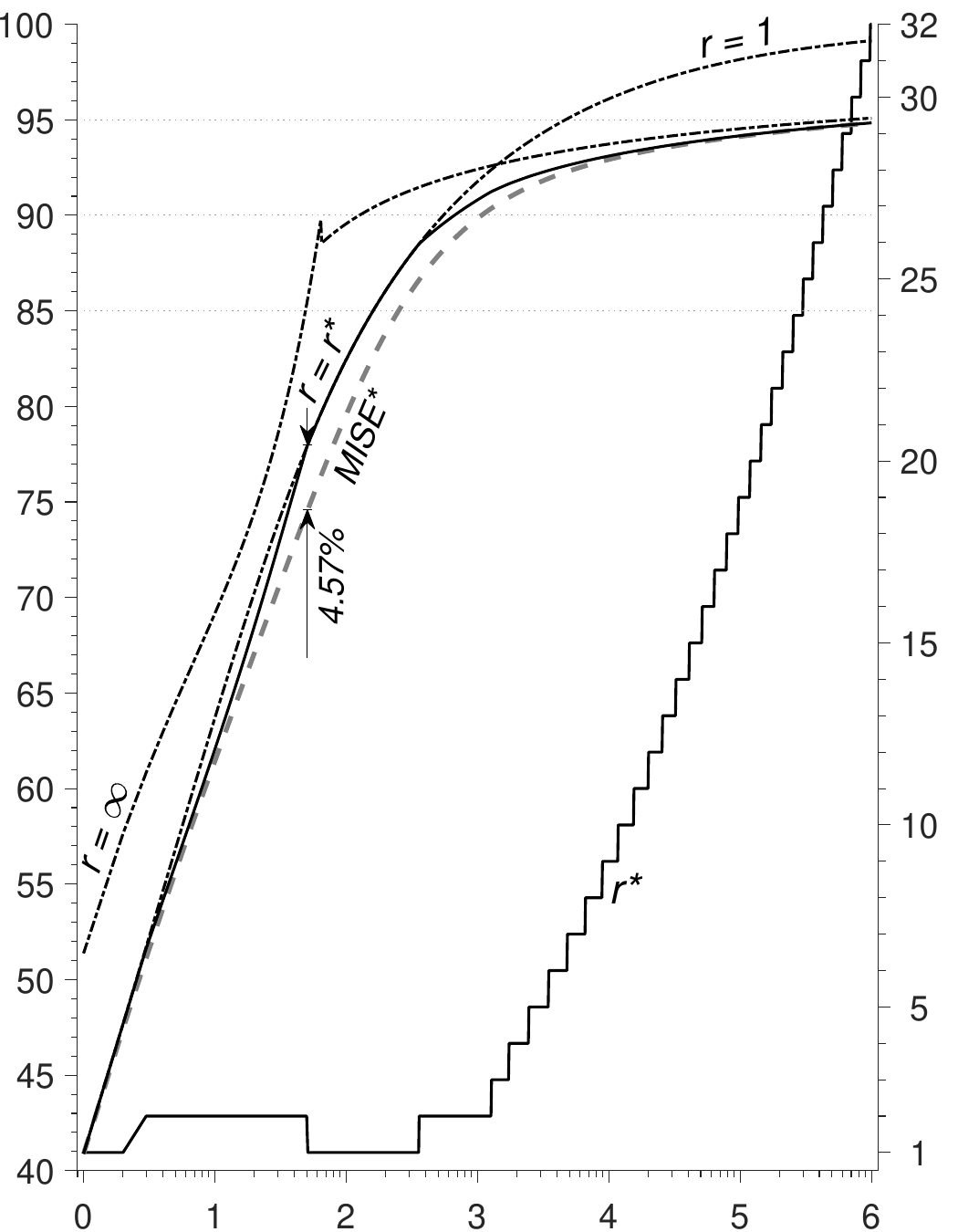} &
\includegraphics[width=0.32\linewidth]{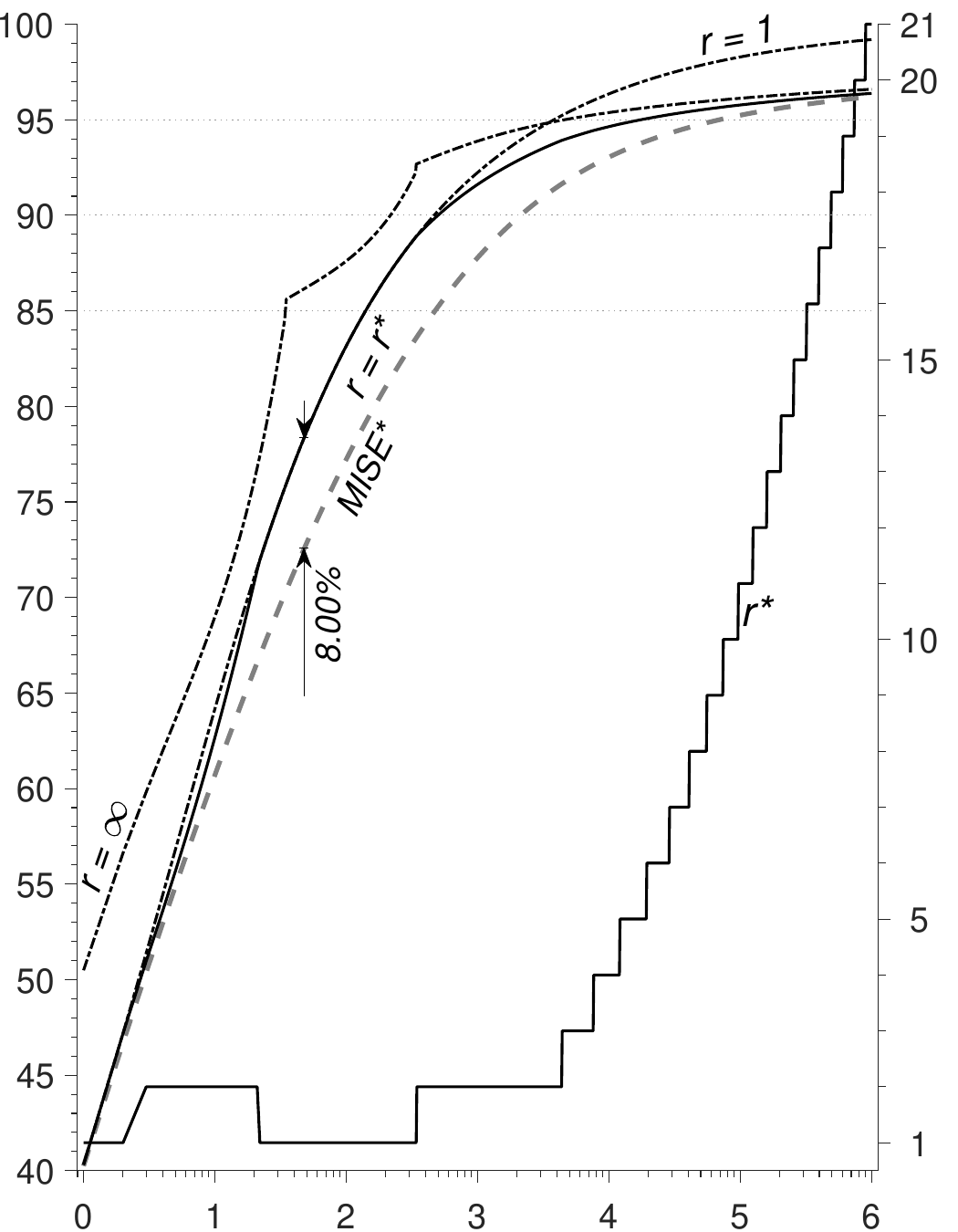} \\[3mm]
\#10: Claw                   & \#11: Double claw           & \#12: Asymmetric claw   \\
\includegraphics[width=0.32\linewidth]{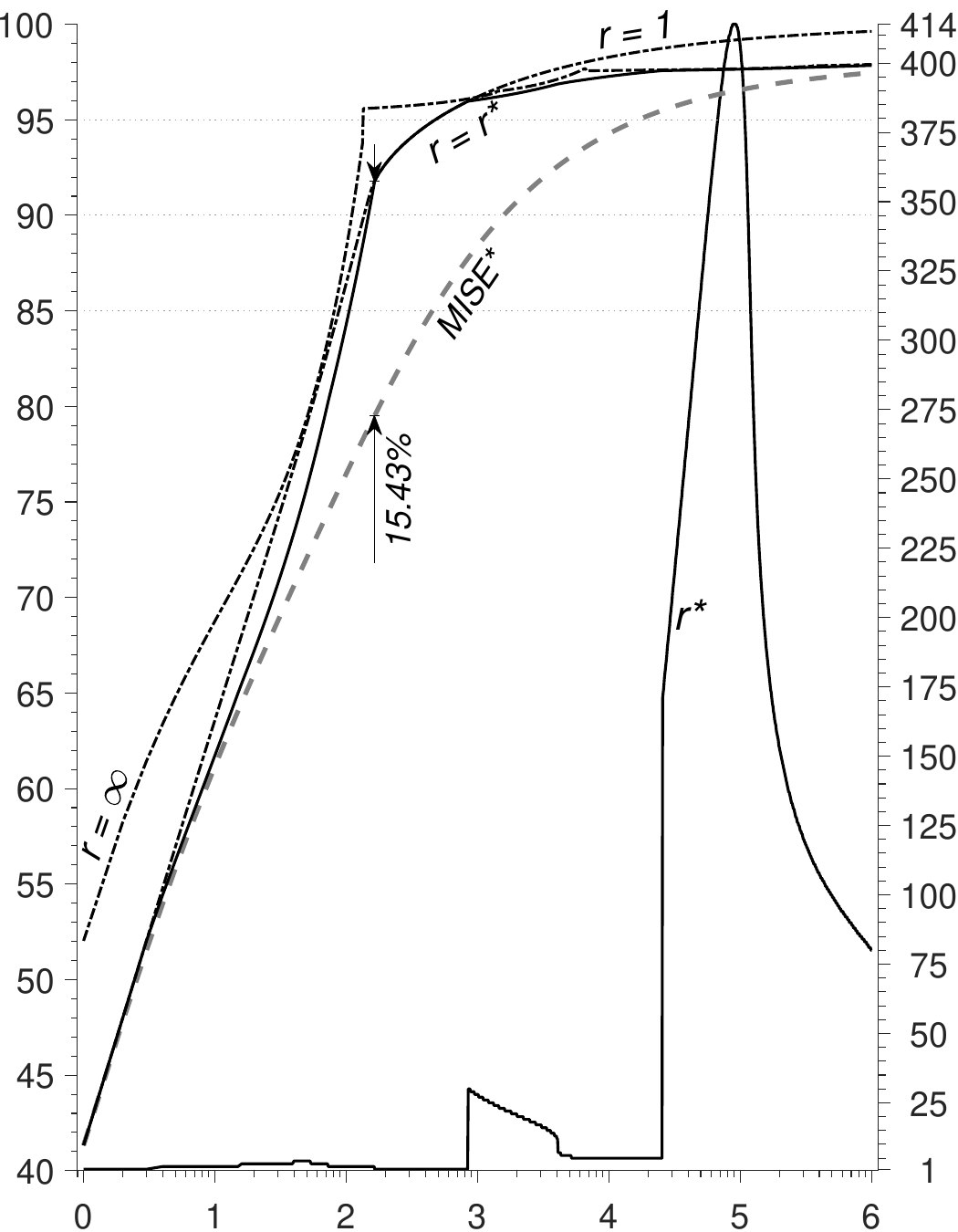} &
\includegraphics[width=0.32\linewidth]{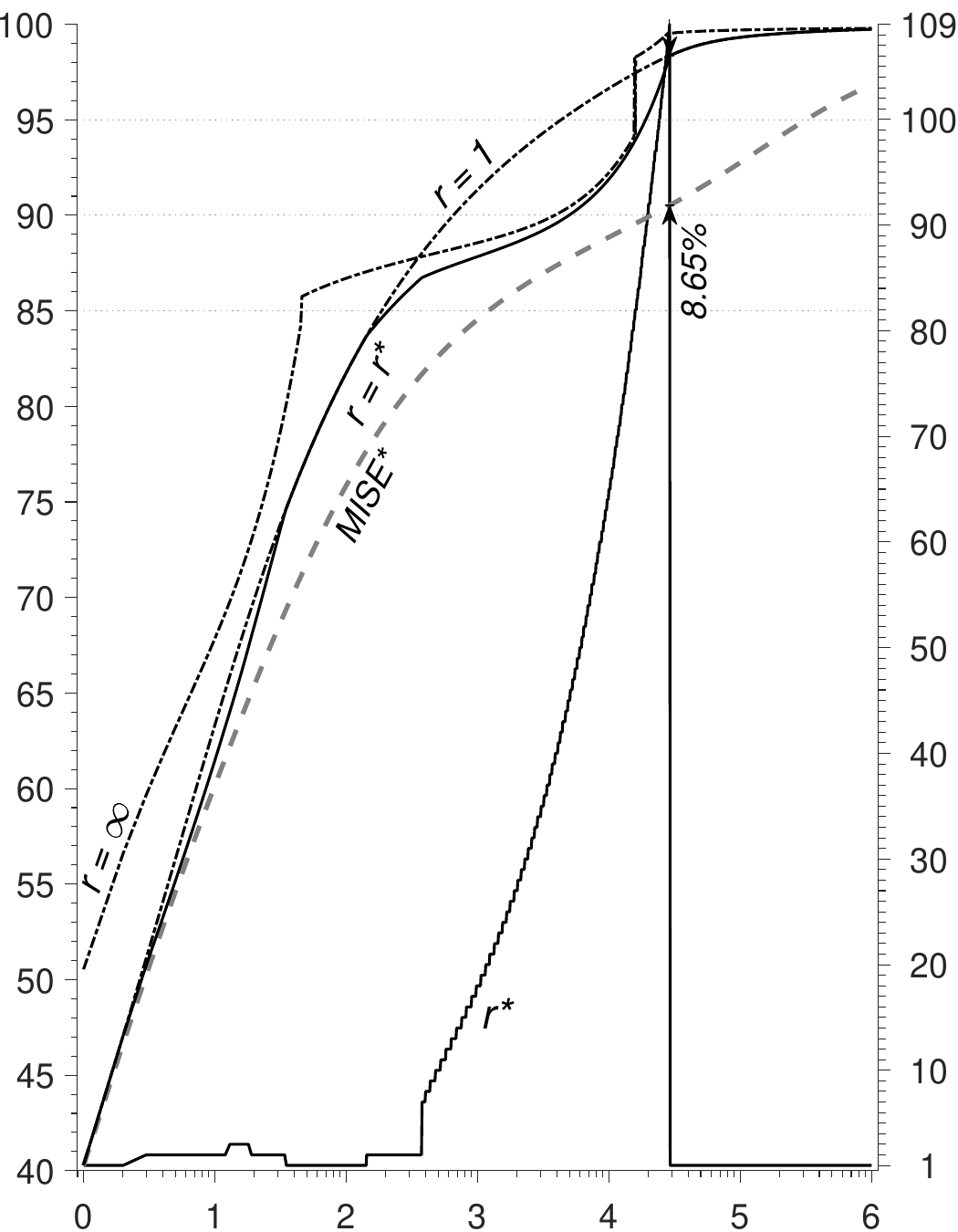} &
\includegraphics[width=0.32\linewidth]{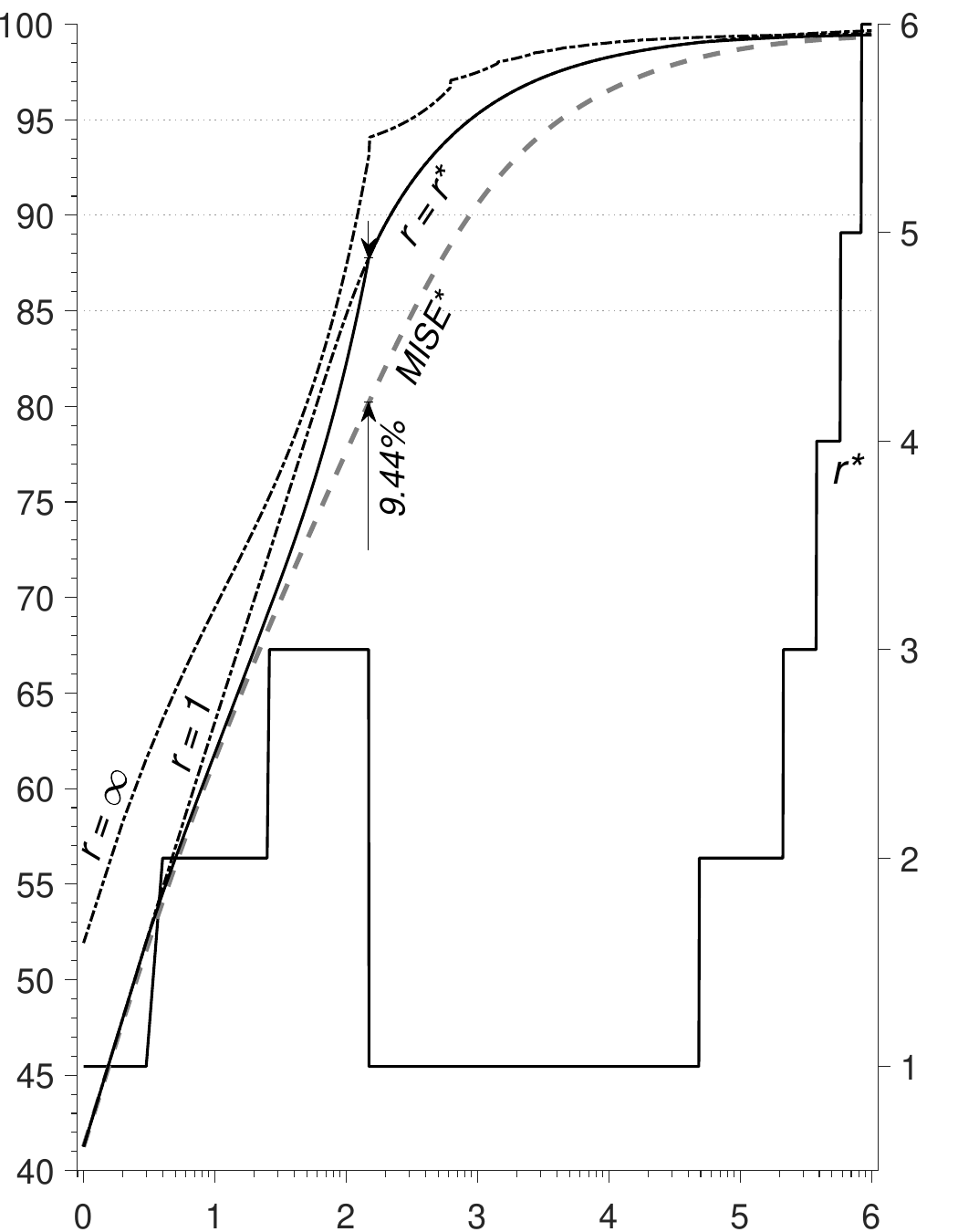} \\[3mm]
\#13: Asymmetric double claw & \#14: Smooth comb            & \#15: Discrete comb  \\
\includegraphics[width=0.32\linewidth]{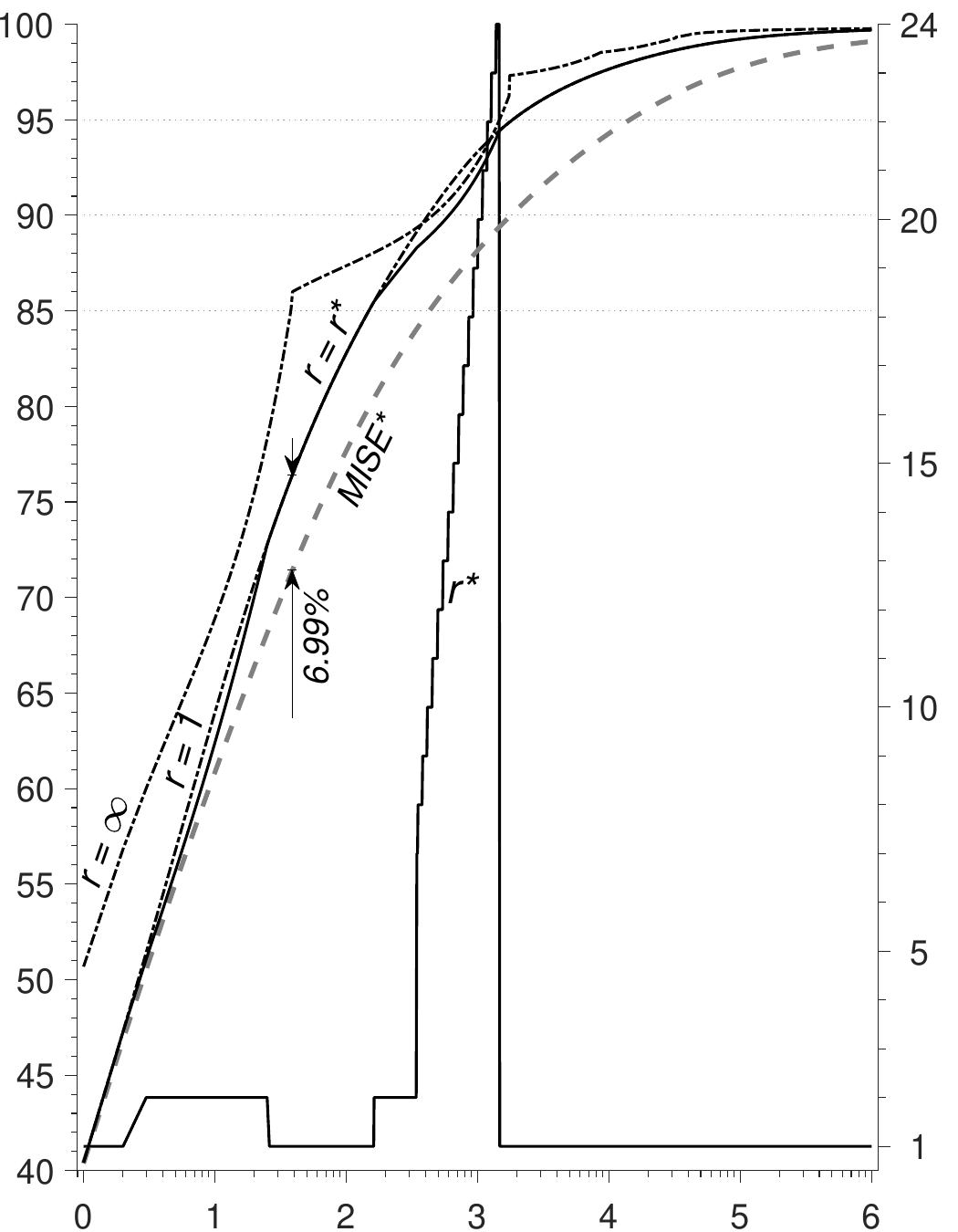} &
\includegraphics[width=0.32\linewidth]{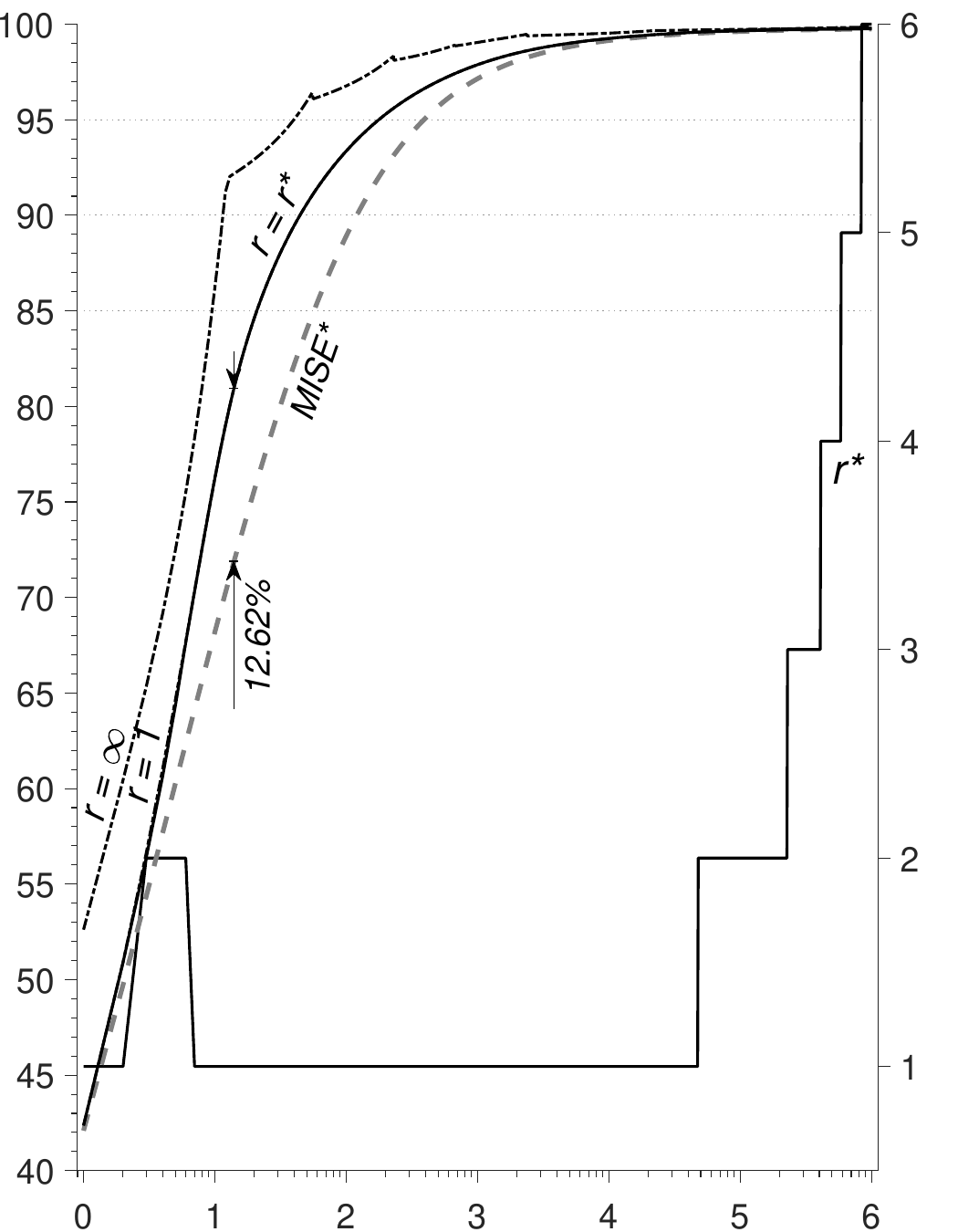} &
\includegraphics[width=0.32\linewidth]{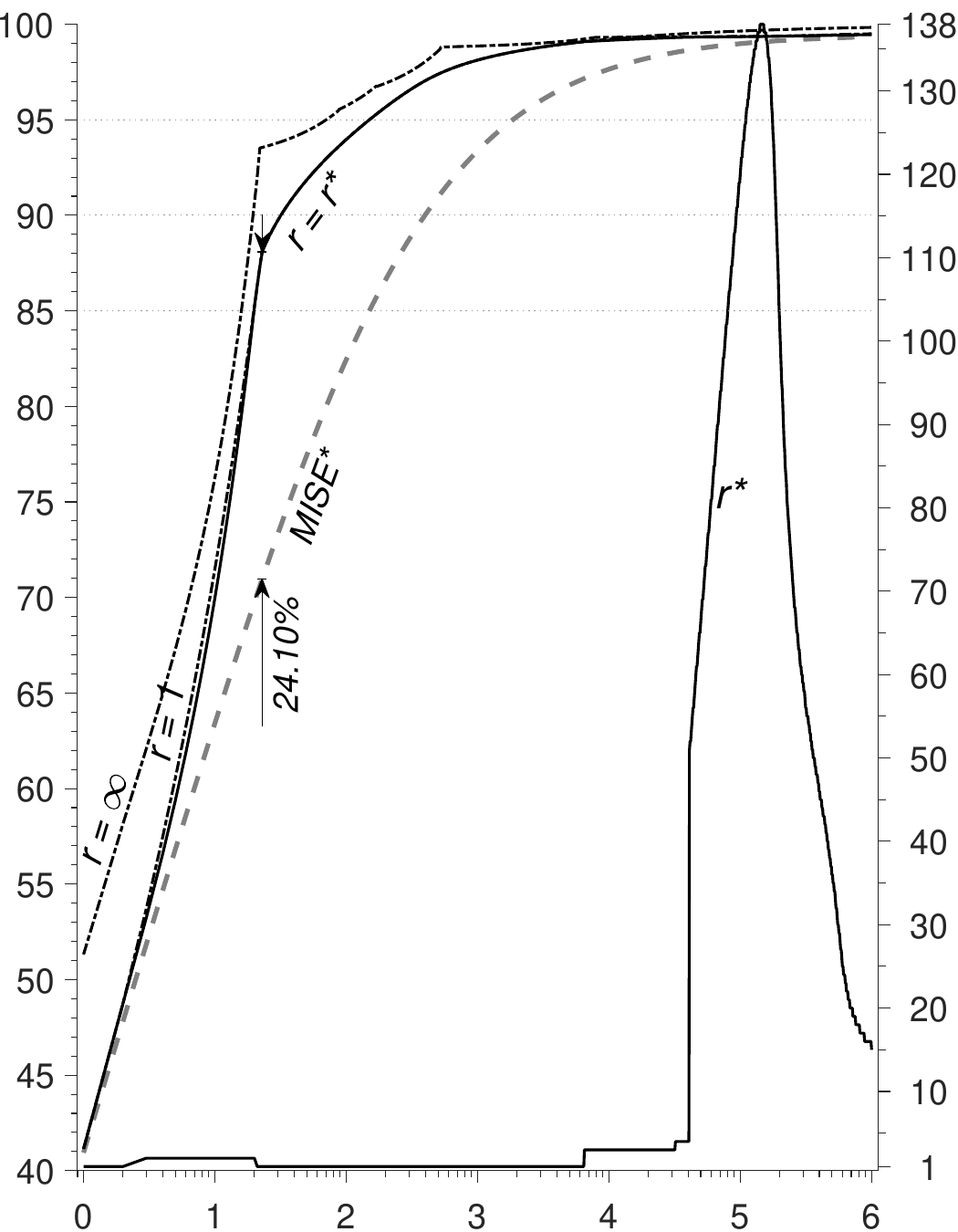} \\
\end{tabular}
\caption{\FigRelOptMISEcaption}
\label{Fig:RelOptMISE.2}
\end{figure}
\renewcommand{\thefigure}{\arabic{figure}}
\renewcommand{\theHfigure}{\arabic{figure}}

The $r^{\ast}$ (solid line, right vertical axes) and $r=r^{\ast}$ (solid line, left vertical axes) in Figure \ref{Fig:RelOptMISE.1} show the optimal $r$ and the resultant relative MISE for the class of Gaussian-based kernels of order $2r$. Optimisation was performed over $h\geq0$ for a given $r$ and then over $r\in\{1,\ldots,r_{\max}\}\cup\{\infty\}$ for a sufficiently large pre-specified $r_{\max}$. The kernel order necessary to achieve the best MISE generally increases with the sample size, but not necessarily in a monotone fashion. 

\begin{figure}[!htbp]\centering
\begin{tabular}{@{}>{\small}c@{\hspace{1mm}}@{\hspace{1mm}}>{\small}c@{\hspace{1mm}}@{\hspace{1mm}}>{\small}c@{}}
\#11: Double claw & \#13: Asymmetric double claw\\
\includegraphics[width=0.48\linewidth]{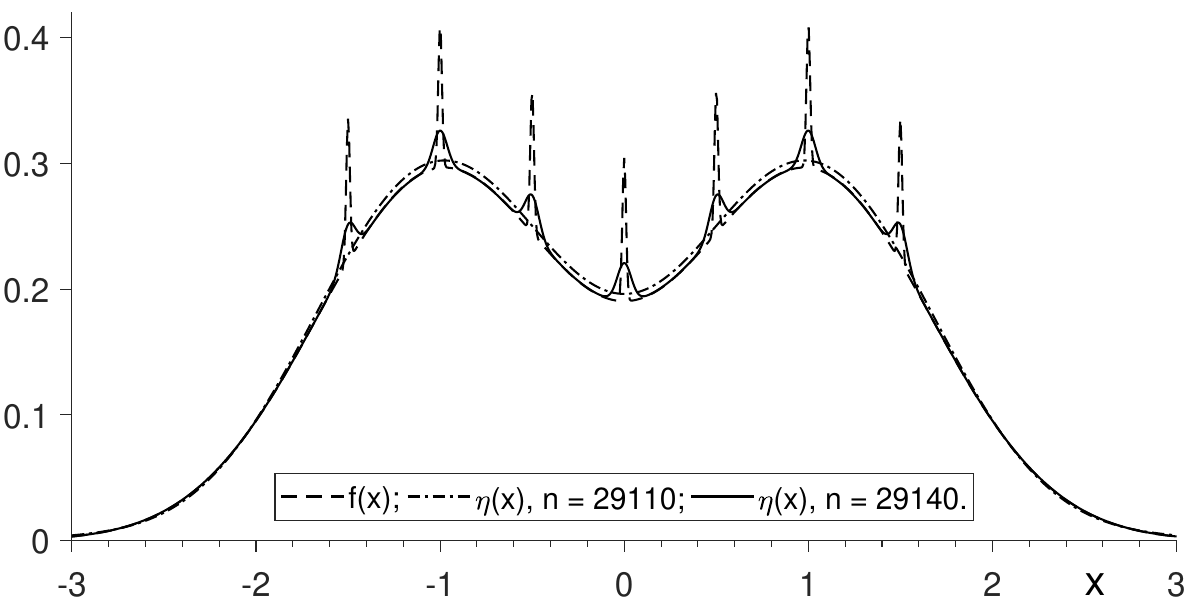} & 
\includegraphics[width=0.48\linewidth]{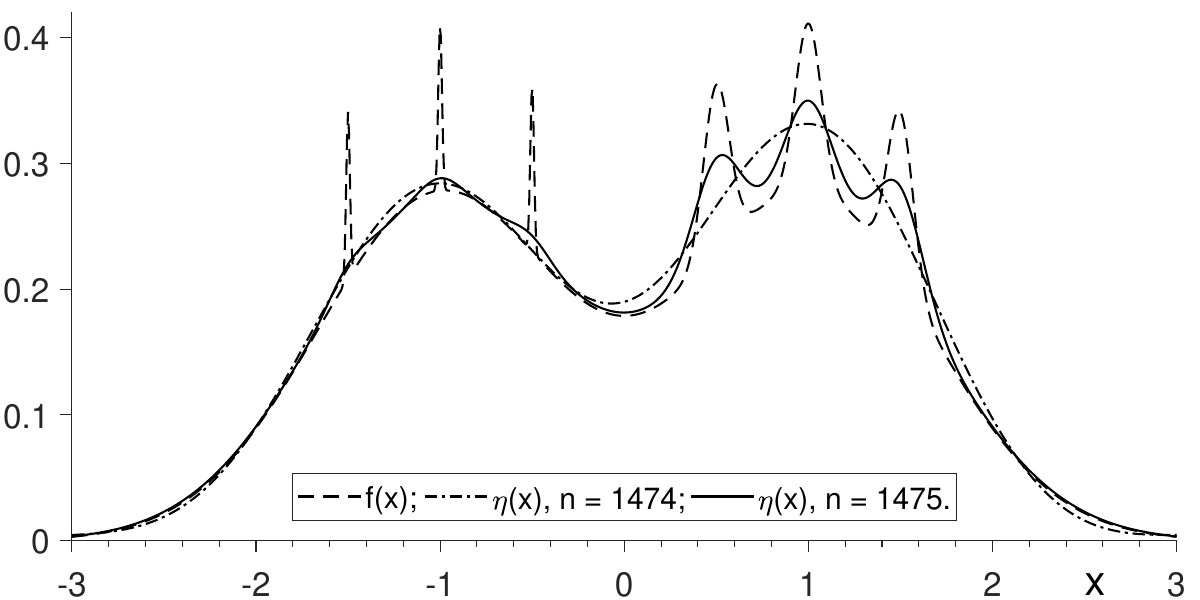} \\
\end{tabular}
\caption{Discontinuities in the optimal MISE}
\label{Fig:gkminEMISEetaDrop}
\end{figure}

The latter phenomenon is related to the discontinuities in the optimal bandwidth discussed in \citet[Section 5]{marron1992} and is a simple consequence of the bias-variance trade-off. For example, for the asymmetric double claw distribution \#13 with $n=1474$ it is optimal to smooth out the claws, and the optimal MISE of $4.384\times10^{-4}$ is achieved with $r^{\ast}=24$; the ISB and IVar are $0.329\times10^{-4}$ and $4.055\times10^{-4}$ respectively. With $n=1475$ it becomes optimal to apply less smoothing so that some of the features in the claws are visible. In this case the optimal MISE of $4.381\times10^{-4}$ is achieved with $r^{\ast}=1$ and ISB is substantially smaller, $0.121\times10^{-4}$, whereas the IVar is larger at $4.260\times10^{-4}$. 
To emphasise the differences, the right panel of Figure \ref{Fig:gkminEMISEetaDrop} shows the resultant expectations, $\E[\widehat{F}_{2r^{\ast}}(x;h_{e}^{\ast})]=\eta^{(-1)}(x)$, as the implied densities, $\eta(x)$, for $n=1474$ and $1475$ alongside the density $f(x)$. Similar discontinuities are observed with other distributions, e.g., with the double claw distribution \#11 between $n=29110$ and $29140$; see left panel of Figure \ref{Fig:gkminEMISEetaDrop}. 

For the relatively `uninteresting', i.e., close to normal distributions (\#1--5), the Gaussian-based kernels offer performance remarkably close to the best achievable (infeasible) MISE; the largest difference (shown by dimension lines) is less than 1\% for the Gaussian and skewed unimodal distributions, and between 2.7 and 4.5\% for the strongly skewed, kurtotic unimodal and outlier distributions. For the distributions with more complicated features (\#6--15) the differences can be as large as 10-20\% at the sample sizes of practical interest. 

There is little surprising about the performance of the second ($r=1$) and infinite ($r=\infty$) order kernels. The former performs well for small $n$, but as the bandwidth converges to zero at the fastest rate, the MISE of the KDFE quickly approaches that of the EDF. In contradistinction, the sinc kernel is expected to deliver best results as the sample size approaches infinity, but underperforms for finite $n$ thus rendering its practical usefulness questionable unless the sample size is very large. 

Importantly, for distributions \#1,2,5, which are close to normal, the benefits of using higher order kernels are realized for sample sizes as small as 10 observations; e.g., for the Gaussian distribution the fourth order kernel becomes optimal when $n=4$. The benefits are still clear for distributions \#3,4,6--9, albeit higher order kernels become optimal at sample sizes of around $1000$. For the remaining distributions (\#10--15) the picture is less clear. While the optimal $r$ is bigger than one over a range of sample sizes, the reduction in MISE it confers is either too small to matter in practice or occurs over a limited range of sample sizes which of course would not be known \textit{a priori}. 

\begin{figure}[!htbp]\centering
\begin{tabular}{@{}>{\small}c@{\hspace{1mm}}@{\hspace{1mm}}>{\small}c@{\hspace{1mm}}@{\hspace{1mm}}>{\small}c@{}}
\#1: Gaussian & \#6: Bimodal &  \#11: Double claw  \\
\includegraphics[width=0.32\linewidth]{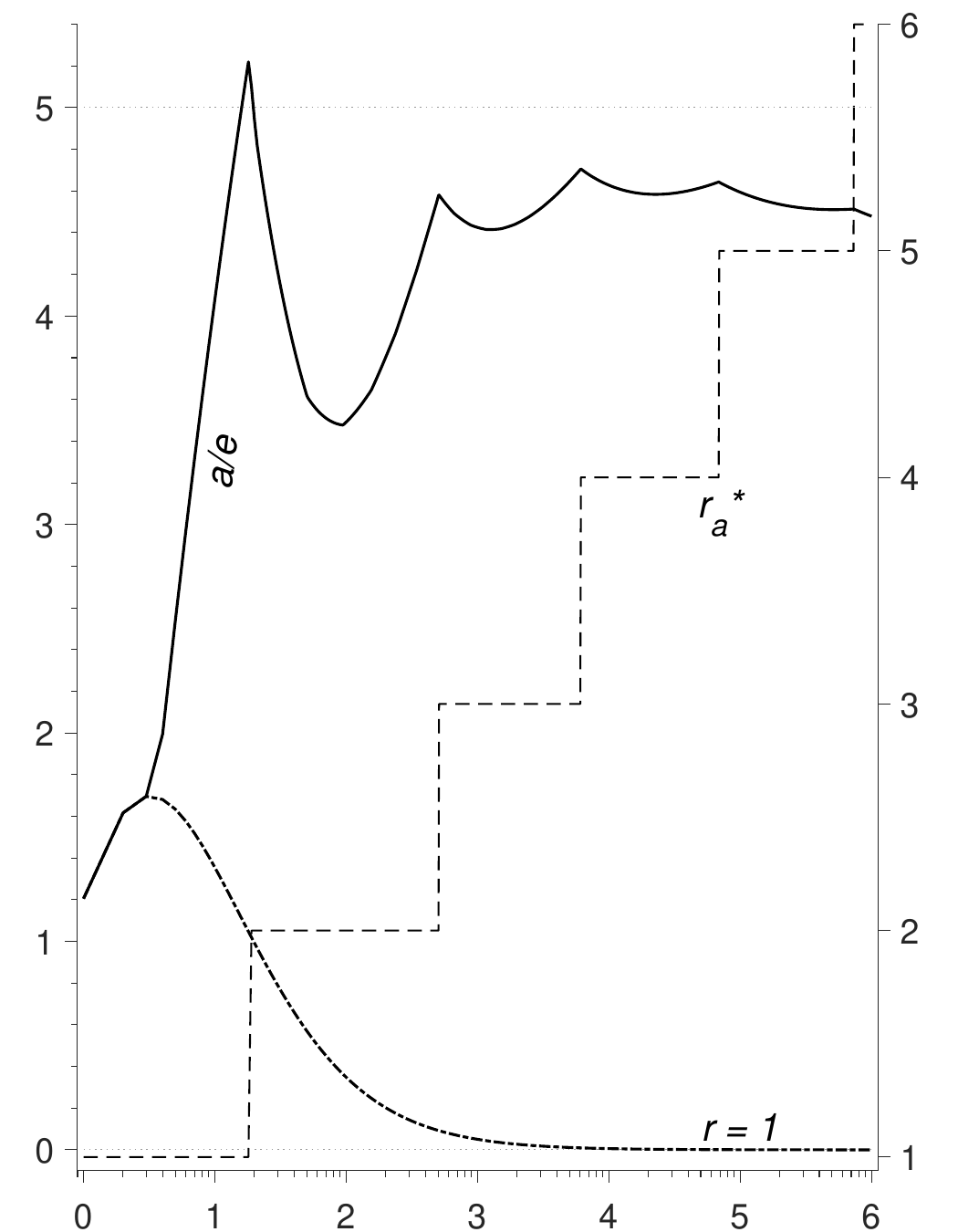} & 
\includegraphics[width=0.32\linewidth]{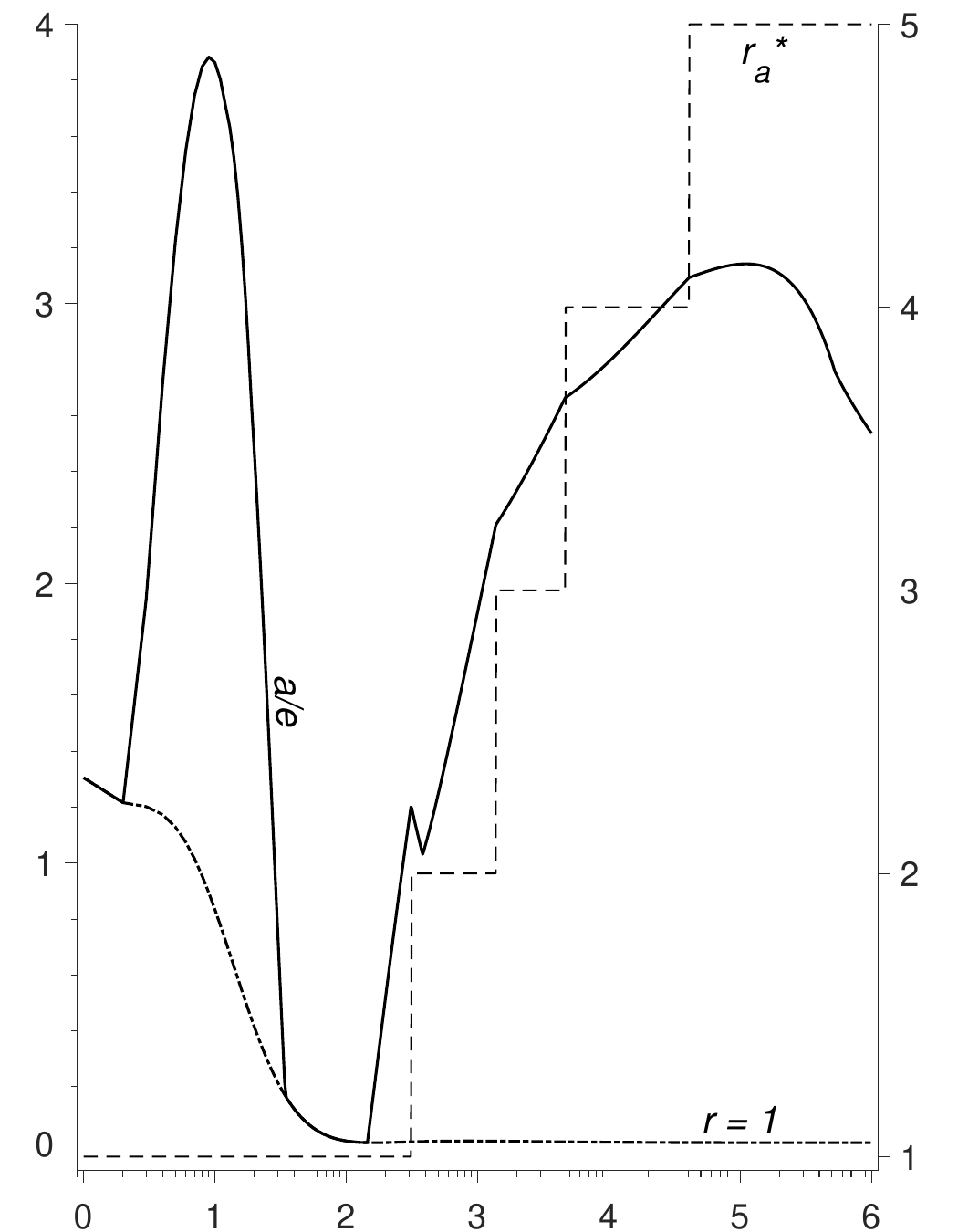} & 
\includegraphics[width=0.32\linewidth]{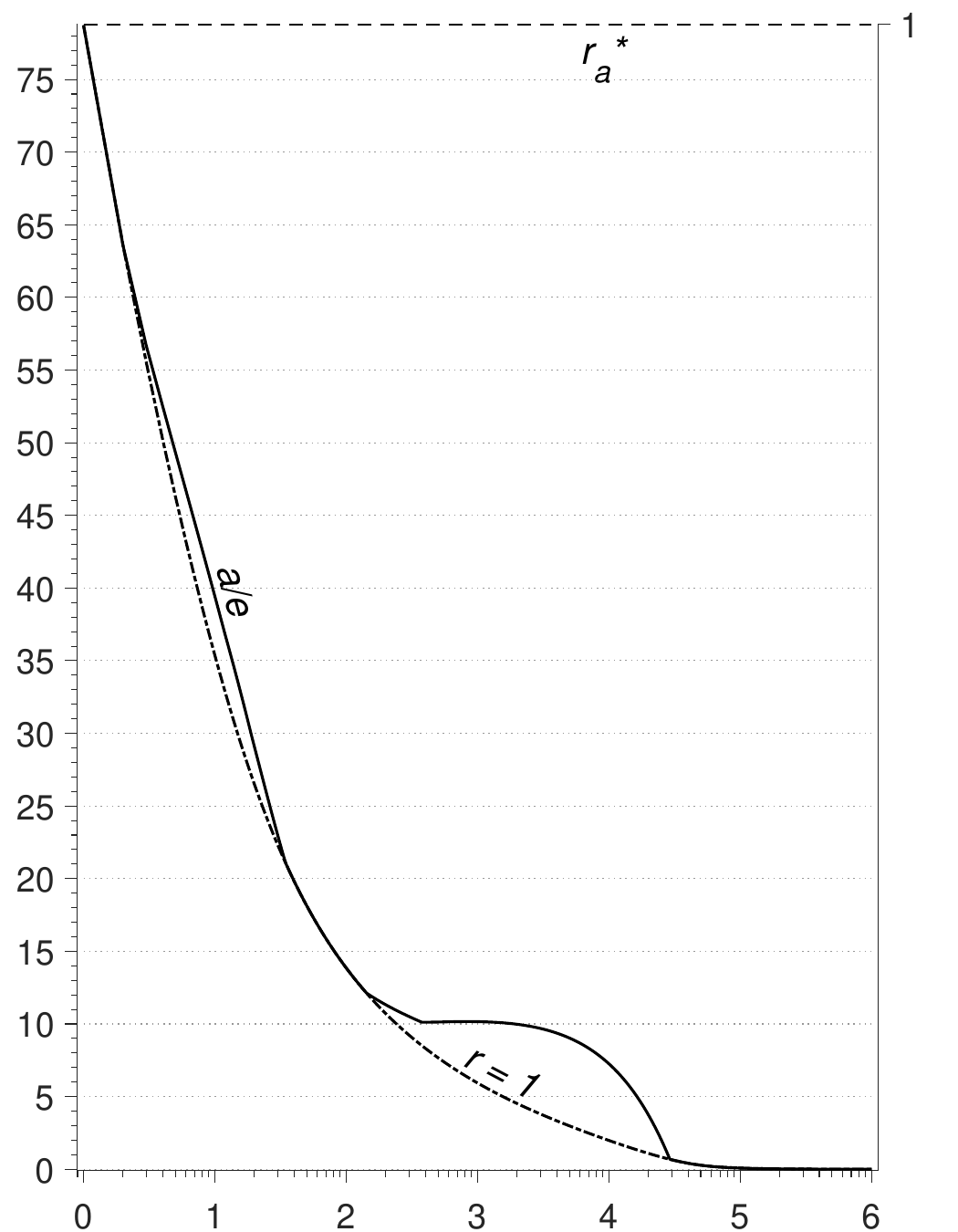} \\ 
\end{tabular}
\begin{flushleft}
\footnotesize     
\vspace*{-0.5\baselineskip}
\noindent\textbf{Legend.} Horizontal axes: common logarithm of the sample size, $\log_{10}(n)$. 
Right vertical axes: \hdashrule[0.5ex][c]{21pt}{0.5pt}{5pt 2pt}$\; r_{a}^{\ast}$, integer $\geq1$: optimal $r$ with $h_{a}^{\ast}$. Left vertical axes, \%: \hdashrule[0.5ex][c]{25pt}{1pt}{5pt 2pt 1pt 2pt}\;$r=1$---relative loss of efficiency if the asymptotically optimal bandwidth is used with $r=1$, $100\{\MISE[\widehat{F}_{2}(\cdot,h_{a}^{\ast})]/\MISE[\widehat{F}_{2}(\cdot,h_{e}^{\ast})]-1\}$; 
\hdashrule[0.5ex][c]{25pt}{1pt}{}\; a/e---relative loss of efficiency with optimal $r$,  
$100\{\MISE[\widehat{F}_{2r_{a}^{\ast}}(\cdot,h_{a}^{\ast})]/\MISE[\widehat{F}_{2r^{\ast}}(\cdot,h_{e}^{\ast})]-1\}$.
\end{flushleft}
\vspace*{-1.2\baselineskip}
\caption{Loss of efficiency with the asymptotically optimal bandwidth}
\label{Fig:RelOptMISE.asy.1}
\end{figure}

Finally, if one were to use the asymptotically optimal bandwidth $h^{\ast}_{a}$ instead of the exact MISE-minimising bandwidth $h_{e}^{\ast}$,  the optimal $r$, $r_{a}^{\ast}$, also generally increases with the sample size, but monotone and much slower; see Figure \ref{Fig:RelOptMISE.asy.1}, right vertical axes. (Results for the distributions not shown in Figure \ref{Fig:RelOptMISE.asy.1} are available upon request.) The corresponding relative MISE is always bigger than MISE with $h^{\ast}_{e}$ and $r^{\ast}$, and usually much bigger in small  samples illustrating the fact that the choice of the asymptotic bandwidth may lead to poor performance. For example, for the double claw distribution, the loss exceeds 5\% for $n\lesssim16,500$. 
The quality of the asymptotic approximation to MISE, shown for $r=1$ in Figure \ref{Fig:RelOptMISE.asy.1}, generally deteriorates as $r$ increases. This can be attributed to the rate of the asymptotically optimal bandwidth slowing as $r$ increases. 

\section{Uniform kernel}\label{sec:uniform.kernel}
If $k$ is a second order kernel, then necessarily $\psi(k)>0$, and with the asymptotically optimal bandwidth $h_{a}^{\ast}$, the MISE \eqref{Eq:Asy.MISE.opt.bw} becomes
\begin{equation*}
\MISE[\widehat{F}(\cdot;h_{a}^{\ast})] = n^{-1}V_{F}
- \tfrac{3}{4}\alpha(k)^{4/3}R(F^{(2)})^{-1/3}n^{-4/3} +o(n^{-4/3}),
\end{equation*} 
where $\alpha(k)=\psi(k)/\sqrt{\mu_{2}(k)}$. 
Following the approach of \citet{marron1988}, \citet{jones1990} shows that maximising $\alpha(k)$ yields the asymptotically optimal kernel for all values of $h$. Due to scale invariance of $\alpha(k)$, i.e., $\alpha(k_{\delta})=\alpha(k)$ for any $\delta>0$, the problem is equivalent to maximising $\psi(k)$ while keeping $\mu_{2}(k)$ constant. The unique second order kernel maximising $\psi(k)$ is the uniform kernel for which $\alpha(k)^{4/3}=3^{-2/3}$. 

Asymptotically, using a different, suboptimal, second order kernel $k$ results in the relative loss in efficiency of
\begin{equation*}
D=\MISE[\widehat{F}(\cdot;h_{a}^{\ast})]/\MISE[\widehat{F}_{u}(\cdot;h_{u,a}^{\ast})]-1
=  \tfrac{3}{4}[3^{-2/3}-\alpha(k)^{4/3}]R(F^{(2)})^{-1/3}V_{F}^{-1}n^{-1/3}  +o(n^{-1/3}),
\end{equation*} 
where the subscript $u$ stands for the uniform kernel. For example, if $k=g_{2}$, the second order Gaussian kernel, $\alpha(g_{2})^{4/3}=\pi^{-2/3}$, 
and if $F$ is the normal cdf, $R(F^{(2)})=1/(2\sqrt{\pi}\sigma^{3})$, $V_{F}=\sigma/\sqrt{\pi}$, and hence 
$D\approx 0.0295 n^{-1/3}$. 
\citet{jones1990} reports the relative values of $\alpha(k)$ for several other popular kernels and concludes that the loss of efficiency resulting from the use of these kernels is generally negligible. 

In finite samples, however, the uniform kernel need not be optimal. 

\begin{theorem}\label{Thm:KCDFE.Exact.MISE.Unif.Kernel}
Let $X_{1},\ldots,X_{n}$ be a random sample from an $m$-component normal mixture distribution \eqref{Eq:Def:NM.dens}, and 
$k(z)=\I\{-1\leq z\leq 1\}/2$. 
Then for $h>0$, 
\begin{subequations}\begin{align}
\label{Thm:KCDFE.Exact.MISE.Unif.Kernel.ISB}
\ISB[\widehat{F}(\cdot;h)]  & = -\frac{1}{2h^{2}}[J(2h;-4)-J(0;-4)] +\frac{2}{h}J(h;-3)
-\frac{\sigma_{f}^{2}}{2h}-\frac{h}{6}-J(0;-2),\\
\label{Thm:KCDFE.Exact.MISE.Unif.Kernel.IVar}
\IVar[\widehat{F}(\cdot;h)] & = -\frac{2h}{3n}
+\frac{1}{2h^{2}n}[J(2h;-4)-J(0;-4)]-\frac{\sigma_{f}^{2}}{2hn},
\end{align}\end{subequations}
where $J(x;p)  = \sum_{i=1}^{m}\sum_{j=1}^{m}w_{i}w_{j}\phi^{(p)}(x;\mu_{i}-\mu_{j},\sigma_{ij,0}^{2})$ and 
$\sigma_{f}^{2}=\sum_{i=1}^{m}w_{i}(\mu_{i}^{2}+\sigma_{i}^{2})-(\sum_{i=1}^{m}w_{i}\mu_{i})^{2}$.
\end{theorem}
Note that $J(0;-2)=V(h;0,0)=U(h;0)$. 
As before, the minimiser of MISE, $h^{\ast}_{u,e}$, can be obtained by standard numerical optimisation techniques.

\begin{figure}[!htbp]\centering
\begin{tabular}{@{}>{\small}c@{\hspace{1mm}}@{\hspace{1mm}}>{\small}c@{\hspace{1mm}}@{\hspace{1mm}}>{\small}c@{}}
\#1: Gaussian & \#3: Strongly skewed  & \#12: Asymmetric claw        \\
\includegraphics[width=0.32\linewidth]{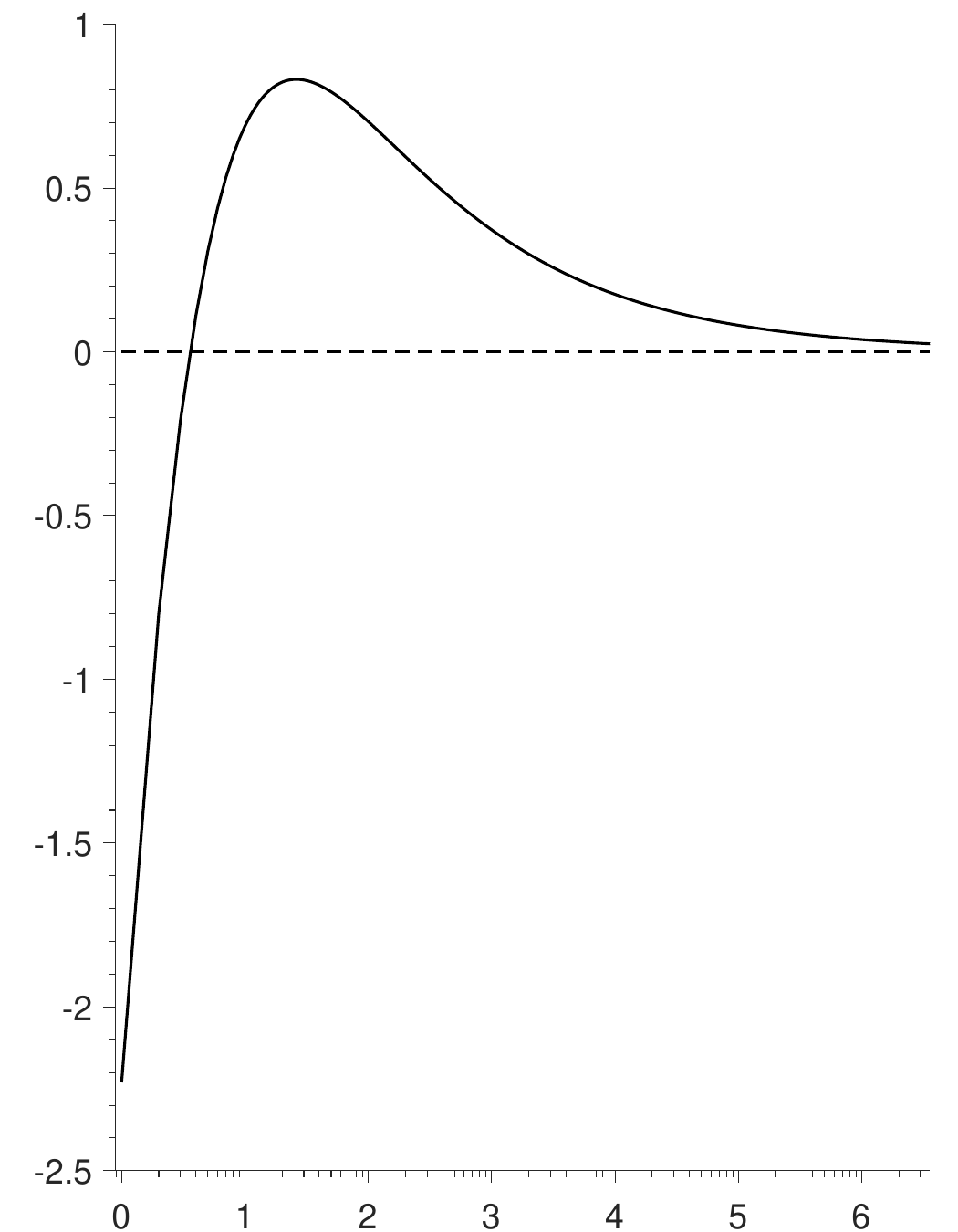} & 
\includegraphics[width=0.32\linewidth]{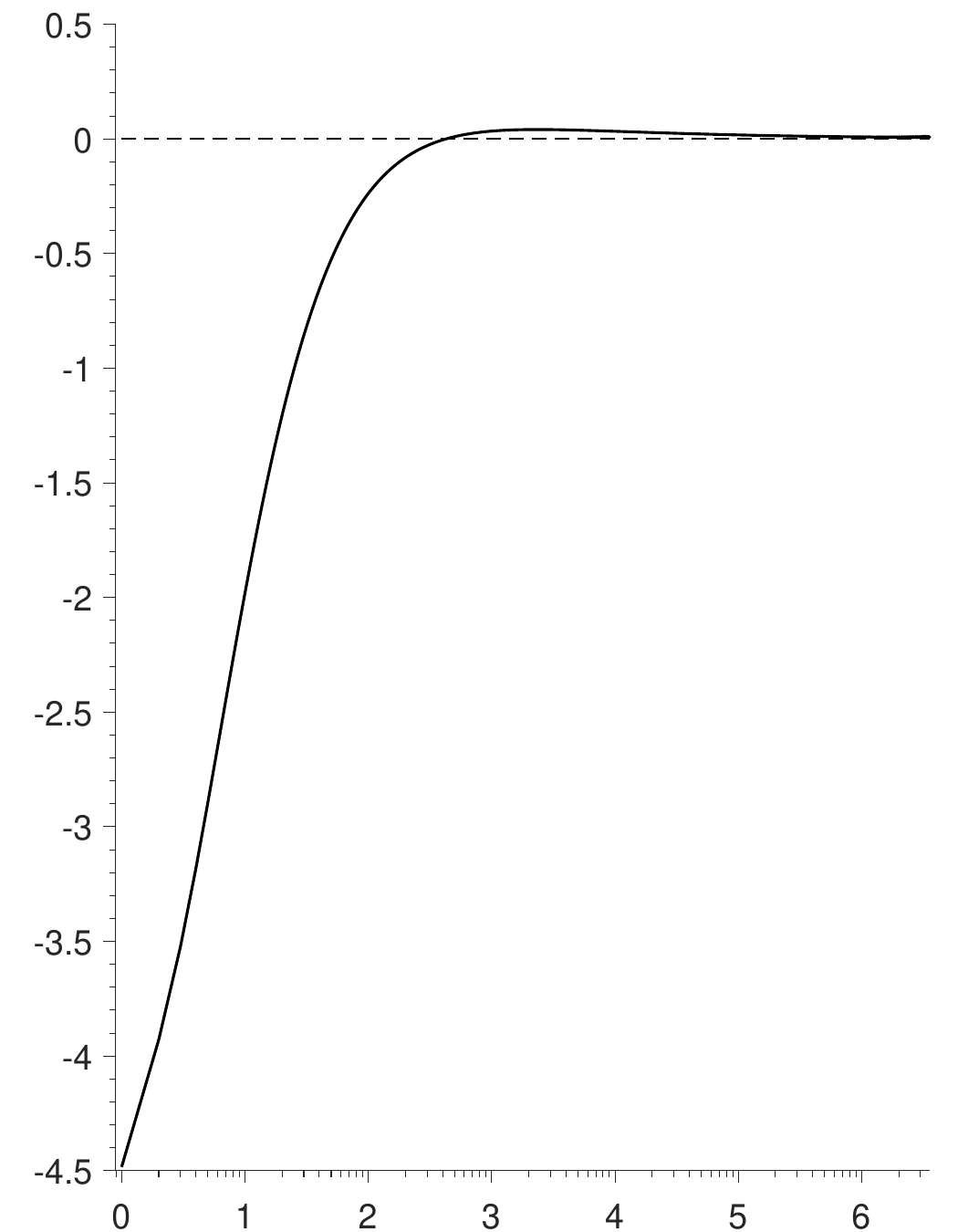} & 
\includegraphics[width=0.32\linewidth]{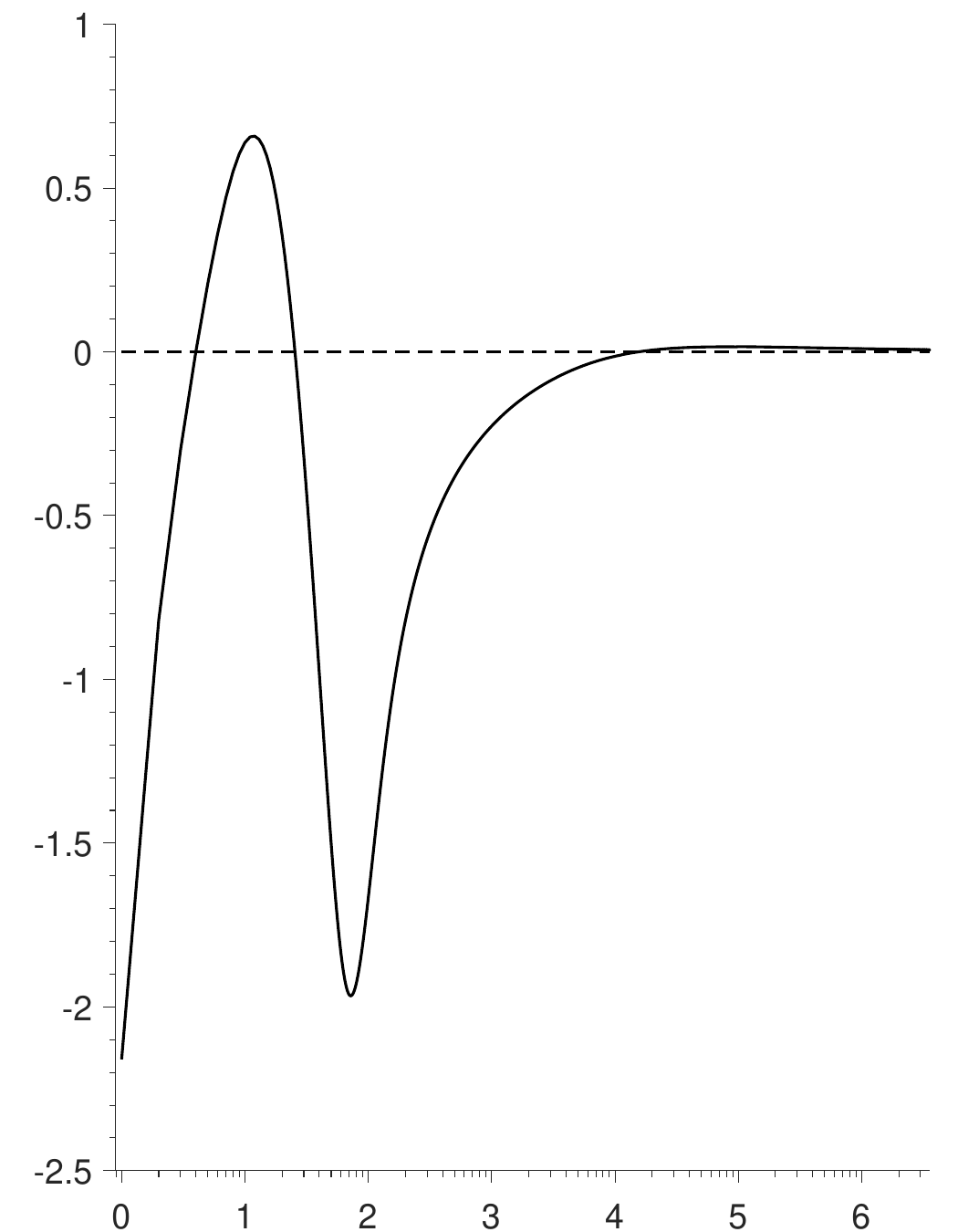} \\
\end{tabular}
\begin{flushleft}
\footnotesize     
\vspace*{-0.5\baselineskip}
\noindent\textbf{Legend.} Horizontal axes: common logarithm of the sample size, $\log_{10}(n)$. 
Vertical axes: percentage relative loss of efficiency,  $100\{\MISE[\widehat{F}_{2}(\cdot;h_{2,e}^{\ast})]/\MISE[\widehat{F}_{u}(\cdot;h_{u,e}^{\ast})]-1\}$.
\end{flushleft}
\vspace*{-1.2\baselineskip}
\caption{Exact percentage relative loss of efficiency resulting from using the second order Gaussian kernel compared to the  uniform kernel}
\label{Fig:RelOptMISE.Uk.G2k}
\end{figure}

Examples shown in Figure \ref{Fig:RelOptMISE.Uk.G2k} illustrate that the differences in exact MISE of $\widehat{F}$ (with exact mise-minimising bandwidth) with the second order Gaussian kernel versus the uniform kernel are generally quite small. For the Gaussian distribution the asymptotic approximation is reliable in samples as small as 50 observations, and only with $n<4$ observations the Gaussian kernel delivers better results than the uniform kernel. The maximum loss of efficiency is $0.83\%$ with $n=26$, and this was the largest loss of efficiency observed in the 15 distributions examined. (Results for the distributions not shown in Figure \ref{Fig:RelOptMISE.Uk.G2k} are available upon request.)

For other densities, the Gaussian kernel can outperform the uniform kernel by 2\% or more at sample sizes of interest. 
Furthermore, large sample sizes may be required for the asymptotic optimality result to manifest itself. 
For example, for the asymmetric claw distribution \#12, uniform kernel performs better than Gaussian for $4\leq n\leq 25$, and then worse until $n$ exceeds approximately $15,600$, at which point the differences become too small to be of practical interest.

\section{Bandwidth selection}\label{Sec:bandwidth.selection}
Exact MISE results for the NM distributions lend themselves naturally to the possibility of estimating the optimal bandwidth \emph{and} kernel order by parametrically fitting a finite normal mixture distribution to the data and plugging this preliminary estimate into the exact MISE expression which can then be minimised over $h$ and $r$. 

The fundamental difference between this approach and the methods based on asymptotic approximations to MISE is the nature of the approximation involved. The proposed method utilises the exact finite sample MISE expressions, but will generally involve a non-asymptotic finite normal mixture approximation to the unknown distribution, unless the latter is in fact a normal mixture. It is thus reasonable to expect that while neither method will be uniformly best, the proposed approach will perform better in situations where the asymptotic approximation to MISE is poor and the distribution can be well approximated by a normal mixture. Extensions to other classes of distributions are also possible. 

The main competitor of the proposed NM plug-in approach is the cross-validation (CV) bandwidth of \citet{bowman1998} which directly minimises an estimate of MISE. (CV criterion is an unbiased estimator of MISE for sample size $n-1$, up to an additive constant which does not depend on $h$). CV approach performs well in simulations and has the advantage that it does not require any preliminary estimates. Comparisons of the CV approach with the methods based on asymptotic approximations can be found in, e.g., \citet{tenreiro2006}. In general, neither method is uniformly better than others. 

The NM plug-in approach yields estimates of $h$ and $r$, and thus can be expected to outcompete CV in cases where higher order kernels provide a substantial improvement in MISE and the underlying distribution can be well approximated by a finite normal mixture. The presence of the approximation error and the need to fit a mixture distribution, including determining the number of components, are the main drawbacks of the proposed procedure. 

In the remainder of this section the performance of the NM plug-in approach is assessed via a simulation study using the same fifteen NM distributions as in section \ref{sec:analysis.of.mise} (Figure \ref{Fig:MWNMdens1-15}) and three non-NM distributions: a Gamma(2,1) distribution  \citep[cf.][Section 4]{bowman1998} and Student $t$ distributions with $3$ and $4$ degrees of freedom. 

At a preliminary stage, an $m$-component normal mixture distribution is fitted to the data using the Expectation Maximization (EM) algorithm; see, e.g., \citet{mclachlan2000}. The number of components is chosen\footnotemark\ by the Akaike (AIC) and Bayesian (BIC) information criteria. The resultant estimates of $m$ are denoted by $\hat{m}_{A}$ and $\hat{m}_{B}$ respectively. There is considerable evidence that supports the use of BIC to select the number of components. In particular, \citet{roeder1997} show that if the goal is to estimate a density $f$ by a univariate normal mixture, choosing $m$ by BIC yields a consistent estimator of $f$; see also \citet[Section 6.9.3]{mclachlan2000}, \citet{fraley2002}, and references therein. 
For the NM distributions the true number of components, $m_{0}$, is also included for comparison. 

\footnotetext{The search is performed over $m\in\{1,\ldots,m_{\max}\}$, where for the NM distributions $m_{\max}$ is set to the true number of components plus 4, and for the non-NM distributions $m_{\max}=10$. 
The fit is repeated 10 times, each time with  new randomly chosen initial parameters, and the fit with the largest likelihood is chosen as the final estimate. If the EM algorithm fails at some  $m^{\prime}<m_{\max}$, the search domain is simply truncated to $\{1,\ldots,m^{\prime}\}$. The cases in which the algorithm failed to fit an $m_{0}$-component mixture were excluded from comparisons involving $m_{0}$ (there were none or very few such cases for distributions \#1,4-9; for distributions \#2,3,10-15 the proportions of such cases were between 6.8--30\% for $n=50$, diminishing to between 0.3--1.5\% for $n=400$).}

At the main stage, the preliminary estimate $\widehat{F}_{\hat{m}}(x) = \sum_{j=1}^{\hat{m}}\hat{w}_{j}\Phi((x-\hat{\mu}_{j})/\hat{\sigma}_{j}^{2})$ is treated as if it were the known true NM distribution to find the optimal bandwidth, $\hat{h}$, and kernel order, $\hat{r}^{\ast}$, by minimising the exact MISE expression.
As in Section \ref{sec:analysis.of.mise}, optimisation was performed over $r\in\{1,\ldots,r_{\max}\}$, where guided by the results presented in Figure \ref{Fig:RelOptMISE.1}, $r_{\max}$ was set to 8, 9, 10, and 13 for sample sizes 50, 100, 200, and 400, respectively. Performance of the resultant KDFE, $\widehat{F}_{2\hat{r}^{\ast}}(x;\hat{h})$, is evaluated by its 
integrated squared error, $\ISE[\widehat{F}(\cdot;\hat{h})]=\int_{-\infty}^{\infty}\{\widehat{F}(x;\hat{h})-F(x)\}^{2}\mathrm{d}x$.
 Comparisons are also made with the case where $r$ is set to one and only the bandwidth is estimated. CV bandwidth is also computed for the second order kernel only. 

Simulation results are reported in Table \ref{Tab:BW.RelMISE.and.pairwise.t.tests}. In all cases these are based on 10,000 random draws. The columns correspond to different combinations of $r$ and $m$ used to construct $\widehat{F}_{2\hat{r}^{\ast}}(x;\hat{h})$. These are compared to the EDF and KDFE with CV bandwidth. The last six columns correspond to comparisons between the KDFE and the parametrically fitted $m$-component NM distribution with the same choice of the number of components, i.e., the preliminary estimate. The entries in the table show the relative reduction in MISE of KDFE versus the benchmark, in percentages. Positive entries, shown in italics, correspond to cases where KDFE performs worse than the benchmark. For comparison, column LB (lower bound) reproduces the best achievable reduction in MISE of KDFE relative to EDF obtainable with the optimal order Gaussian-based kernel when $F$ is known and the infeasible exact MISE-minimising bandwidth is used (line $r=r^{\ast}$ in Figure \ref{Fig:RelOptMISE.1}, minus 100).

\renewcommand{\arraystretch}{0.92}
\setlength{\tabcolsep}{4pt}
\newcommand{\TabRelMISEcaption}{Relative reduction in MISE of KDFE and p-values for the two-sided paired $t$-tests for equality of the considered ISE means}
\begin{sidewaystable}[htp]\centering\small
\caption{\TabRelMISEcaption}
\label{Tab:BW.RelMISE.and.pairwise.t.tests}
\begin{tabular}{l|rrrrrrrr|rrrr|rrrrrr}\hline\hline
\multirowcell{2}{\diaghead(4,-3){\hskip2.5em}{$n$\\~}{$r,m$}}& 
LB &1, $m_{0}$&1, $\hat{m}_{A}$&1, $\hat{m}_{B}$ & CV &$\hat{r}^{\ast}$, $m_{0}$&$\hat{r}^{\ast}$, $\hat{m}_{A}$&$\hat{r}^{\ast}$, $\hat{m}_{B}$& 1, $\hat{m}_{A}$&1, $\hat{m}_{B}$&$\hat{r}^{\ast}$, $\hat{m}_{A}$&$\hat{r}^{\ast}$, $\hat{m}_{B}$& 1, $m_{0}$&1, $\hat{m}_{A}$&1, $\hat{m}_{B}$  &$\hat{r}^{\ast}$, $m_{0}$&$\hat{r}^{\ast}$, $\hat{m}_{A}$&$\hat{r}^{\ast}$, $\hat{m}_{B}$\\
&\multicolumn{8}{c}{versus EDF} \vline&\multicolumn{4}{c}{versus CV} \vline& \multicolumn{6}{c}{vs. parametric estimate $\widehat{F}_{\hat{m}}(x)$} with the same $\hat{m}$\\\hline\hline
\multicolumn{19}{c}{NM distribution \#1, Gaussian ($m_{0}=1$)}\\ \hline
50&-30.13&-22.41&-20.46&-21.96&-19.82&-26.93&-22.53&\textbf{-26.06}&-0.80&-2.67&-3.38&\textbf{-7.79}&\textit{23.80}&-0.03~\textbf{\ddag}&\textit{19.74}&\textit{16.60}&-2.64&\textit{13.44}\\ 
100&-27.55&-19.11&-17.89&-19.02&-17.35&-25.61&-21.76&\textbf{-25.35}&-0.65&-2.02&-5.34&\textbf{-9.68}&\textit{28.48}&\textit{5.58}&\textit{26.90}&\textit{18.16}&\textit{0.61}~\textbf{\dag}&\textit{16.98}\\ 
200&-25.47&-15.71&-15.01&-15.67&-14.51&-23.89&-20.68&\textbf{-23.75}&-0.59&-1.36&-7.22&\textbf{-10.81}&\textit{34.36}&\textit{12.52}&\textit{33.46}&\textit{21.32}&\textit{5.02}&\textit{20.68}\\ 
400&-23.77&-12.67&-12.30&-12.67&-11.85&-22.75&-20.26&\textbf{-22.73}&-0.51&-0.92&-9.54&\textbf{-12.34}&\textit{38.80}&\textit{19.63}&\textit{38.65}&\textit{22.78}&\textit{8.77}&\textit{22.67}\\ 
\hline\hline 
\multicolumn{19}{c}{NM distribution \#2, Skewed unimodal ($m_{0}=3$)}\\ \hline
50&-25.58&-18.27&-17.93&\textbf{-17.95}&-17.49&-18.16&-17.35&-17.32&-0.52&\textbf{-0.55}&\textit{0.17}~\textbf{\ddag}&\textit{0.21}~\textbf{\ddag}&-10.67&-11.65&-15.24&-10.55&-11.03&-14.59\\ 
100&-22.77&-15.79&-15.52&-15.39&-15.27&-16.65&\textbf{-15.91}&-14.51&-0.29&-0.14&\textbf{-0.76}&\textit{0.90}&-7.01&-10.13&-25.55&-7.95&-10.55&-24.77\\ 
200&-20.54&-13.15&-12.96&-12.85&-12.57&-15.54&\textbf{-15.27}&-13.94&-0.44&-0.32&\textbf{-3.09}&-1.56&-2.81&-3.44&-28.01&-5.48&-6.00&-28.91\\ 
400&-18.81&-11.25&-11.25&-11.44&-10.97&-15.03&-15.21&\textbf{-16.17}&-0.31&-0.53&-4.76&\textbf{-5.85}&\textit{0.50}&\textit{2.21}&-5.34&-3.78&-2.35&-10.40\\ 
\hline\hline 
\multicolumn{19}{c}{NM distribution \#3, Strongly skewed ($m_{0}=8$)}\\ \hline
50&-8.64&-6.78&\textbf{-6.22}&-6.12&-5.17&-6.78&-6.18&-5.77&\textbf{-1.11}&-1.00&-1.06&-0.63&-4.36&-5.91&-16.12&-4.36&-5.87&-15.81\\ 
100&-6.36&-5.31&\textbf{-4.82}&-4.75&-4.41&-5.31&-4.82&-4.51&\textbf{-0.43}&-0.36&-0.43&-0.10&-2.94&-4.37&-17.40&-2.94&-4.37&-17.19\\ 
200&-4.79&-3.84&-3.98&\textbf{-3.99}&-3.83&-3.84&-3.98&-3.87&-0.16&\textbf{-0.16}&-0.16&-0.04&-1.71&-3.41&-15.06&-1.71&-3.41&-14.96\\ 
400&-3.68&-3.32&\textbf{-3.16}&-3.16&-3.09&-3.32&-3.15&-2.95&\textbf{-0.08}&-0.08&-0.07&\textit{0.14}&-0.64&-2.07&-14.89&-0.63&-2.06&-14.70\\ 
\hline\hline 
\multicolumn{19}{c}{NM distribution \#4, Kurtotic unimodal ($m_{0}=2$)}\\ \hline
50&-9.25&-1.57&\textbf{-4.60}&\textit{0.29}~\textbf{\ddag}&-3.01&\textit{0.39}~\textbf{\dag}&-3.93&\textit{4.85}&\textbf{-1.64}&\textit{3.40}&-0.95&\textit{8.11}&\textit{12.37}&\textit{5.87}&\textit{6.80}&\textit{14.61}&\textit{6.61}&\textit{11.66}\\ 
100&-6.99&-3.41&\textbf{-4.81}&-4.06&-4.26&-2.19&-4.73&-3.17&\textbf{-0.57}&\textit{0.21}&-0.49&\textit{1.14}&\textit{24.71}&\textit{10.48}&\textit{24.59}&\textit{26.29}&\textit{10.57}&\textit{25.74}\\ 
200&-5.42&-4.20&-4.24&\textbf{-4.26}&-4.04&-4.04&-4.15&-4.16&-0.20&\textbf{-0.22}&-0.11&-0.12&\textit{32.28}&\textit{14.45}&\textit{32.01}&\textit{32.50}&\textit{14.56}&\textit{32.14}\\ 
400&-4.26&-3.54&-3.52&\textbf{-3.54}&-3.45&-2.42&-2.55&-2.42&-0.08&\textbf{-0.10}&\textit{0.93}&\textit{1.06}&\textit{33.11}&\textit{17.18}&\textit{32.74}&\textit{34.66}&\textit{18.36}&\textit{34.28}\\ 
\hline\hline 
\multicolumn{19}{c}{NM distribution \#5, Outlier ($m_{0}=2$)}\\ \hline
50&-14.32&-8.86&-9.11&\textbf{-9.40}&-9.32&-10.91&-4.93&-5.94&\textit{0.22}~\textbf{\ddag}&\textbf{-0.10}~\textbf{\ddag}&\textit{4.84}&\textit{3.72}&\textit{19.82}&-12.03&-6.89&\textit{17.13}&-7.98&-3.34\\ 
100&-12.97&-9.03&-8.59&-8.93&-8.25&-11.57&-9.84&\textbf{-11.16}&-0.37&-0.74&-1.73&\textbf{-3.16}&\textit{22.81}&\textit{8.64}&\textit{20.17}&\textit{19.38}&\textit{7.16}&\textit{17.23}\\ 
200&-11.92&-7.69&-7.44&-7.68&-7.19&-10.95&-9.77&\textbf{-10.91}&-0.28&-0.53&-2.79&\textbf{-4.02}&\textit{25.78}&\textit{12.97}&\textit{25.23}&\textit{21.33}&\textit{10.13}&\textit{20.84}\\ 
400&-11.09&-6.32&-6.17&-6.31&-5.96&-10.51&-9.56&\textbf{-10.50}&-0.22&-0.37&-3.82&\textbf{-4.82}&\textit{28.81}&\textit{17.07}&\textit{28.60}&\textit{23.04}&\textit{12.84}&\textit{22.86}\\ 
\hline\hline 
\multicolumn{19}{c}{NM distribution \#6, Bimodal ($m_{0}=2$)}\\ \hline
50&-22.66&-20.05&-19.08&\textbf{-19.78}&-19.47&-19.45&-18.28&-16.92&\textit{0.48}&\textbf{-0.38}&\textit{1.49}&\textit{3.17}&-9.00&-15.77&-20.73&-8.32&-14.93&-17.91\\ 
100&-18.25&-16.65&-15.93&\textbf{-16.47}&-15.99&-15.63&-14.99&-10.24&\textit{0.07}~\textbf{\ddag}&\textbf{-0.57}&\textit{1.19}&\textit{6.84}&-4.59&-9.70&-25.58&-3.43&-8.69&-20.03\\ 
200&-15.15&-13.70&-13.26&\textbf{-13.67}&-13.11&-13.21&-12.66&-11.14&-0.16&\textbf{-0.64}&\textit{0.52}&\textit{2.28}&\textit{0.28}~\textbf{\dag}&-3.66&-11.10&\textit{0.85}&-3.00&-8.49\\ 
400&-13.24&-11.09&-10.83&-11.08&-10.65&-11.30&-10.71&\textbf{-11.26}&-0.21&-0.49&-0.07~\textbf{\ddag}&\textbf{-0.69}&\textit{4.14}&\textit{0.45}&\textit{3.94}&\textit{3.89}&\textit{0.59}&\textit{3.73}\\ 
\hline\hline 
\multicolumn{19}{l}{~}\\[-1.5ex]
\multicolumn{19}{l}{\footnotesize\textbf{\dag}---p-value of the two-sided paired $t$-test for equality of the respective ISE means is between 1 and 5\%; \textbf{\ddag}---more than 5\%. In all other cases p-value is less than 1\%.}\\
\multicolumn{19}{l}{\footnotesize LB: best achievable reduction in relative MISE when $F$ is known, using the optimal order Gaussian-based kernel and the infeasible exact MISE-minimising bandwidth.}
\end{tabular}
\end{sidewaystable}

\renewcommand{\thetable}{\arabic{table} (Continued)}
\renewcommand{\theHtable}{\thetable.\theContdTable} 
\addtocounter{table}{-1}
\begin{sidewaystable}[htp]\centering\small
\caption{\TabRelMISEcaption}
\label{Tab:BW.RelMISE.and.pairwise.t.tests.2}
\begin{tabular}{l|rrrrrrrr|rrrr|rrrrrr}\hline\hline
\multirowcell{2}{\diaghead(4,-3){\hskip2.5em}{$n$\\~}{$r,m$}}& 
LB &1, $m_{0}$&1, $\hat{m}_{A}$&1, $\hat{m}_{B}$ & CV &$\hat{r}^{\ast}$, $m_{0}$&$\hat{r}^{\ast}$, $\hat{m}_{A}$&$\hat{r}^{\ast}$, $\hat{m}_{B}$& 1, $\hat{m}_{A}$&1, $\hat{m}_{B}$&$\hat{r}^{\ast}$, $\hat{m}_{A}$&$\hat{r}^{\ast}$, $\hat{m}_{B}$& 1, $m_{0}$&1, $\hat{m}_{A}$&1, $\hat{m}_{B}$  &$\hat{r}^{\ast}$, $m_{0}$&$\hat{r}^{\ast}$, $\hat{m}_{A}$&$\hat{r}^{\ast}$, $\hat{m}_{B}$\\
&\multicolumn{8}{c}{versus EDF} \vline&\multicolumn{4}{c}{versus CV} \vline& \multicolumn{6}{c}{vs. parametric estimate $\widehat{F}_{\hat{m}}(x)$} with the same $\hat{m}$\\\hline\hline
\multicolumn{19}{c}{NM distribution \#7, Separated bimodal ($m_{0}=2$)}\\ \hline
50&-11.51&-10.46&-9.94&\textbf{-10.34}&-9.90&-10.39&-9.91&-10.28&-0.04~\textbf{\ddag}&\textbf{-0.49}&-0.01~\textbf{\ddag}&-0.42&\textit{0.30}&-4.01&-0.40&\textit{0.37}&-3.97&-0.32\\ 
100&-9.87&-8.67&-8.35&-8.64&-8.25&-9.54&-8.60&\textbf{-9.46}&-0.11&-0.42&-0.38&\textbf{-1.32}&\textit{3.11}&-1.46&\textit{2.82}&\textit{2.13}&-1.73&\textit{1.88}\\ 
200&-8.85&-6.96&-6.77&-6.95&-6.66&-8.41&-7.63&\textbf{-8.38}&-0.12&-0.32&-1.04&\textbf{-1.85}&\textit{4.96}&\textit{0.78}&\textit{4.86}&\textit{3.32}&-0.15&\textit{3.25}\\ 
400&-8.03&-5.74&-5.65&-5.73&-5.53&-7.72&-7.22&\textbf{-7.71}&-0.13&-0.21&-1.78&\textbf{-2.30}&\textit{6.59}&\textit{2.99}&\textit{6.55}&\textit{4.35}&\textit{1.29}&\textit{4.31}\\ 
\hline\hline 
\multicolumn{19}{c}{NM distribution \#8, Skewed bimodal ($m_{0}=2$)}\\ \hline
50&-22.16&-19.32&-18.43&\textbf{-18.95}&-18.62&-18.75&-17.91&-17.50&\textit{0.23}&\textbf{-0.41}&\textit{0.87}&\textit{1.37}&-8.73&-14.47&-17.69&-8.09&-13.92&-16.22\\ 
100&-17.55&-15.79&-15.15&\textbf{-15.49}&-15.20&-14.80&-14.32&-11.00&\textit{0.05}~\textbf{\ddag}&\textbf{-0.34}&\textit{1.03}&\textit{4.95}&-3.06&-9.02&-23.05&-1.92&-8.14&-18.97\\ 
200&-13.93&-12.61&-12.28&\textbf{-12.55}&-12.22&-11.68&-11.59&-9.49&-0.07~\textbf{\dag}&\textbf{-0.38}&\textit{0.71}&\textit{3.11}&\textit{3.86}&-1.86&-10.99&\textit{4.97}&-1.08&-7.88\\ 
400&-11.20&-10.30&-10.14&\textbf{-10.30}&-10.00&-9.69&-9.59&-9.59&-0.16&\textbf{-0.34}&\textit{0.45}&\textit{0.45}&\textit{9.34}&\textit{3.81}&\textit{8.53}&\textit{10.08}&\textit{4.44}&\textit{9.38}\\ 
\hline\hline 
\multicolumn{19}{c}{NM distribution \#9, Trimodal ($m_{0}=3$)}\\ \hline
50&-21.32&-18.19&-17.92&\textbf{-18.64}&-18.54&-17.80&-17.34&-14.97&\textit{0.77}&\textbf{-0.12}~\textbf{\dag}&\textit{1.48}&\textit{4.39}&-12.25&-14.89&-24.08&-11.84&-14.29&-20.66\\ 
100&-16.84&-15.06&-14.80&\textbf{-15.45}&-15.03&-14.67&-14.29&-11.20&\textit{0.27}&\textbf{-0.50}&\textit{0.88}&\textit{4.51}&-8.79&-10.51&-22.78&-8.37&-9.96&-18.89\\ 
200&-13.30&-12.09&-11.89&\textbf{-12.30}&-11.81&-11.72&-11.33&-10.78&-0.09&\textbf{-0.55}&\textit{0.54}&\textit{1.17}&-4.91&-7.38&-11.12&-4.50&-6.79&-9.58\\ 
400&-10.62&-9.89&-9.76&\textbf{-9.94}&-9.64&-9.50&-9.19&-7.61&-0.13&\textbf{-0.33}&\textit{0.50}&\textit{2.25}&-2.10&-4.97&-12.71&-1.67&-4.37&-10.45\\ 
\hline\hline 
\multicolumn{19}{c}{NM distribution \#10, Claw ($m_{0}=6$)}\\ \hline
50&-24.28&-11.82&-15.49&-17.11&-16.33&-11.82&-16.20&\textbf{-19.41}&\textit{1.01}&-0.93&\textit{0.16}~\textbf{\ddag}&\textbf{-3.67}&-12.04&-8.88&\textit{2.27}&-12.04&-9.65&-0.56~\textbf{\ddag}\\ 
100&-15.71&-9.28&-9.80&-11.07&-10.34&-9.20&-9.80&\textbf{-11.81}&\textit{0.61}&-0.80&\textit{0.61}&\textbf{-1.63}&-8.71&-8.15&-1.79&-8.63&-8.15&-2.61\\ 
200&-7.30&-6.05&\textbf{-5.80}&-3.52&-5.23&-5.88&-5.30&\textit{0.86}&\textbf{-0.60}&\textit{1.80}&-0.07~\textbf{\ddag}&\textit{6.43}&-7.91&-7.05&-12.24&-7.75&-6.56&-8.26\\ 
400&-5.36&-3.87&\textbf{-4.52}&\textit{2.12}&\textit{3.71}~\textbf{\ddag}&-3.66&-4.31&\textit{13.05}&\textbf{-7.94}~\textbf{\ddag}&-1.53~\textbf{\ddag}&-7.74~\textbf{\ddag}&\textit{9.01}~\textbf{\ddag}&-10.95&-4.90&-25.23&-10.76&-4.69&-17.23\\ 
\hline\hline 
\multicolumn{19}{c}{NM distribution \#11, Double claw ($m_{0}=9$)}\\ \hline
50&-22.67&-18.05&-18.72&\textbf{-19.56}&-19.27&-17.67&-18.08&-16.90&\textit{0.68}&\textbf{-0.36}&\textit{1.48}&\textit{2.94}&-16.09&-15.44&-20.26&-15.70&-14.77&-17.62\\ 
100&-18.24&-15.45&-15.93&\textbf{-16.60}&-16.12&-15.02&-14.97&-10.32&\textit{0.23}&\textbf{-0.57}&\textit{1.38}&\textit{6.92}&-12.91&-10.12&-26.18&-12.47&-9.10&-20.62\\ 
200&-15.13&-12.31&-13.23&\textbf{-13.70}&-13.14&-11.98&-12.67&-11.26&-0.10&\textbf{-0.64}&\textit{0.54}&\textit{2.17}&-9.55&-3.99&-11.37&-9.21&-3.37&-8.86\\ 
400&-13.18&-10.22&-10.63&-10.90&-10.44&-10.07&-10.47&\textbf{-11.02}&-0.22&-0.51&-0.04~\textbf{\ddag}&\textbf{-0.65}&-6.20&\textit{0.12}~\textbf{\ddag}&\textit{3.79}&-6.04&\textit{0.30}&\textit{3.64}\\ 
\hline\hline 
\multicolumn{19}{c}{NM distribution \#12, Asymmetric claw ($m_{0}=6$)}\\ \hline
50&-25.13&-16.03&-16.87&-18.38&-16.77&-15.93&-16.64&\textbf{-19.65}&-0.12~\textbf{\ddag}&-1.94&\textit{0.16}~\textbf{\ddag}&\textbf{-3.47}&-13.10&-13.97&-10.14&-13.00&-13.73&-11.54\\ 
100&-17.76&-11.92&-12.41&-12.92&-12.00&-11.70&-12.04&\textbf{-12.94}&-0.47&-1.04&-0.04~\textbf{\ddag}&\textbf{-1.07}&-9.37&-12.26&-21.44&-9.14&-11.89&-21.46\\ 
200&-10.43&-9.13&\textbf{-8.86}&-8.08&-8.83&-8.91&-8.50&-4.66&\textbf{-0.04}~\textbf{\ddag}&\textit{0.83}&\textit{0.36}&\textit{4.58}&-6.66&-8.28&-32.07&-6.43&-7.92&-29.54\\ 
400&-7.33&-6.49&-6.52&-4.80&\textbf{-6.53}&-6.27&-6.36&\textit{0.85}&\textit{0.00}~\textbf{\ddag}&\textit{1.85}&\textit{0.18}&\textit{7.90}&-6.06&-4.97&-27.29&-5.83&-4.81&-22.98\\ 
\hline\hline 
\multicolumn{19}{l}{~}\\[-1.5ex]
\multicolumn{19}{l}{\footnotesize\textbf{\dag}---p-value of the two-sided paired $t$-test for equality of the respective ISE means is between 1 and 5\%; \textbf{\ddag}---more than 5\%. In all other cases p-value is less than 1\%.}\\
\multicolumn{19}{l}{\footnotesize LB: best achievable reduction in relative MISE when $F$ is known, using the optimal order Gaussian-based kernel and the infeasible exact MISE-minimising bandwidth.}
\end{tabular}
\end{sidewaystable}

\addtocounter{table}{-1}
\addtocounter{ContdTable}{1}
\begin{sidewaystable}[htp]\centering\small
\caption{\TabRelMISEcaption}
\label{Tab:BW.RelMISE.and.pairwise.t.tests.3}
\begin{tabular}{l|rrrrrrrr|rrrr|rrrrrr}\hline\hline
\multirowcell{2}{\diaghead(4,-3){\hskip2.5em}{$n$\\~}{$r,m$}}& 
LB &1, $m_{0}$&1, $\hat{m}_{A}$&1, $\hat{m}_{B}$ & CV &$\hat{r}^{\ast}$, $m_{0}$&$\hat{r}^{\ast}$, $\hat{m}_{A}$&$\hat{r}^{\ast}$, $\hat{m}_{B}$& 1, $\hat{m}_{A}$&1, $\hat{m}_{B}$&$\hat{r}^{\ast}$, $\hat{m}_{A}$&$\hat{r}^{\ast}$, $\hat{m}_{B}$& 1, $m_{0}$&1, $\hat{m}_{A}$&1, $\hat{m}_{B}$  &$\hat{r}^{\ast}$, $m_{0}$&$\hat{r}^{\ast}$, $\hat{m}_{A}$&$\hat{r}^{\ast}$, $\hat{m}_{B}$\\
&\multicolumn{8}{c}{versus EDF} \vline&\multicolumn{4}{c}{versus CV} \vline& \multicolumn{6}{c}{vs. parametric estimate $\widehat{F}_{\hat{m}}(x)$} with the same $\hat{m}$\\\hline\hline
\multicolumn{19}{c}{NM distribution \#13, Asymmetric double claw ($m_{0}=8$)}\\ \hline
50&-21.73&-16.79&-18.08&\textbf{-18.74}&-18.50&-16.42&-17.31&-15.18&\textit{0.52}&\textbf{-0.29}&\textit{1.45}&\textit{4.07}&-15.11&-15.11&-21.88&-14.74&-14.32&-18.46\\ 
100&-17.22&-14.49&-14.86&\textbf{-15.51}&-15.08&-14.19&-14.07&-9.29&\textit{0.26}&\textbf{-0.51}&\textit{1.19}&\textit{6.81}&-11.79&-9.14&-24.37&-11.47&-8.30&-18.80\\ 
200&-13.72&-11.17&-11.84&\textbf{-12.41}&-11.83&-10.96&-11.23&-10.38&-0.02~\textbf{\ddag}&\textbf{-0.66}&\textit{0.67}&\textit{1.64}&-8.49&-3.88&-6.49&-8.27&-3.21&-4.33\\ 
400&-11.29&-9.08&-9.29&\textbf{-9.62}&-9.13&-8.82&-9.00&-9.58&-0.17&\textbf{-0.54}&\textit{0.15}~\textbf{\dag}&-0.49&-5.67&-0.76&\textit{3.52}&-5.40&-0.44&\textit{3.57}\\ 
\hline\hline 
\multicolumn{19}{c}{NM distribution \#14, Smooth comb ($m_{0}=6$)}\\ \hline
50&-9.41&-8.16&-7.83&\textbf{-8.04}&-7.52&-8.16&-7.83&-8.02&-0.33&\textbf{-0.56}&-0.33&-0.54&-5.66&-5.20&-5.47&-5.66&-5.19&-5.45\\ 
100&-6.64&-6.10&-5.98&\textbf{-6.12}&-5.92&-6.09&-5.96&-6.03&-0.06&\textbf{-0.21}&-0.04~\textbf{\dag}&-0.11&-3.54&-3.16&-2.14&-3.52&-3.15&-2.04\\ 
200&-4.70&-4.62&-4.44&\textbf{-4.48}&-4.42&-4.62&-4.39&-4.25&-0.02~\textbf{\dag}&\textbf{-0.05}&\textit{0.03}&\textit{0.18}&-1.70&-1.38&-0.51&-1.69&-1.34&-0.27\\ 
400&-3.34&-3.29&\textbf{-3.24}&-3.24&-3.24&-3.23&-3.23&-3.08&\textbf{-0.01}~\textbf{\ddag}&-0.00~\textbf{\ddag}&\textit{0.01}~\textbf{\ddag}&\textit{0.16}&-0.67&\textit{0.21}&\textit{1.33}&-0.62&\textit{0.23}&\textit{1.49}\\ 
\hline\hline 
\multicolumn{19}{c}{NM distribution \#15, Discrete comb ($m_{0}=6$)}\\ \hline
50&-8.15&-6.79&\textbf{-7.12}&-7.07&-7.04&-6.79&-7.12&-6.99&\textbf{-0.09}&-0.03~\textbf{\ddag}&-0.09&\textit{0.05}~\textbf{\dag}&-4.86&-4.87&-7.38&-4.86&-4.87&-7.30\\ 
100&-6.03&-5.49&-5.43&\textbf{-5.58}&-5.47&-5.49&-5.43&-5.58&\textit{0.04}&\textbf{-0.11}&\textit{0.04}&-0.11&-2.99&-2.91&-3.00&-2.99&-2.91&-3.00\\ 
200&-4.33&-4.17&-4.06&\textbf{-4.14}&-4.02&-4.16&-4.00&-3.84&-0.04&\textbf{-0.12}&\textit{0.02}&\textit{0.19}&-1.61&-1.56&-0.30&-1.59&-1.50&\textit{0.01}~\textbf{\ddag}\\ 
400&-2.95&-2.78&\textbf{-2.86}&-2.80&-2.82&-2.55&-2.78&-2.62&\textbf{-0.03}&\textit{0.02}&\textit{0.05}&\textit{0.22}&-1.27&-0.67&\textit{0.02}~\textbf{\ddag}&-1.03&-0.59&\textit{0.21}\\ 
\hline\hline 
\multicolumn{18}{l}{~}\\[-1.5ex]
\multicolumn{18}{c}{Non-NM distributions}\\
\multicolumn{18}{l}{~}\\[-1.5ex]
\multicolumn{18}{c}{Gamma(2,1) distribution}\\ \hline
50&&~&\textbf{-15.41}&-15.28&-15.36&~&-14.12&-11.30&\textbf{-0.06}~\textbf{\ddag}&\textit{0.09}~\textbf{\ddag}&\textit{1.47}&\textit{4.80}&&&&&&\\ 
100&&~&-12.43&\textbf{-12.75}&-12.30&~&-11.74&-10.89&-0.14~\textbf{\dag}&\textbf{-0.50}&\textit{0.65}&\textit{1.61}&&&&&&\\ 
200&&~&-10.75&\textbf{-11.10}&-10.88&~&-10.23&-10.07&\textit{0.15}&\textbf{-0.25}&\textit{0.73}&\textit{0.91}&&&&&&\\ 
400&&~&-9.07&\textbf{-9.33}&-9.20&~&-8.63&-8.95&\textit{0.15}&\textbf{-0.14}&\textit{0.63}&\textit{0.28}&&&&&&\\ 
\hline\hline 
\multicolumn{18}{c}{$t_{3}$ distribution}\\ \hline
50&&~&-10.86&-10.02&\textbf{-14.16}&~&-4.61&-1.66~\textbf{\dag}&\textit{3.84}&\textit{4.82}&\textit{11.12}&\textit{14.56}&&&&&&\\ 
100&&~&-10.62&-10.07&\textbf{-12.15}&~&-5.04&-1.95~\textbf{\ddag}&\textit{1.74}&\textit{2.36}&\textit{8.09}&\textit{11.61}&&&&&&\\ 
200&&~&-10.10&-10.07&\textbf{-10.46}&~&-8.22&-7.03&\textit{0.40}~\textbf{\dag}&\textit{0.44}&\textit{2.51}&\textit{3.84}&&&&&&\\ 
400&&~&-8.95&-9.01&-8.98&~&\textbf{-9.51}&-8.70&\textit{0.03}~\textbf{\ddag}&-0.03~\textbf{\ddag}&\textbf{-0.58}~\textbf{\ddag}&\textit{0.31}~\textbf{\ddag}&&&&&&\\ 
\hline\hline 
\multicolumn{18}{c}{$t_{4}$ distribution}\\ \hline
50&&~&-14.32&-13.99&\textbf{-15.51}&~&-11.71&-10.31&\textit{1.40}&\textit{1.80}&\textit{4.50}&\textit{6.15}&&&&&&\\ 
100&&~&-12.87&-12.33&\textbf{-13.41}&~&-10.87&-7.91&\textit{0.62}&\textit{1.25}&\textit{2.94}&\textit{6.36}&&&&&&\\ 
200&&~&-11.60&-11.38&-11.68&~&\textbf{-11.87}&-9.57&\textit{0.09}~\textbf{\ddag}&\textit{0.34}&\textbf{-0.21}~\textbf{\ddag}&\textit{2.39}&&&&&&\\ 
400&&~&-9.86&-9.93&-9.87&~&\textbf{-11.08}&-10.96&\textit{0.02}~\textbf{\ddag}&-0.06~\textbf{\ddag}&\textbf{-1.33}~\textbf{\ddag}&-1.20~\textbf{\ddag}&&&&&&\\ 
\hline\hline 
\multicolumn{19}{l}{~}\\[-1.5ex]
\multicolumn{19}{l}{\footnotesize\textbf{\dag}---p-value of the two-sided paired $t$-test for equality of the respective ISE means is between 1 and 5\%; \textbf{\ddag}---more than 5\%. In all other cases p-value is less than 1\%.}\\
\multicolumn{19}{l}{\footnotesize LB: best achievable reduction in relative MISE when $F$ is known, using the optimal order Gaussian-based kernel and the infeasible exact MISE-minimising bandwidth.}
\end{tabular}
\end{sidewaystable}

\renewcommand{\thetable}{\arabic{table}}
\renewcommand{\theHtable}{\thetable} 
\setlength{\tabcolsep}{6pt} 
\renewcommand{\arraystretch}{1} 

Two-sided paired $t$-tests for equality of the considered ISE means were also performed. In the majority of cases the mean of  ISE differences is significantly different from zero at less than 1\% and, hence, only those cases where p-values are more than 1\% are labelled with \dag\ if p-value is between 1 and 5\% and with \ddag\ if p-value is more than 5\%. 

With very few exceptions, all considered methods of bandwidth selection result in estimators with significantly smaller MISE than EDF, and the reduction in MISE achieved by the best out of the feasible estimators (emphasised in bold) is very close to LB. Exceptions occur for distribution \#4 with $n=50$ and \#10 with $n=200,400$; for Student $t_{3}$ distribution with $n=50,100$  selecting the bandwidth and $\hat{r}^{\ast}$ based on $\hat{m}_{B}$ improves on EDF slightly, but not significantly so. 

Another immediate observation is that for NM distributions, with the exception of \#12 with $n=400$, CV is never the best method. CV does outperform other methods for the Student $t$ distributions in small samples though.  Unfortunately, no other bandwidth selection method is uniformly best either. Selecting the number of mixture components for the preliminary estimator by BIC is superior to the AIC-based procedure for distributions \#1, 5, 7, and---if attention is restricted to second order kernel---\#6, 8, 9, 11, and 13. AIC-based procedure delivers better results with optimal order kernels for distributions \#3, 8, 9, and Student $t_{3}$ and $t_{4}$ distributions, but in many of these cases it is better to use second order kernel.
Interestingly, knowing the true number of mixture components is not necessarily advantageous to bandwidth selection for many NM distributions. This is most notable for the kurtotic unimodal distribution with $n=50$. 

The conclusion about the benefits of using higher order kernels is similar to that reached in Section \ref{sec:analysis.of.mise}, but the potential reduction in MISE higher order kernels can confer is achieved in fewer cases. The benefits are very clear for the normal distribution, as well as for the outlier and separated bimodal distributions with $n\geq 100$. Overall, if one were to chose a single method, the combination of the BIC-based procedure and second order kernel would deliver good results.

Finally, it is worth noting that smoothing often improves on parametric normal mixture cdf estimators in terms of their MISE when  sample sizes are small, and even when the true number of mixture components is known; see last six columns in Table \ref{Tab:BW.RelMISE.and.pairwise.t.tests}. 

\section{Concluding remarks}\label{sec:conclusions}
The exact MISE expressions derived in this paper can usefully complement asymptotic analysis and simulation studies to investigate the finite sample performance of kernel estimators of a broad variety of distribution functions. In the examples considered here, the Gaussian-based kernels are found to perform well in general, and remarkably so for the regularly shaped distributions. The analysis also offers a guide on when to use higher order kernels in distribution function estimation.

As in the case of density estimation, the asymptotic approximation to MISE can be poor in finite samples, and  bandwidth selection methods based on such approximations, including simple rule of thumb bandwidths popular in applied work (see Appendix \ref{Sec:NRR.bandwidth}), should be applied with some caution.  

The normal-mixture plug-in method of jointly selecting the optimal bandwidth and  kernel order proposed in this paper offers a simple practical alternative to existing bandwidth selection procedures. Using BIC to determine the number of mixture components to fit at a preliminary stage delivers good results, but does not uniformly outperform other methods. Fine-tuning the rules for selecting the number of components is one issue that future research could usefully address.

Asymptotic optimality of arbitrary order kernels for cdf estimation remains a partially open question. A potentially fruitful approach to derive optimal smooth kernels could be to combine the results in  \citet{falk1983} and \citet{mammitzsch1984} with the restrictions on  the behaviour of $k$ at the end points of its support and the number of sign changes that $k$ has on the real line considered in \citet[Section 4]{granovsky1991}.

\addcontentsline{toc}{section}{Acknowledgements}
\section*{Acknowledgements}
The author would like to thank anonymous referees for helpful comments. 

\setlength{\bibsep}{0pt plus 0.3ex}
\small
\addcontentsline{toc}{section}{References}
\addcontentsline{toc}{section}{Appendices}
\section*{Appendices}
\appendix
\footnotesize

Throughout the  Appendices, AMPW refers to  \citet{aldershof1995}, and DLMF to the NIST Digital Library of Mathematical Functions, an online companion to \citet{Olver2010}, release 1.0.10, available at \href{http://dlmf.nist.gov/}{dlmf.nist.gov}. 

\section{Proofs}\label{App:Proofs}
\noindent\textbf{Proof of Theorem \ref{Thm:KCDFE.Exact.MISE.0}}. By definition of $K$, 
$K((x-z)/h)=\int_{-\infty}^{x}k_{h}(u-z)\mathrm{d}u$. Hence, with $\eta(u)=(k_{h}\ast f)(u)=\E[k_{h}(u-X_{1})]$, 
 interchanging the order of integration, 
$\E[\widehat{F}(x;h)]=\E[K((x-X_{1})/h)]
=\eta^{(-1)}(x)$. 
Thus, 
\begin{subequations}\begin{align}\label{Thm:KCDFE.Exact.MISE.0.Proof.ISB.lA}
\ISB[\widehat{F}(\cdot;h)] & = \int_{-\infty}^{\infty}\{\E[\widehat{F}(z;h)]-F(z)\}^{2}\mathrm{d}z 
 = \int_{-\infty}^{\infty}[\eta^{(-1)}(z)-f^{(-1)}(z)]^{2}\mathrm{d}z \\
\label{Thm:KCDFE.Exact.MISE.0.Proof.ISB.lB}
& = \int_{-\infty}^{\infty}\mathrm{d}z\int_{-\infty}^{0}\mathrm{d}x\int_{-\infty}^{0}\mathrm{d}y
[\eta(x+z)\eta(y+z)-2\eta(x+z)f(y+z)+f(x+z)f(y+z)]\\
\notag
& = \frac{1}{2}\int_{-\infty}^{\infty}\mathrm{d}w \int_{-\infty}^{0}\mathrm{d}v\int_{v}^{-v}\mathrm{d}u
\left[\eta\left(w+\frac{v+u}{2}\right)\eta\left(w+\frac{v-u}{2}\right)-2\eta\left(w+\frac{v+u}{2}\right)f\left(w+\frac{v-u}{2}\right)
\right. \\  \label{Thm:KCDFE.Exact.MISE.0.Proof.ISB.lC} & \left. 
\mkern480mu+f\left(w+\frac{v+u}{2}\right)f\left(w+\frac{v-u}{2}\right)\right] \\
\label{Thm:KCDFE.Exact.MISE.0.Proof.ISB.lD}
& = \frac{1}{2}\int_{-\infty}^{0}\mathrm{d}v\int_{v}^{-v}\mathrm{d}u~ \xi(u),
\end{align}\end{subequations}
where $\xi(u)  = \int_{-\infty}^{\infty}\left[\eta(t)\eta(t-u)-2\eta(t)f(t-u)+f(t)f(t-u)\right]\mathrm{d}t$. Line \eqref{Thm:KCDFE.Exact.MISE.0.Proof.ISB.lC} follows by rotating about the $z$-axis counterclockwise by $\pi/4$ and stretching the resultant $x$ and $y$ axes by $\sqrt{2}$.
Substituting $t=w+(v+u)/2$ gives line \eqref{Thm:KCDFE.Exact.MISE.0.Proof.ISB.lD}. Since both $f$ and $\eta$ integrate to unity over the real line, $\int_{-\infty}^{\infty}\xi(u)\mathrm{d}u=0$. Exploiting the symmetry of $k$, 
\begin{equation*}
 \int_{-\infty}^{\infty}\eta(t)f(t-u)\mathrm{d}t = \int_{-\infty}^{\infty}\int_{-\infty}^{\infty}k_{h}(s)f(t-s)f(t-u)\mathrm{d}s\mathrm{d}t
= \int_{-\infty}^{\infty}\int_{-\infty}^{\infty}k_{h}(s)f(t+u)f(t-s)\mathrm{d}s\mathrm{d}t
= \int_{-\infty}^{\infty}\eta(t)f(t+u)\mathrm{d}t;
\end{equation*}
and since $(g\star g)(u)=(g\star g)(-u)$, it follows that $\xi$ is symmetric about the origin. Therefore $\frac{1}{2}\int_{v}^{-v} \xi(u) \mathrm{d}u=-\int_{-\infty}^{v}\xi(u) \mathrm{d}u$, which gives \eqref{Thm:KCDFE.Exact.MISE.0.ISB}. 

The derivation of the integrated variance of $\widehat{F}$ follows the same steps as the derivation of the ISB. 
\begin{subequations}\begin{align}\label{Thm:KCDFE.Exact.MISE.0.Proof.IVar.lA}
\IVar[\widehat{F}(\cdot;h)] & = \frac{1}{n}\int_{-\infty}^{\infty}\left\{ \E[K((z-X_{1})/h)^{2}]- \E[K((z-X_{1})/h)]^{2} \right\}\mathrm{d}z \\
\label{Thm:KCDFE.Exact.MISE.0.Proof.IVar.lB}
& = \frac{1}{n}\int_{-\infty}^{\infty}\mathrm{d}z\int_{-\infty}^{0}\mathrm{d}x\int_{-\infty}^{0}\mathrm{d}y 
\left[  \int_{-\infty}^{\infty}k_{h}(x+z-t)k_{h}(y+z-t)f(t)\mathrm{d}t - \eta(x+z)\eta(y+z) \right]\\
\label{Thm:KCDFE.Exact.MISE.0.Proof.IVar.lC}
& = \frac{1}{2n}\int_{-\infty}^{0}\mathrm{d}v\int_{v}^{-v}\mathrm{d}u~\zeta(u),
\end{align}\end{subequations}
where $\zeta(u)  = \int_{-\infty}^{\infty} \left[  \int_{-\infty}^{\infty}k_{h}(s-t)k_{h}(s -u-t)f(t)\mathrm{d}t 
- \eta(s)\eta(s -u) \right]\mathrm{d}s
= (k_{h}\star k_{h})(u) - (\eta\star\eta)(u)$
is symmetric about the origin and integrates to zero over the real line. Therefore, 
$\IVar[\widehat{F}(\cdot;h)]=-n^{-1}\zeta^{(-2)}(0) = -hn^{-1}(k\star k)^{(-2)}(0)+n^{-1}(\eta\star\eta)^{(-2)}(0)$. 
Since $k$ is symmetric, $(k\star k)(u)=(k\ast k)(u)$. 
Finally, since $k$ integrates to unity, and $\lim_{t\to\infty}t^{2}k(t)=0$, integration by parts gives $\int_{-\infty}^{\infty}K(x)K(-x)\mathrm{d}x=\int_{-\infty}^{\infty}2xK(x)k(x)\mathrm{d}x=\psi(k)$. But, changing the order of integration, 
\begin{equation*}\tag*{$\blacksquare$}
\int_{-\infty}^{\infty}K(z)K(-z)\mathrm{d}z
= \int_{-\infty}^{0}\mathrm{d}x\int_{-\infty}^{0}\mathrm{d}y\int_{-\infty}^{\infty}k(x-z)k(y+z)\mathrm{d}z
= \int_{-\infty}^{0}\mathrm{d}x\int_{-\infty}^{0}\mathrm{d}y(k\ast k)(y+x)
= (k\ast k)^{(-2)}(0).
\end{equation*} 

\vspace*{0.5\baselineskip}

In what follows, derivatives of $\phi(x;\mu,\sigma^{2})$ with respect to $x$ are denoted by $\phi^{(r)}(x;\mu,\sigma^{2})=\phi^{(r)}\left((x-\mu)/\sigma\right)/\sigma^{r+1}$, $r=0,1,2,\ldots$. The formula is also valid for the antiderivatives, $r=-1,-2,\ldots$.
Note that even order derivatives of $\phi$ are even functions, i.e., $\phi^{(2r)}(x;z,\sigma^{2})=\phi^{(2r)}(z;x,\sigma^{2})$, $r=0,1,2,\ldots$, and $\phi^{(-2)}(x;z,\sigma^{2})= \phi^{(-2)}(z;x,\sigma^{2})+x-z$.
Repeated use is made of the results in AMPW, in particular, Corollary 5.2, \textit{viz.}
\begin{equation}\label{Eq:AMPW.C.5.2}
\int_{-\infty}^{\infty}\phi^{(r_{1})}(x;\mu_{1},\sigma_{1}^{2})\phi^{(r_{2})}(x;\mu_{2},\sigma_{2}^{2})\mathrm{d}x 
= (-1)^{r_{1}}\phi^{(r_{1}+r_{2})}(\mu_{1}-\mu_{2};0,\sigma_{1}^{2}+\sigma_{2}^{2}), 
\quad r_{1},r_{2}=0,1,2,\ldots
\end{equation}

\vspace*{0.5\baselineskip}

\noindent\textbf{Proof of Theorem \ref{Thm:Exact.MISE}}. As in Theorem \ref{Thm:KCDFE.Exact.MISE.0}, set $\eta(u)=(k_{h}\ast f)(u)$ and $\xi(u)=(\eta\star\eta)(u)-2(\eta\star f)(u)+(f\star f)(u)$. With $K(x)=G_{2r}(x)$, eq. \eqref{Eq:Def:Gaussian.Kernels.CDF}, and $f(x)$ the normal mixture density \eqref{Eq:Def:NM.dens}, for a fixed $x$ and $h>0$, by \eqref{Eq:AMPW.C.5.2},
\begin{equation}\label{A.T1.P.Expectation}
\eta(u)= \sum_{j=1}^{m}w_{j}\sum_{s=0}^{r-1}\frac{(-1)^{s}h^{2s}}{2^{s}s!}
\int_{-\infty}^{\infty}\phi^{(2s)}(z;u,h^{2})\phi(z;\mu_{j},\sigma_{j}^{2})\mathrm{d}z
=\sum_{j=1}^{m}w_{j}\sum_{s=0}^{r-1}\frac{(-1)^{s}h^{2s}}{2^{s}s!}\phi^{(2s)}(u;\mu_{j},\sigma_{j}^{2}+h^{2}).
\end{equation}
Also by \eqref{Eq:AMPW.C.5.2}, 
\begin{align}\notag
\xi(u) & = \sum_{i=1}^{m}\sum_{j=1}^{m}w_{i}w_{j}\sum_{s=0}^{r-1}\sum_{t=0}^{r-1}\frac{(-1)^{s+t}h^{2s+2t}}{2^{s+t}s!t!}\phi^{(2s+2t)}(u;\mu_{i}-\mu_{j},\sigma_{ij,2}^{2}) \\
\label{A.T1.P.ISB.xi}
&\quad -2\sum_{i=1}^{m}\sum_{j=1}^{m}w_{i}w_{j} \sum_{s=0}^{r-1}\frac{(-1)^{s}h^{2s}}{2^{s}s!}\phi^{(2s)}(u;\mu_{i}-\mu_{j},\sigma_{ij,1}^{2})
+ \sum_{i=1}^{m}\sum_{j=1}^{m}w_{i}w_{j}\phi(u;\mu_{i}-\mu_{j},\sigma_{ij,0}^{2}),
\end{align}
where the first summand is $(\eta\star\eta)(u)$, which also appears in the expression for the integrated variance. 
Eq. \eqref{Thm1.ISB} is a rearrangement of $-\xi^{(-2)}(0)$ using $\phi^{(r)}(0;\mu,\sigma^{2})=\phi^{(r)}\left(-\mu/\sigma\right)/\sigma^{r+1}$, $r=-2,-1,0,1,2,\ldots$.

The expression for $\psi(g_{2r})$ is easy to obtain using the relation $\psi(k)=(k\star k)^{(-2)}(0)$. Using the second definition of $g_{2r}$ in eq. \eqref{Eq:Def:Gaussian.Kernels.PDF}, integrating using \eqref{Eq:AMPW.C.5.2}, and substituting $\phi^{(2r)}(0)=(-1)^{r}(2\pi)^{-1/2}\OF(2r)$ (AMPW eq. 2.13, and by verification for $r=-1$), gives
\begin{equation*}
\psi(g_{2r})
=(g_{2r}\star g_{2r})^{(-2)}(0) 
= \sum_{s=0}^{r-1}\sum_{t=0}^{r-1}\frac{(-1)^{s+t}}{2^{s+t}s!t!}\phi^{(2s+2t-2)}(0;0,2)
= -\frac{1}{\sqrt{\pi}}\sum_{s=0}^{r-1}\sum_{t=0}^{r-1}\frac{\OF(2s+2t-2)}{2^{2s+2t}s!t!}.
\end{equation*}
(The same expression can be obtained from definition $\psi(g_{2r})=2\int_{-\infty}^{\infty}xG_{2r}(x)g_{2r}(x)\mathrm{d}x$ using AMPW Corollary 6.2.2.)
The second expression for $\psi(g_{2r})$  in eq. \eqref{Thm1.C1.def} can be derived by changing the summation over rows to summation over the diagonals and using $\OF(2r)=\pi^{-1/2}2^{r}\Gamma(r+1/2)$ to obtain
\begin{equation*}
\sqrt{\pi}\psi(g_{2r})
= -\frac{1}{\sqrt{\pi}}\sum_{p=0}^{2r-2}\frac{\Gamma(p-1/2)}{2^{p+1}\Gamma(p+1)}\sum_{q=\max(0,p-r+1)}^{\min(r-1,p)}\binom{p}{q}
= 1 -\frac{1}{\pi}\sum_{p=1}^{2r-2}\mathrm{B}(p-1/2,3/2)
+\frac{1}{\sqrt{\pi}}\sum_{p=r}^{2r-2}\frac{\Gamma(p-1/2)}{\Gamma(p+1)}\frac{1}{2^{p}}\sum_{q=0}^{p-r}\binom{p}{q},
\end{equation*}
where $\mathrm{B}(a,b)$ is the beta function. 
Using the integral representation of  $\mathrm{B}(a,b)$ (DLMF \href{http://dlmf.nist.gov/5.12.E1}{5.12.1}) it is easy to see that $\sum_{p=1}^{2r-2}\mathrm{B}(p-1/2,3/2)= \pi-\mathrm{B}(2r-3/2,1/2)$. Finally, substituting $I_{1/2}(r,p-r+1)=2^{-p}\sum_{q=0}^{p-r}\binom{p}{q}$ (DLMF \href{http://dlmf.nist.gov/8.17.E4}{8.17.4}, \href{http://dlmf.nist.gov/8.17.E5}{8.17.5}) gives the required expression. 

The large $r$ approximation \eqref{Thm1.C1.large.r.approx} follows by applying the Euler-Maclaurin sum formula (DLMF \href{http://dlmf.nist.gov/2.10.E1}{2.10.1}), approximating $I_{1/2}(r,s-r+1)$ by $\Phi((s-2r+1)/s^{1/2})$ (normal approximation to the binomial distribution), and expanding the ratio of gamma functions as $\Gamma(z+\alpha)/\Gamma(z+\beta)=z^{\alpha-\beta}[1+O(z^{-1})]$ (DLMF \href{http://dlmf.nist.gov/5.11.E13}{5.11.13}). This gives 
\begin{equation*}
B(r) = \sum_{s=r}^{2r-2}\frac{\Gamma(s-1/2)}{\pi\Gamma(s+1)}I_{1/2}(r,s-r+1) 
= \frac{1}{\pi}\int_{r}^{2r-2}s^{-3/2}[1+O(s^{-1})]\Phi\left(\frac{s-2r+1}{s^{1/2}}\right)\mathrm{d}s+O(r^{-3/2}).
\end{equation*}
Integrating by parts, changing the variables as $t=(s-2r+1)/s^{1/2}$, $s(t)=t^{2}/2+2r-1+t(t^{2}+8r-4)^{1/2}/2$, and expanding $s(t)^{-1/2}$ into a Taylor series around $-(2r-2)^{-1/2}$ gives the leading term as 
$B(r) = \sqrt{2}/[\pi^{3/2}(4r-3)]+O(r^{-3/2})$, and thus eq. \eqref{Thm1.C1.large.r.approx}. 
\hfill $\blacksquare$

\vspace*{0.5\baselineskip}

\noindent\textbf{Proof of Theorem \ref{Thm:KCDFE.Exact.MISE.Unif.Kernel}.} By Theorem \ref{Thm:KCDFE.Exact.MISE.0} with $k(z)=\I\{-1\leq z\leq 1\}/2$, 
$\eta(u) = \frac{1}{2}\int_{-1}^{1}f(u-ht)\mathrm{d}t$, 
$(\eta\star\eta)(u)=\frac{1}{4}\int_{-1}^{1}\int_{-1}^{1}(f\star f)(u+h(t-s))\mathrm{d}s\mathrm{d}t$, 
$(\eta\star f)(u)=\frac{1}{2}\int_{-1}^{1}(f\star f)(u+ht)\mathrm{d}t$, and with $f$ given by \eqref{Eq:Def:NM.dens}, 
$(f\star f)(u)=J(u;0)$, as in \eqref{A.T1.P.ISB.xi}. 
Hence, 
$\ISB[\widehat{F}(\cdot;h)]  = 
-\frac{1}{4}\int_{-1}^{1}\int_{-1}^{1}J(h(t-s);-2)\mathrm{d}s\mathrm{d}t
+\int_{-1}^{1}J(ht;-2)\mathrm{d}t
-J(0;-2)$.
Verifying directly that $J(0;-3)=\sigma_{f}^{2}/2$ and $J(-u;-3)=-J(u;-3)+u^{2}/2+\sigma_{f}^{2}$ gives
$\int_{-1}^{1}J(ht;-2)\mathrm{d}t = h^{-1}[J(h;-3)-J(-h;-3)]= h^{-1}[2J(h;-3)-h^{2}/2-\sigma_{f}^{2}]$ and  
\begin{equation*}
-\frac{1}{4}\int_{-1}^{1}\int_{-1}^{1}J(h(t-s);-2)\mathrm{d}s\mathrm{d}t
= -\frac{1}{4h^{2}}\int_{0}^{2h}\left[J(s;-3) -J(-s;-3)\right]\mathrm{d}s
= -\frac{1}{2h^{2}}[J(2h;-4)-J(0;-4)]
+\frac{h}{3}+\frac{\sigma_{f}^{2}}{2h}.
\end{equation*}
This gives eq. \eqref{Thm:KCDFE.Exact.MISE.Unif.Kernel.ISB}. Finally, by direct integration, $\psi(k)=1/3$,  and  eq. \eqref{Thm:KCDFE.Exact.MISE.Unif.Kernel.IVar} follows immediately. \hfill$\blacksquare$


\section{Alternative expressions for Gaussian kernels and exact MISE}\label{App:computational.aspects}
For computational reasons, especially when $r$ is large, it is convenient to express the kernels \eqref{Eq:Def:Gaussian.Kernels.CDF} and the exact MISE formulae in Theorem \ref{Thm:Exact.MISE} using the Kummer confluent hypergeometric function, $\Kummer(\alpha,\beta;z)$. Specifically, since for $s=1,2,3,\ldots$, 
$\phi^{(2s-1)}(x)=\pi^{-1}(-1)^{s}2^{s-1/2}\Gamma(s+1/2)x\Kummer(s+1/2,3/2;-x^{2}/2)$ 
(DLMF \href{http://dlmf.nist.gov/13.6.E17}{13.6.17}, \href{http://dlmf.nist.gov/13.2.E39}{13.2.39}), 
\eqref{Eq:Def:Gaussian.Kernels.CDF} can be written as 
\begin{equation}
\label{Eq:Def:Gaussian.Kernels.CDF.c}
G_{2r}(x) = \Phi(x) +\frac{x}{\sqrt{2\pi}}\sum_{s=1}^{r-1}\frac{\Gamma(s+1/2)}{\sqrt{\pi}\Gamma(s+1)}\Kummer\left(s+\frac{1}{2},\frac{3}{2};-\frac{x^{2}}{2}\right).
\end{equation}

Similarly, using $\phi^{(2s-2)}(x)/\phi^{(2s-2)}(0)=\Kummer(s-1/2,1/2;-x^{2}/2)$, $s=1,2,3,\ldots$
(DLMF \href{http://dlmf.nist.gov/13.6.E16}{13.6.16}), and changing the double summation over $s,t$ in  \eqref{Thm1.ISB} and \eqref{Thm1.IVar} to summation over diagonals, the exact MISE can be evaluated as
\begin{equation}\label{Thm1.MISE.2}
\MISE[\widehat{F}_{2r}(\cdot;h)]
 =  \frac{1}{\sqrt{2\pi}}\left(\frac{n-1}{2n}A_{2} - A_{1}\right)
-\frac{h}{n}\psi(g_{2r}) -V_{F},
\end{equation}
where $V_{F}$ is defined in eq. \eqref{Thm1.Cor2.V0}, and, with $U(h;q)$ defined in eq. \eqref{Thm1.Cor.2ndOrderKenr.U.def}, 
$R_{s}=\Gamma(s-1/2)/(\sqrt{\pi}\Gamma(s+1))$, and $\omega_{r,s} = 1-\I\{s\geq r\}2I_{1/2}(r,s-r+1)$,
\begin{align}\label{Thm1.MISE.2.A1}
A_{1} &=-2\sqrt{2\pi}U(h;1)
+\sum_{i=1}^{m}\sum_{j=1}^{m}w_{i}w_{j}\sigma_{ij,1}
 \left[\sum_{s=1}^{r-1}R_{s}\left(\frac{h^{2}}{\sigma_{ij,1}^{2}}\right)^{s}
\Kummer\left(s-\frac{1}{2},\frac{1}{2},-\frac{1}{2}\frac{(\mu_{j}-\mu_{i})^{2}}{\sigma_{ij,1}^{2}}\right)\right],\\\label{Thm1.MISE.2.A2}
A_{2}&= -2\sqrt{2\pi}U(h;2)
 + \sum_{i=1}^{m}\sum_{j=1}^{m}w_{i}w_{j}\sigma_{ij,2}
\left[\sum_{s=1}^{2r-2}R_{s}\omega_{r,s}\left(\frac{h^{2}}{\sigma_{ij,2}^{2}/2}\right)^{s}
\Kummer\left(s-\frac{1}{2},\frac{1}{2};-\frac{1}{2}\frac{(\mu_{j}-\mu_{i})^{2}}{\sigma_{ij,2}^{2}}\right)\right].
\end{align}
For the normal distribution, $m=1$, expressions \eqref{Thm1.MISE.2.A1}-\eqref{Thm1.MISE.2.A2} simplify to $V_{F}=\sigma/\sqrt{\pi}$, 
\begin{equation*}
A_{1} =\sqrt{h^{2}+2\sigma^{2}}\sum_{s=0}^{r-1}R_{s}\left(\frac{h^{2}}{h^{2}+2\sigma^{2}}\right)^{s}, \quad\text{and}\quad
A_{2}= \sqrt{2h^{2}+2\sigma^{2}}
\sum_{s=0}^{2r-2}R_{s}\omega_{r,s}\left(\frac{h^{2}}{h^{2}+\sigma^{2}}\right)^{s}.
\end{equation*}

The ratios of gamma functions can be evaluated either recursively, or as $\Gamma(a)/\Gamma(b)=\exp(\ln\Gamma(a)-\ln\Gamma(b))$ to avoid overflows with large positive $a,b$. The Kummer confluent hypergeometric function can be evaluated recursively in $s$ (DLMF \href{http://dlmf.nist.gov/13.3.E1}{13.3.1}). Thus, using the recurrence DLMF \href{http://dlmf.nist.gov/8.17.E17}{8.17.17} for the incomplete beta function appearing in $\omega_{r,s}$, the quantities $A_{1}$ and $A_{2}$  can be computed recursively in $s$. 
A MATLAB (\href{http://www.mathworks.com/}{www.mathworks.com}) implementation is available from the author upon request. Results presented in this paper were computed with Advanpix Multiprecision Computing Toolbox for MATLAB (\href{http://www.advanpix.com/}{www.advanpix.com}). In the multi-precision implementation, a backward recursion is used to compute $\omega_{r,s}$ starting with $\omega_{r,2r-1}=0$ and $\omega_{r,2r-2}=\Gamma(r-1/2)/(\sqrt{\pi}\Gamma(r))$. In the standard double precision version, it is better to use the MATLAB built-in incomplete beta function instead, as the errors accumulate fast. Same applies to computation of $\psi(g_{2r})$ in \eqref{Thm1.C1.def} and $G_{2r}$ in \eqref{Eq:Def:Gaussian.Kernels.CDF.c}.


\section{Normal reference rule bandwidth}\label{Sec:NRR.bandwidth}
In practice it is common to choose the bandwidth by a simple plug-in or reference rule, such as the normal reference rule (NRR). For example, with the second order Gaussian kernel, the asymptotically optimal bandwidth for the normal distribution with variance $\sigma^{2}$  is  $h_{a}^{\ast}=\sigma 4^{1/3}n^{-1/3}$. 

An exact MISE NRR bandwidth can be defined in a similar fashion. For the normal distribution, the exact MISE-minimising bandwidth  is of the form $h_{e}^{\ast}=\sigma h_{1}^{\ast}$, where $h_{1}^{\ast}$ is the bandwidth optimal for the standard normal distribution, which is straightforward to compute. 
Following \citet{silverman1986}, let $\zeta_{F}=\IQR_{F}/(2\Phi^{-1}(0.75))$, where $\IQR_{F}$ is the interquartile range of the distribution $F$ and $\Phi^{-1}$ is the Gaussian quantile function; $2\Phi^{-1}(0.75)\approx1.349$. 
Then the version of the NRR (or Silverman's rule of thumb) bandwidth based on the exact MISE can be defined as 
$h^{\ast}_{nrr} = \min(\sigma_{F},\zeta_{F})h_{1}^{\ast}$,
where $\sigma_{F}$ is the standard deviation of $F$. 
Analogous definition for the infinite order kernel is simply $h^{\ast}_{nrr}=\min(\sigma_{F},\zeta_{F})/\sqrt{\ln(n+1)}$. 

\begin{figure}[!htbp]\centering
\begin{tabular}{@{}>{\small}c@{\hspace{1mm}}@{\hspace{1mm}}>{\small}c@{\hspace{1mm}}@{\hspace{1mm}}>{\small}c@{}}
 \#2:  Skewed unimodal  & \#3:  Strongly skewed & \#4:  Kurtotic unimodal\\
\includegraphics[width=0.31\linewidth]{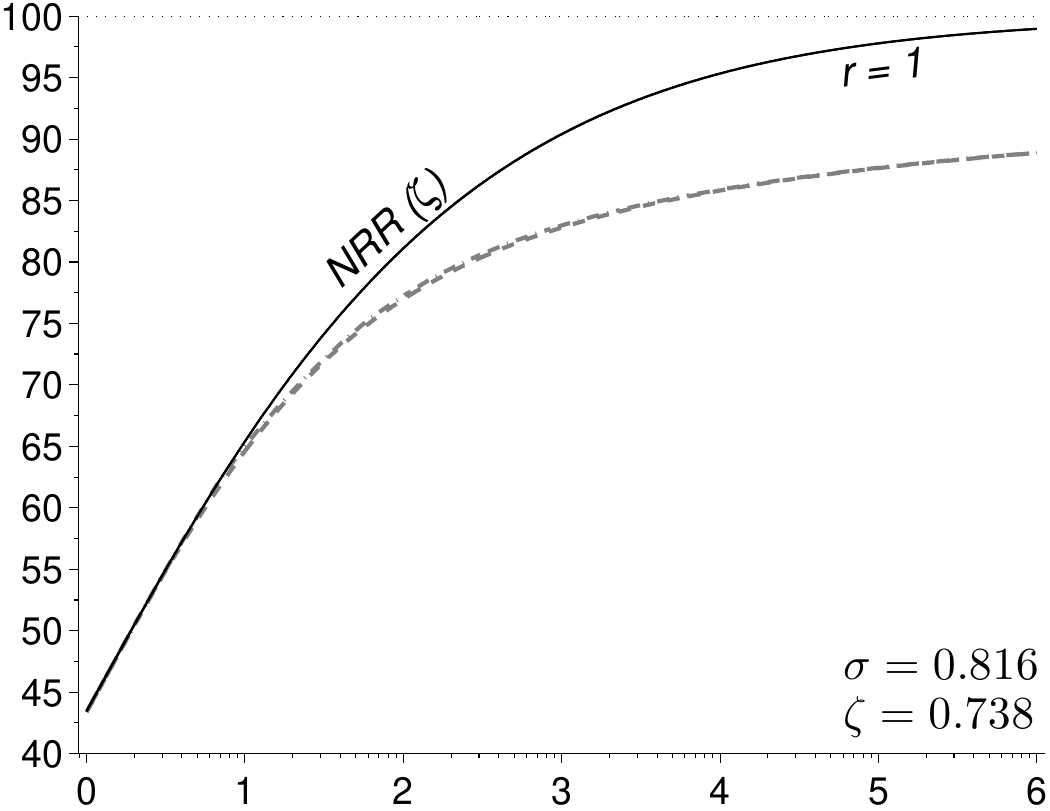}&
\includegraphics[width=0.31\linewidth]{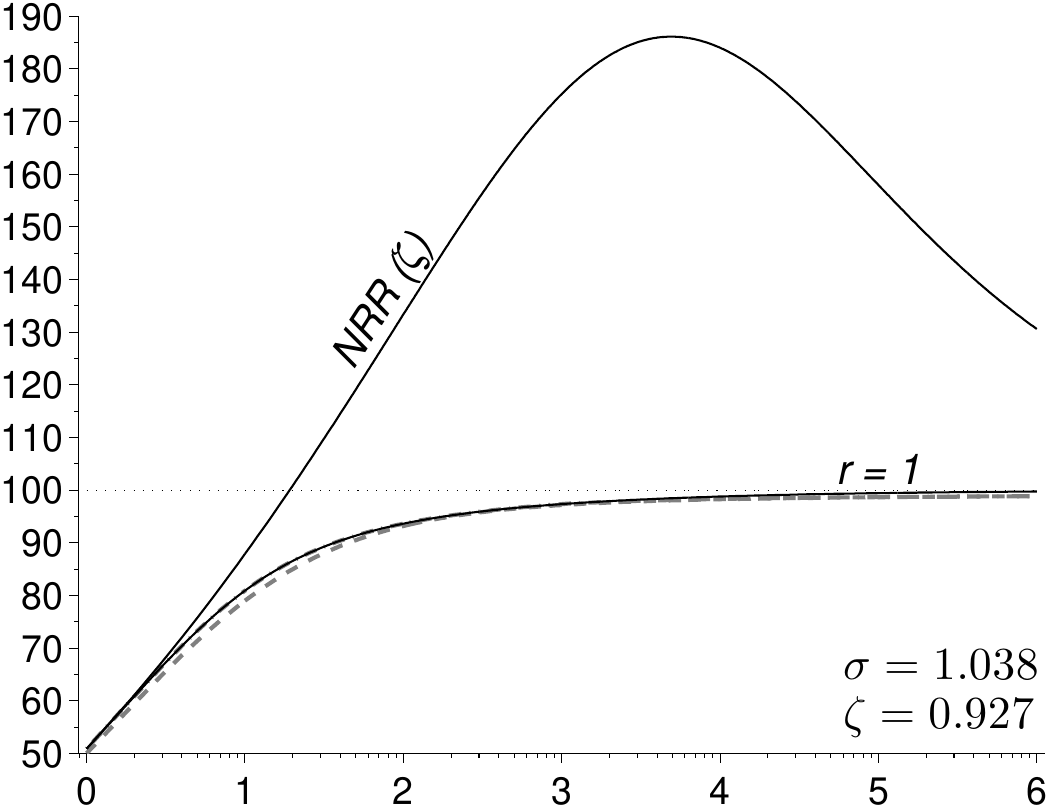}&
\includegraphics[width=0.31\linewidth]{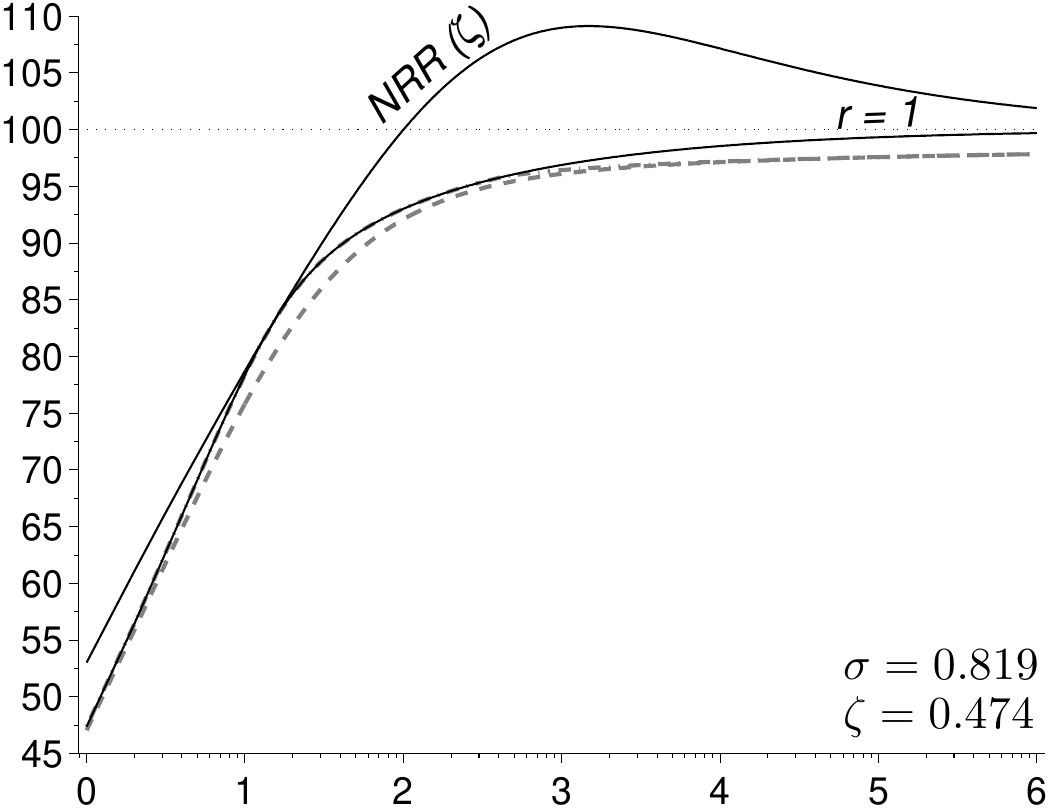}\\
\#7:  Separated bimodal     & \#14: Smooth comb      & \#15: Discrete comb\\
\includegraphics[width=0.31\linewidth]{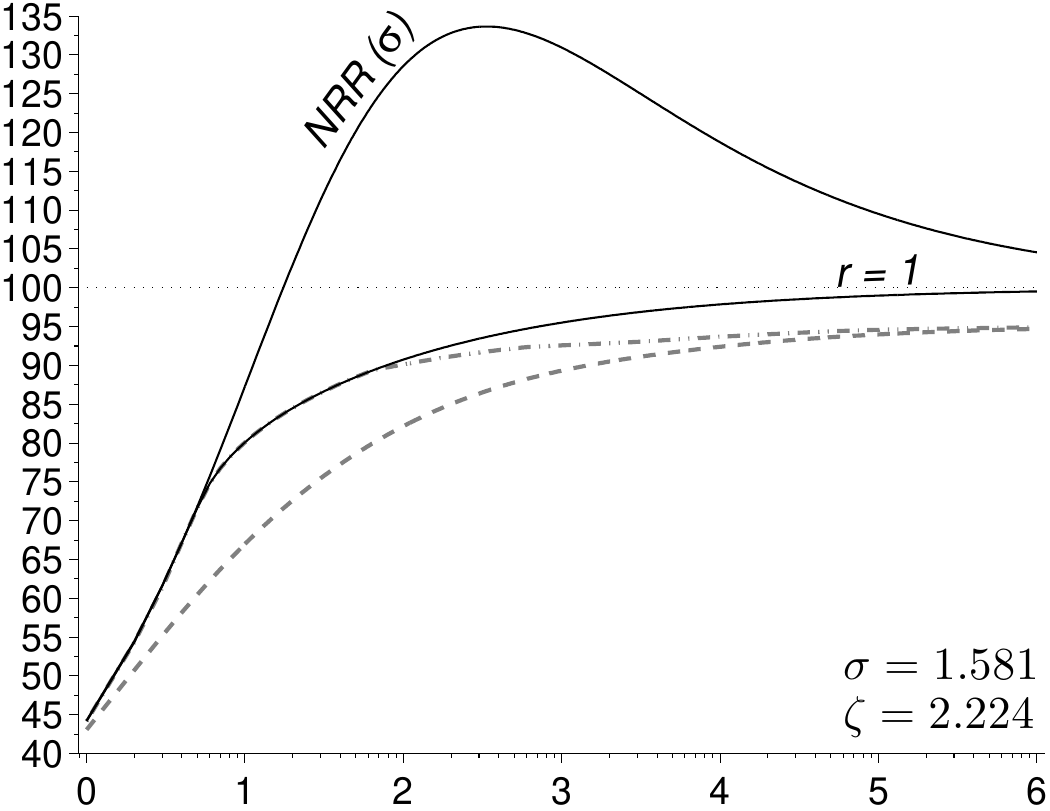}&
\includegraphics[width=0.31\linewidth]{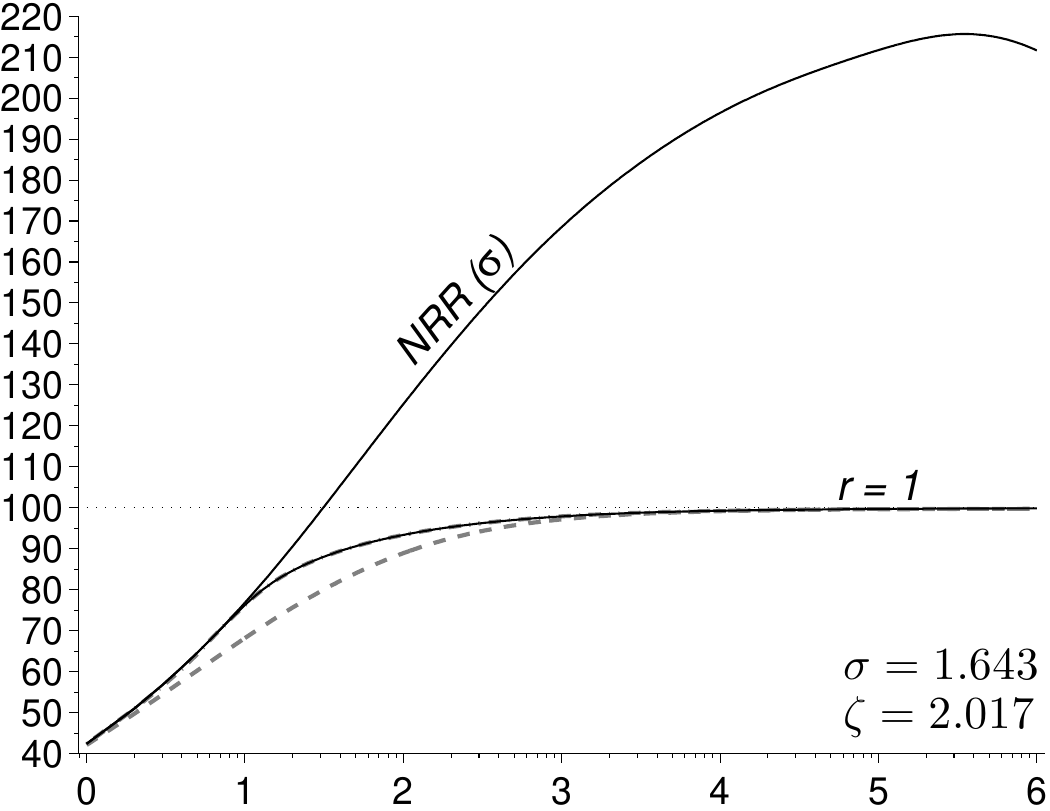}&
\includegraphics[width=0.31\linewidth]{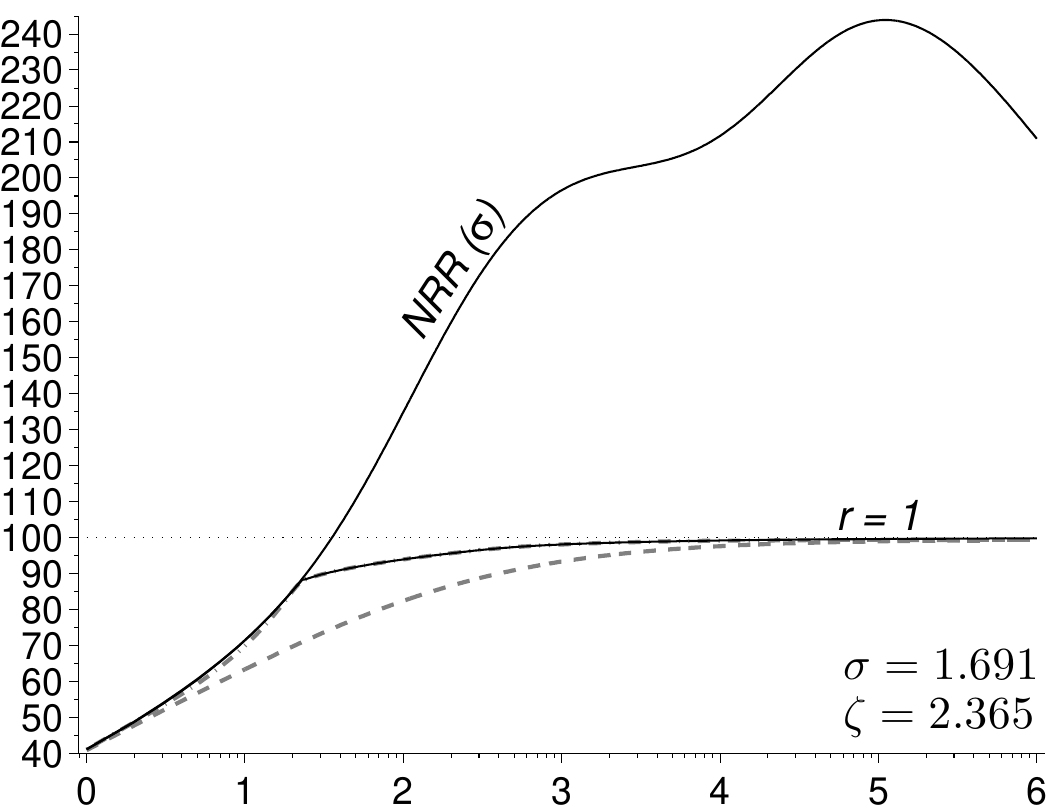}\\
\end{tabular}
\begin{flushleft}
\footnotesize     
\vspace*{-0.5\baselineskip}
\noindent\textbf{Legend.}  Horizontal axes: 
$\log_{10}(n)$. 
Vertical axes, \%: minimum relative MISE with the second order kernel and (i) the 
optimal bandwidth ($r=1$), (ii) the NRR bandwidth (the smaller of $\sigma$ and $\zeta$ in parentheses). 
Grey dashed and dash-dot lines show the $\MISE^{\ast}$ and the minimum relative $\MISE$ with the optimal order kernel as in Figure \ref{Fig:RelOptMISE.1}. 
\end{flushleft}
\vspace*{-1.2\baselineskip}
\caption{Performance of the normal reference rule bandwidth}
\label{Fig:nrrMISE}
\end{figure}

Figure \ref{Fig:nrrMISE} shows the relative MISE achievable with the second order kernel and the NRR bandwidth. (Results for the distributions not shown in Figure \ref{Fig:nrrMISE} are available upon request.) 
Using the $\min(\sigma_{F},\zeta_{F})$ rather than either $\sigma_{F}$ or $\zeta_{F}$ alone turns out to better in virtually all examples and sample sizes considered. As expected, the NRR bandwidth performs well for the moderately skewed unimodal distribution \#2 (it coincides with the optimal bandwidth for the normal distribution). Surprisingly, it also performs well for the outlier, bimodal, skewed bimodal, and trimodal distributions, as well as generally for very small sample sizes. KDFE with the NRR bandwidth will also level off with the EDF in terms of MISE asymptotically. 
However, as is clearly seen for the strongly skewed and comb-like distributions, performance in samples as large as a million observations can be  extremely poor. 

Note that in practice $\sigma_{F}$ and $\zeta_{F}$ will be replaced by estimates, thus increasing the best achievable MISE. The resultant differences in MISE can be quantified by simulation, but as the conclusion about poor performance of NRR bandwidth will remain unchanged, such a simulation is not pursued here.

Performance of the NRR bandwidth with higher order kernels and/or asymptotic NRR bandwidth is generally much worse and is therefore not shown.

\end{document}